\documentclass[orivec]{llncs}
\usepackage{amsmath,amssymb,amsthm,bbm}
\usepackage{mathpartir}
\usepackage{stmaryrd}
\usepackage{soul}
\usepackage{listings}
\usepackage{proof}
\usepackage{color}
\usepackage{hyperref}
\usepackage{float}
\usepackage{rotating}
\usepackage{mdframed}
\usepackage{enumitem}
\usepackage{xifthen}
\usepackage{tikz}
\usepackage{multirow}
\usepackage[misc]{ifsym}

\newif\ifdraft
\draftfalse

\newcommand{\rname}[1]{[\textsf{#1}]}
\newcommand{\jrhol}[7]{#1 \mid #2 \vdash #3 : #4 \sim #5 : #6 \mid #7}

\newcommand{\juhol}[5]{#1 \mid #2 \vdash #3 : #4 \mid #5}

\newcommand{\jhol}[3]{#1 \mid #2 \vdash #3}

\newcommand{\jlc}[3]{#1 \vdash #2 : #3}

\newcommand{\subst}[2]{[#2/ #1]}
\newcommand{\ltag}{_1}
\newcommand{\rtag}{_2}
\newcommand{\res}{\mathbf{r}}
\newcommand{\sem}[1]{\llbracket #1 \rrbracket}

\newcommand{\nat}{\mathbb{N}}
\newcommand{\zint}{\mathbb{Z}}
\newcommand{\real}{\mathbb{R}}

\newcommand{\bool}{\mathbb{B}}

\newcommand{\cons}[2]{#1\! ::\! #2}

\newcommand{\casenat}[3]{{\rm case}\ #1\ {\sf of}\ 0 \mapsto #2 ; S \mapsto #3}
  
\newcommand{\defsubst}[2]{\subst{\res\ltag}{#1}\subst{\res\rtag}{#2}}

\newcommand{\defeq}{\triangleq}

\newcommand{\reftype}[3]{\ifthenelse{\isempty{#1}}{\{\res : #2 \mid #3\}}{\{#1 : #2 \mid #3\}}}
\newcommand{\ertype}[2]{\ifthenelse{\isempty{#2}}{\lfloor #1 \rfloor}{\lfloor #1 \rfloor(#2)}}
\newcommand{\errtype}[2]{\ifthenelse{\isempty{#2}}{\llfloor #1 \rrfloor}{\llfloor #1 \rrfloor(#2)}}

\newcommand{\emfor}[2]{\ifthenelse{\isempty{#2}}{#1}{#1\subst{\res}{#2}}}
\newcommand{\emterm}[2]{\ifthenelse{\isempty{#2}}{#1}{#1\subst{\res}{#2}}}

\ifdraft
\newcommand{\aag}[1]{\textit{\color{blue}[AA]: #1}}
\newcommand{\gb}[1]{\textit{\color{red}[GB]: #1}}
\newcommand{\mg}[1]{\textit{\color{cyan}[MG]: #1}}
\newcommand{\dg}[1]{\textit{\color{brown}[DG]: #1}}
\newcommand{\ales}[1]{\textit{\color{green}[AB]: #1}}
\newcommand{\lb}[1]{\textit{\color{green}[LB]: #1}}
\else
\newcommand{\aag}[1]{}
\newcommand{\gb}[1]{}
\newcommand{\mg}[1]{}
\newcommand{\dg}[1]{}
\newcommand{\ales}[1]{}
\newcommand{\lb}[1]{}
\fi

\theoremstyle{definition}
\newtheorem{thm}{Theorem}
\newtheorem{lem}[thm]{Lemma}

\DeclareDocumentCommand{\later}{ o m}{
  \IfNoValueTF{#1}
  {\triangleright #2}
  {\triangleright\hrt{#1} . #2}}
\DeclareDocumentCommand{\latern}{ o m}{
  \IfNoValueTF{#1}
  {\triangleright #2}
  {\triangleright #1 . #2}}

\newcommand{\hrt}[1]{\left[ #1 \right]}
\newcommand{\dsubst}[3]{\triangleright[#1 \leftarrow #2].#3}
\newcommand{\nextt}[2]{\triangleright[#1].#2}
\newcommand{\prev}[1]{{\rm prev}\ #1}
\newcommand{\boxx}[1]{{\rm box}\ #1}
\newcommand{\jguhol}[7]{#1 \mid #2 \mid #3 \mid #4 \vdash #5 : #6 \mid #7}
\newcommand{\jguholsc}[7]{\Omega \vdash #5 : #6 \mid #7}
\newcommand{\letbox}[3]{{\rm letb}\ #1 \leftarrow #2\ {\rm in}\ #3}

\newcommand{\letconst}[3]{{\rm letc}\ #1 \leftarrow #2\ {\rm in}\ #3}
\newcommand{\conshat}[2]{#1\hat{::}#2}
\newcommand{\fix}[2]{{\rm fix}\ #1.\ #2}

\newcommand{\app}{\circledast}
\newcommand{\jgrhol}[9]{#1 \mid #2 \mid #3 \mid #4 \vdash #5 : #6 \sim #7 : #8 \mid #9}
\newcommand{\jgrholsc}[9]{\Omega \vdash #5 : #6 \sim #7 : #8 \mid #9}
\newcommand{\jgrholnoc}[5]{\vdash #1 : #2 \sim #3 : #4 \mid #5}
\newcommand{\jgrholnocnot}[3]{\vdash #1 \sim #2 \mid #3}
\newcommand{\jghol}[5]{#1 \mid #2 \mid #3 \mid #4 \vdash #5}
\newcommand{\jgholsc}[5]{\Omega \vdash #5}
\newcommand{\conj}{\mathrel{\wedge}}
\newcommand{\trees}{\mathcal{S}}
\newcommand{\ot}{\leftarrow}

\newcommand{\To}{\Rightarrow}

\newcommand{\inj}[1]{{\rm inj}_{#1}}
\newcommand{\CASE}{\operatorname{\mathrm{case}}}
\newcommand{\OF}{\operatorname{\mathrm{of}}}

\newcommand{\hd}{\operatorname{\mathrm{hd}}}
\newcommand{\tl}{\operatorname{\mathrm{tl}}}

\newcommand{\Distr}{\mathsf{D}}
\newcommand{\Str}[1]{\ifstrempty{#1}{\operatorname{\mathrm{Str}}}{\operatorname{\mathrm{Str}}_{#1}}}
\newcommand{\munit}[1]{\mathsf{munit}(#1)}
\newcommand{\mlet}[3]{\mathsf{let}~#1=#2~\mathsf{in}~#3}
\newcommand{\carac}[1]{\mathbbm{1}_{#1}}
\newcommand{\supp}{\mathsf{supp}}

\newcommand{\markov}{\operatorname{\mathsf{markov}}}

\newcommand{\isetsep}{\;\ifnum\currentgrouptype=16 \middle\fi|\;}

\newcommand{\bern}[1]{\mathcal{B}(#1)}
\newcommand{\unif}[1]{\mathcal{U}_{#1}}
\newcommand{\rfrac}[2]{{}^{#1}\!/\!_{#2}}
\newcommand{\All}{\operatorname{All}}
\newcommand{\coupling}[3]{#2 \mathrel{\mathcal{L}\left(#1\right)} #3}
\newcommand{\coupl}[3]{\diamond_{#1, #2}. #3}

\newcommand{\shrinkcaption}{\vspace{-1em}}

\begin{document}

\title{Relational Reasoning for Markov Chains in a Probabilistic Guarded Lambda Calculus}
\author{Alejandro Aguirre\inst{1}$^\text{(\Letter)}$
        \and Gilles Barthe\inst{1}
        \and Lars Birkedal\inst{2}
        \and Ale\u s Bizjak\inst{2}
        \and \\ Marco Gaboardi\inst{3}
        \and Deepak Garg\inst{4}}
\institute{IMDEA Software Institute
           \\ \texttt{alejandro.aguirre@imdea.org}
           \and Aarhus University
           \and University at Buffalo, SUNY
           \and MPI-SWS}


\maketitle

\begin{abstract}
We extend the simply-typed guarded $\lambda$-calculus with discrete
probabilities and endow it with a program logic for reasoning about
relational properties of guarded probabilistic computations.  This
provides a framework for programming and reasoning about infinite
stochastic processes like Markov chains.  We demonstrate the logic
sound by interpreting its judgements in the topos of trees and by
using probabilistic couplings for the semantics of relational
assertions over distributions on discrete types.

The program logic is designed to support syntax-directed proofs in the
style of relational refinement types, but retains the expressiveness
of higher-order logic extended with discrete distributions, and the
ability to reason relationally about expressions that have different
types or syntactic structure. In addition, our proof system leverages
a well-known theorem from the coupling literature to justify better
proof rules for relational reasoning about probabilistic
expressions. We illustrate these benefits with a broad range of
examples that were beyond the scope of previous systems, including
shift couplings and lump couplings between random walks.
\end{abstract}

\section{Introduction}

Stochastic processes are often used in mathematics, physics, biology
or finance to model evolution of systems with uncertainty. In
particular, Markov chains are ``memoryless'' stochastic processes, in
the sense that the evolution of the system depends only on the current
state and not on its history. Perhaps the most emblematic example of a
(discrete time) Markov chain is the simple random walk over the
integers, that starts at 0, and that on each step moves one position
either left or right with uniform probability.  Let $p_i$ be the
position at time $i$. Then, this Markov chain can be described as:
\[
     p_0 = 0 \quad\quad
     p_{i+1} = \begin{cases}
       p_i + 1\ \text{with probability}\ 1/2 \\
       p_i - 1\ \text{with probability}\ 1/2
     \end{cases}
\]

The goal of this paper is to develop a programming and reasoning
framework for probabilistic computations over infinite objects, such
as Markov chains. Although programming and reasoning frameworks for
infinite objects and probabilistic computations are well-understood in
isolation, their combination is challenging. In particular, one must
develop a proof system that is powerful enough for proving interesting
properties of probabilistic computations over infinite objects, and
practical enough to support effective verification of these
properties.

\paragraph*{Modelling probabilistic infinite objects}
A first challenge is to model probabilistic infinite objects. We focus
on the case of Markov chains, due to its importance. A (discrete-time)
Markov chain is a sequence of random variables $\{X_i\}$ over some
fixed type $T$ satisfying some independence property. Thus, the
straightforward way of modelling a Markov chain is as a \emph{stream
  of distributions} over $T$. Going back to the simple example
outlined above, it is natural to think about this kind of
\emph{discrete-time} Markov chain as characterized by the sequence of
positions $\{p_i\}_{i \in \nat}$, which in turn can be described as an
infinite set indexed by the natural numbers. This suggests that a
natural way to model such a Markov chain is to use \emph{streams} in
which each element is produced \emph{probabilistically} from the
previous one. However, there are some downsides to this
representation. First of all, it requires explicit reasoning about
probabilistic dependency, since $X_{i+1}$ depends on $X_i$. Also, we
might be interested in global properties of the executions of the
Markov chain, such as ``The probability of passing through the initial
state infinitely many times is 1''. These properties are naturally
expressed as properties of the whole stream.  For these reasons, we
want to represent Markov chains as \emph{distributions over
  streams}. Seemingly, one downside of this representation is that the
set of streams is not countable, which suggests the need for
introducing heavy measure-theoretic machinery in the semantics of the
programming language, even when the underlying type is discrete or
finite.

Fortunately, measure-theoretic machinery can be avoided
(for discrete distributions) by developing a probabilistic extension
of the simply-typed guarded $\lambda$-calculus and giving a semantic
interpretation in the topos of trees~\cite{CBGB16}. Informally, the
simply-typed guarded $\lambda$-calculus~\cite{CBGB16} extends the
simply-typed lambda calculus with a \emph{later} modality, denoted by
$\later$. The type $\later{A}$ ascribes expressions that are available
one unit of logical time in the future. The $\later$ modality allows
one to model infinite types by using \lq\lq
finite\rq\rq\ approximations. For example, a stream of natural numbers
is represented by the sequence of its (increasing) prefixes in the
topos of trees. The prefix containing the first $i$ elements has the
type $S_i \defeq \nat \times \later{\nat} \times \ldots \times
\later^{(i-1)}{\nat}$, representing that the first element is
available now, the second element a unit time in the future, and so
on. This is the key to representing probability distributions over
infinite objects without measure-theoretic semantics: We model
probability distributions over non-discrete sets as discrete
distributions over their (the sets') approximations. For example, a
distribution over streams of natural numbers (which a priori would
be non-discrete since the set of streams is uncountable) would be
modelled by a \emph{sequence of distributions} over the finite
approximations $S_1, S_2, \ldots$ of streams. Importantly, since each
$S_i$ is countable, each of these distributions can be
discrete.

\dg{As Alejandro also noted after his example that I commented out,
  it's unclear what's so specific or special about the $\later$
  modality. It seems that we could use \emph{any} approximating
  sequence of discrete sets to avoid the issue with
  measure-theory. What's special about $\later$ and guards in our
  context?}
\lb{the use of $\later$ and guards just allow us to use types to
  ensure productivity (and contractiveness to ensure existence of
  fixed points)}
\mg{My understanding is that there is nothing special but we need in the logic to have a way to talk about these approximations, and
  $\later$ and guards gives us a natural way to do this. no?}

\paragraph*{Reasoning about probabilistic computations}
Probabilistic computations exhibit a rich set of properties. One
natural class of properties is related to probabilities of events,
saying, for instance, that the probability of some event $E$ (or of an
indexed family of events) increases at every iteration. However,
several interesting properties of probabilistic computation, such as
stochastic dominance or convergence (defined below) are relational, in
the sense that they refer to two runs of two processes. In principle,
both classes of properties can be proved using a higher-order logic
for probabilistic expressions, e.g.\, the internal logic of the topos
of trees, suitably extended with an axiomatization of finite
distributions. However, we contend that an alternative approach
inspired from refinement types is desirable and provides better
support for effective verification. More specifically, reasoning in a
higher-order logic, e.g.\, in the internal logic of the topos of trees,
does not exploit the \emph{structure of programs} for non-relational
reasoning, nor the \emph{structural similarities} between programs for
relational reasoning. As a consequence, reasoning is more involved.
To address this issue, we define a relational proof system that
exploits the structure of the expressions and supports syntax-directed
proofs, with necessary provisions for escaping the syntax-directed
discipline when the expressions do not have the same structure. The
proof system manipulates judgements of the form:
\begin{equation*}
\jgrhol{\Delta}{\Sigma}{\Gamma}{\Psi}{t_1}{A_1}{t_2}{A_2}{\phi}
\label{eq:judgement-2}
\end{equation*}
where $\Delta$ and $\Gamma$ are two typing contexts, $\Sigma$ and
$\Psi$ respectively denote sets of assertions over variables in these
two contexts, $t_1$ and $t_2$ are well-typed expressions of type $A_1$
and $A_2$, and $\phi$ is an assertion that may contain the special
variables $\res\ltag$ and $\res\rtag$ that respectively correspond to
the values of $t_1$ and $t_2$. The context $\Delta$ and $\Gamma$, the
terms $t_1$ and $t_2$ and the types $A_1$ and $A_2$ provide a
specification, while $\Sigma$, $\Psi$, and $\phi$ are useful for
reasoning about relational properties over $t_1,t_2$, their inputs and
their outputs. This form of judgement is similar to that of Relational
Higher-Order Logic \cite{ABGGS17}, from which our system draws inspiration.

In more detail, our relational logic comes with typing rules that
allow one to reason about relational properties by exploiting as much
as possible the syntactic similarities between $t_1$ and $t_2$, and to
fall back on pure logical reasoning when these are not available. In
order to apply relational reasoning to guarded computations the logic
provides relational rules for the later modality $\later{}$ and for a
related modality $\square{}$, called ``constant''. These rules allow
the relational verification of general relational properties that go
beyond the traditional notion of program equivalence and, moreover,
they allow the verification of properties of guarded computations over
different types. The ability to reason about computations of different
types provides significant benefits over alternative formalisms for
relational reasoning. For example, it enables reasoning about
relations between programs working on different data structures,
e.g. a relation between a program working on a stream of natural
numbers, and a program working on a stream of pairs of natural
numbers, or having different structures, e.g. a relation between
an application and a case expression.

Importantly, our approach for reasoning formally about probabilistic
computations is based on \emph{probabilistic couplings}, a standard
tool from the analysis of Markov chains~\cite{Lindvall02,Thorisson00}.
From a verification perspective, probabilistic couplings go beyond
equivalence properties of probabilistic programs, which have been
studied extensively in the verification literature, and yet support
compositional reasoning~\cite{BartheEGHSS15,BartheGHS17}. The main
attractive feature of coupling-based reasoning is that it limits the
need of explicitly reasoning about the probabilities---this avoids
complex verification conditions. We provide sound proof rules for
reasoning about probabilistic couplings. Our rules make several
improvements over prior relational verification logics based on
couplings. First, we support reasoning over probabilistic processes of
different types. Second, we use Strassen's
theorem~\cite{strassen1965existence} a remarkable result about
probabilistic couplings, to achieve greater expressivity. Previous
systems required to prove a bijection between the sampling spaces to
show the existence of a coupling~\cite{BartheEGHSS15,BartheGHS17},
Strassen's theorem gives a way to show their existence which is
applicable in settings where the bijection-based approach cannot be
applied.  And third, we support reasoning with what are called shift
couplings, coupling which permits to relate the states of two Markov
chains at possibly different times (more explanations below).


\paragraph*{Case studies}
We show the flexibility of our formalism by verifying several examples
of relational properties of probabilistic computations, and Markov
chains in particular. These examples cannot be verified with existing
approaches.

First, we verify a classic example of probabilistic non-interference
which requires the reasoning about computations at different
types. Second, in the context of Markov chains, we verify an example
about stochastic dominance which exercises our more general rule for
proving the existence of couplings modelled by expressions of
different types. Finally, we verify an example involving shift
relations in an infinite computation. This style of reasoning is
motivated by \lq\lq shift\rq\rq\ couplings in Markov chains. In
contrast to a standard coupling, which relates the states of two
Markov chains at the same time $t$, a shift coupling relates the
states of two Markov chains at possibly different times. Our specific
example relates a standard random walk (described earlier) to a
variant called a lazy random walk; the verification requires relating
the state of standard random walk at time $t$ to the state of the lazy
random walk at time $2t$. We note that this kind of reasoning is
impossible with conventional relational proof rules even in a
non-probabilistic setting. Therefore, we provide a novel family of
proof rules for reasoning about shift relations. At a high level, the
rules combine a careful treatment of the later and constant modalities
with a refined treatment of fixpoint operators, allowing us to relate
different iterates of function bodies.

\subsection*{Summary of contributions}
With the aim of providing a general framework for programming and
reasoning about Markov chains, the three main contributions of this work are:
\begin{enumerate}
  \item A probabilistic extension of the guarded $\lambda$-calculus,
    that enables the definition of Markov chains as discrete
    probability distributions over streams.
  \item A relational logic based on coupling to reason in a
    syntax-directed manner about (relational) properties of Markov
    chains. This logic supports reasoning about programs that have
    different types and structures. Additionally, this logic uses results from the
    coupling literature to achieve greater expressivity than previous
    systems.
  \item An extension of the relational logic that allows to relate the
    states of two streams at possibly different times. This extension
    supports reasoning principles, such as shift couplings, that escape
    conventional relational logics.
  \end{enumerate}

\section{Mathematical preliminaries}
This section reviews the definition of discrete probability
sub-distributions and introduces mathematical couplings.
\begin{definition}[Discrete probability distribution]
Let $C$ be a discrete (i.e., finite or countable) set. A (total)
distribution over $C$ is a function $\mu : C \to [0,1]$ such that $
\sum_{x\in C} \mu(x) = 1 .$ The support of a distribution $\mu$ is the
set of points with non-zero probability, $ \supp\ \mu \defeq \{x \in C
\mid \mu(x) > 0 \} .$ We denote the set of distributions over $C$ as
$\Distr(C)$. Given a subset $E \subseteq C$, the probability of
sampling from $\mu$ a point in $E$ is denoted $\Pr_{x\leftarrow \mu}[x
  \in E]$, and is equal to $\sum_{x \in E} \mu(x)$.
\end{definition}

\begin{definition}[Marginals]
Let $\mu$ be a distribution over a product space $C_1\times C_2$. The
first (second) marginal of $\mu$ is another distribution
$\Distr(\pi_1)(\mu)$ ($\Distr(\pi_2)(\mu)$) over $C_1$ ($C_2$) defined
as:
\[\Distr(\pi_1)(\mu)(x) = \sum_{y \in C_2} \mu(x,y) \qquad
  \left(\Distr(\pi_2)(\mu)(y) = \sum_{x \in C_1} \mu(x,y) \right)
\]
\end{definition}

\subsubsection{Probabilistic couplings}
Probabilistic couplings are a fundamental tool in the analysis of
Markov chains. When analyzing
a relation between two probability distributions it is sometimes useful to
consider instead a distribution over the product space that somehow ``couples''
the randomness in a convenient manner.

Consider for instance the case of the following Markov chain, which
counts the total amount of tails observed when
tossing repeatedly a biased coin with probability of tails $p$:
\[
     n_0 = 0 \quad\quad
     n_{i+1} = \left\{\begin{array}{l}
       n_i + 1\ \text{with probability}\ p \\
       n_i\ \text{with probability}\ (1-p)
     \end{array}\right.
\]
If we have two biased coins with probabilities of tails $p$ and $q$ with $p\leq q$ and we
respectively observe $\{n_i\}$ and $\{m_i\}$ we would expect that, in
some sense, $n_i \leq m_i$ should hold for all $i$ (this property is
known as stochastic dominance).  A formal proof of this fact using
elementary tools from probability theory would require to compute the
cumulative distribution functions for $n_i$ and $m_i$ and then to
compare them. The coupling method reduces this proof to showing a way
to pair the coin flips so that if the first coin shows tails, so does
the second coin.

%
%

We now review the definition of couplings and state relevant
properties.
\begin{definition}[Couplings]
Let $\mu_1\in\Distr(C_1)$ and $\mu_2\in\Distr(C_2)$, and 
$R\subseteq C_1\times C_2$.
\begin{itemize}
\item A distribution $\mu\in\Distr(C_1\times C_2)$ is a coupling for
  $\mu_1$ and $\mu_2$ iff its first and second marginals coincide with
  $\mu_1$ and $\mu_2$ respectively, i.e.\, $\Distr(\pi_1)(\mu)=\mu_1$
  and $\Distr(\pi_2)(\mu)=\mu_2$.

\item A distribution $\mu\in\Distr(C_1\times C_2)$ is a $R$-coupling
  for $\mu_1$ and $\mu_2$ if it is a coupling for $\mu_1$ and $\mu_2$
  and, moreover, $\Pr_{(x_1,x_2)\leftarrow \mu} [R~x_1~x_2]=1$, i.e., if
  the support of the distribution $\mu$ is included in $R$.
\end{itemize}
Moreover, we write $\coupl{\mu_1}{\mu_2}{R}$ iff there exists a
$R$-coupling for $\mu_1$ and $\mu_2$.

\end{definition}
Couplings always exist. For instance, the product distribution of two
distributions is always a coupling.
%
%
Going back to the example about the two coins, it can be proven by computation
that the following is a coupling that lifts the less-or-equal relation
($0$ indicating heads and $1$ indicating tails):
\[\left\{\begin{array}{lll}
   &(0,0) \ \text{w/ prob}\ (1-q)\quad\quad
   &(0,1) \ \text{w/ prob}\ (q-p) \\
   &(1,0) \ \text{w/ prob}\ 0\quad\quad
   &(1,1) \ \text{w/ prob}\ p
\end{array}\right.\]

The following theorem in~\cite{strassen1965existence} gives a
necessary and sufficient condition for the existence of $R$-couplings
between two distributions. The theorem is remarkable in the sense that
it proves an equivalence between an existential property (namely the
existence of a particular coupling) and a universal property
(checking, for each event, an inequality between probabilities).
\begin{theorem}[Strassen's theorem]
  Consider $\mu_1\in\Distr(C_1)$ and $\mu_2\in\Distr(C_2)$, and $R\subseteq
  C_1 \times C_2$. Then $\coupl{\mu_1}{\mu_2}{R}$ iff for every
  $X \subseteq C_1$, $\Pr_{x_1\leftarrow \mu_1}[x_1\in X] \leq
  \Pr_{x_2\leftarrow \mu_2}[x_2\in R(X)]$, where $R(X)$ is the image
  of $X$ under $R$, i.e.\, $R(X) =\{ y \in C_2 \mid \exists x \in
  X.~R~x~y\}$.
\end{theorem}


An important property of couplings is closure under sequential
composition.


\begin{lemma}[Sequential composition couplings]\label{lem:sequential-composition-of-couplings}
Let $\mu_1\in\Distr(C_1)$, $\mu_2\in\Distr(C_2)$, $M_1:C_1\rightarrow
\Distr(D_1)$ and $M_2:C_2\rightarrow \Distr(D_2)$. Moreover, let
$R\subseteq C_1\times C_2$ and $S\subseteq D_1\times D_2$. Assume:
$(1)$ $\coupl{\mu_1}{\mu_2}{R}$; and
$(2)$ for every $x_1\in C_1$ and $x_2\in C_2$ such that $R~x_1~x_2$,
  we have $\coupl{M_1(x_1)}{M_2(x_2)}{S}$.
Then $\coupl{(\mathsf{bind}~\mu_1~M_1)}{(\mathsf{bind}~\mu_2~M_2)}{S}$,
where $\mathsf{bind}~\mu~M$ is defined as
$$(\mathsf{bind}~\mu~M)(y) =\sum_x \mu(x) \cdot M(x)(y)$$

\end{lemma}
We conclude this section with the following lemma, which follows from
Strassen's theorem:
\begin{lemma}[Fundamental lemma of couplings]\label{lem:fun-coup}
Let $R\subseteq C_1 \times C_2$, $E_1\subseteq C_1$ and $E_2\subseteq
C_2$ such that for every $x_1\in E_1$ and $x_2\in C_2$, $R~x_1~x_2$
implies $x_2\in E_2$, i.e.\, $R(E_1)\subseteq E_2$. Moreover, let
$\mu_1\in\Distr(C_1)$ and $\mu_2\in\Distr(C_2)$ such that
$\coupl{\mu_1}{\mu_2}{R}$.  Then
$$\Pr_{x_1\leftarrow \mu_1} [x_1\in E_1] \leq \Pr_{x_2\leftarrow\mu_2} [x_2\in E_2]$$
\end{lemma}
This lemma can be used to prove probabilistic inequalities
from the existence of suitable couplings:

\begin{corollary}\label{cor:fundamental}
Let $\mu_1,\mu_2\in\Distr(C)$:
\begin{enumerate}
  \item If $\coupl{\mu_1}{\mu_2}{(=)}$, then for all $x\in C$, $\mu_1(x) = \mu_2(x)$.
  \item If $C = \nat$ and $\coupl{\mu_1}{\mu_2}{(\geq)}$, then for all $n\in \nat$,
    $\Pr_{x\leftarrow \mu_1}[x\geq n] \geq \Pr_{x\leftarrow \mu_2}[x\geq n]$
\end{enumerate}
\end{corollary}

In the example at the beginning of the section, the property we want
to prove is precisely that, for every $k$ and $i$, the following
holds:
\[
\Pr_{x_1\leftarrow n_i} [x_1 \geq k] \leq \Pr_{x_2\leftarrow m_i} [x_2 \geq k]
\]
Since we have a $\leq$-coupling, this proof is immediate.
This example is formalized in \autoref{sec:proba-ex}.

\section{Overview of the system}
\label{sec:overview}

In this section we give a high-level overview of our system, with the
details on sections \ref{sec:syntax}, \ref{sec:ghol} and \ref{sec:grhol}.
We start by presenting the base logic, and then we show how to extend it with
probabilities and how to build a relational reasoning system on top of it.

\subsection{Base logic: Guarded Higher-Order Logic}

Our starting point is the Guarded Higher-Order Logic~\cite{CBGB16}
(Guarded HOL)
inspired by the topos of
trees. In addition to the usual constructs of HOL to reason about lambda
terms, this logic features the $\later$ and $\square$ modalities to reason
about infinite terms, in particular streams. The $\later$ modality is used
to reason about objects that will be available in the future, such as tails
of streams. For instance, suppose we want to define an $\All(s,\phi)$ predicate, expressing that
all elements of a stream $s \equiv \cons{n}{xs}$ satisfy a property $\phi$. This can be axiomatized as
follows:
\[\forall (xs: \later \Str{\nat}) (n : \nat). \phi\ n \Rightarrow \later[s\ot
  xs]{\All(s, x. \phi)} \Rightarrow \All(\cons{n}{xs}, x. \phi)\]
We use $x. \phi$ to denote that the formula $\phi$ depends on a free variable
$x$, which will get replaced by the first argument of $\All$.
We have two antecedents. The first one states that the head $n$ satisfies
$\phi$. The second one, $\later[s\ot xs]{\All(s, x. \phi)}$, states that all
elements of $xs$ satisfy $\phi$. Formally, $xs$ is the tail of the stream and will
be available in the future, so it has type $\later\Str{\nat}$.
The \emph{delayed substitution} $\triangleright[s\ot xs]$ replaces $s$ of type
$\Str{\nat}$ with $xs$ of type $\later\Str{\nat}$ inside $\All$ and shifts the
whole formula one step into the future.
In other words, $\later[s\ot xs]{\All(s, x. \phi)}$
states that $\All(-, x.\phi)$ will be satisfied by $xs$ in the future, once
it is available.



\subsection{A system for relational reasoning}
\label{sec:overview:grhol}

When proving relational properties it
is often convenient to build proofs guided by the syntactic structure
of the two expressions to be related. This style of reasoning is
particularly appealing when the two expressions have the same
structure and control-flow, and is appealingly close to the
traditional style of reasoning supported by refinement types. At the
same time, 
a strict
adherence to the syntax-directed discipline is detrimental to the
expressiveness of the system; for instance, it makes it difficult or
even impossible to reason about structurally dissimilar terms. To
achieve the best of both worlds, we present a relational proof system built
on top of Guarded HOL, which we call Guarded RHOL. Judgements have the shape:
\[\jgrhol{\Delta}{\Sigma}{\Gamma}{\Psi}{t_1}{A_1}{t_2}{A_2}{\phi}\]
where $\phi$ is a logical formula that may contain two distinguished
variables $\res\ltag$ and $\res\rtag$ that respectively represent the
expressions $t_1$ and $t_2$. This judgement subsumes two typing
judgements on $t_1$ and $t_2$ and a relation $\phi$ on these two
expressions. However, this form of judgement does not tie the logical
property to the type of the expressions, and is key to achieving
flexibility while supporting syntax-directed proofs whenever needed.
The proof system combines rules of two different flavours: two-sided
rules, which relate expressions with the same top-level constructs,
and one-sided rules, which operate on a single expression.

We then extend Guarded HOL with a modality $\diamond$
that lifts assertions over discrete types $C_1$ and $C_2$ to assertions over
$\Distr(C_1)$ and $\Distr(C_2)$.
%
%
Concretely, we define for every assertion $\phi$, variables
$x_1$ and $x_2$ of type $C_1$ and $C_2$ respectively, and expressions
$t_1$ and $t_2$ of type $\Distr(C_1)$ and $\Distr (C_2)$ respectively,
the modal assertion $\diamond_{ [x_1\leftarrow t_1,x_2\leftarrow t_2]}
\phi$ which holds iff the interpretations of $t_1$ and $t_2$ are
related by the probabilistic lifting of the interpretation of
$\phi$. We call this new logic Probabilistic Guarded HOL.

We accordingly extend the relational proof system to support reasoning
about probabilistic expressions by adding judgements of the form:
\[\jgrhol{\Delta}{\Sigma}{\Gamma}{\Psi}{t_1}{\Distr(C_1)}
          {t_2}{\Distr(C_2)}{
            \diamond_{[x_1\leftarrow \res\ltag, x_2\leftarrow \res\rtag]} \phi}\]
expressing that $t_1$ and $t_2$ are distributions related by a
$\phi$-coupling. We call this proof system Probabilistic Guarded RHOL.
These judgements can be built by using the following rule,
that lifts relational judgements over discrete types
$C_1$ and $C_2$ to judgements over distribution types $\Distr(C_1)$
and $\Distr(C_2)$ when the premises of Strassen's theorem are
satisfied.
\[
\infer[\sf COUPLING]{
       \jgrhol{\Delta}{\Sigma}{\Gamma}{\Psi}{t_1}{\Distr(C_1)}{t_2}{\Distr(C_2)}{
         \diamond_{[y_1\leftarrow \res_1, y_2\leftarrow
             \res_2]}\phi}}{
       \jghol{\Delta}{\Sigma}{\Gamma}{\Psi}{\forall X_1 \subseteq C_1.
        \Pr_{y_1\ot t_1} [y_1\in X_1] \leq \Pr_{y_2\ot t_2} [\exists y_1 \in X_1. \phi]}}
\]
%

Recall that (discrete time) Markov chains are ``memoryless''
probabilistic processes, whose specification is given by a (discrete)
set $C$ of states, an initial state $s_0$ and a probabilistic
transition function $\mathsf{step}:C \rightarrow \Distr(C)$, where
$\Distr(S)$ represents the set of discrete distributions over $C$.  As
explained in the introduction, a convenient modelling of Markov chains
is by means of probabilistic streams, i.e.\, to model a Markov chain
as an element of $\Distr(\Str{S})$, where $S$ is its underlying state
space. To model Markov chains, we introduce a $\markov$ operator with
type $C \to (C \to \Distr(C)) \to \Distr(\Str{C})$ that, given an
initial state and a transition function, returns a Markov chain. We
can reason about Markov chains by the \rname{Markov} rule (the context, omitted,
does not change):
\begin{gather*}
\inferrule*[right=\sf Markov]
      {\jgrholnoc{t_1}{C_1}{t_2}{C_2}{\phi} \\\\
       \jgrholnoc{h_1}{C_1 \to \Distr(C_1)}{h_2}{C_2 \to \Distr(C_2)}{\psi_3} \\\\
         \vdash \psi_4}
      {\jgrholnoc{\operatorname{markov}(t_1,h_1)}{\Distr(\Str{D_1})}{\operatorname{markov}(t_2,h_2)}{\Distr(\Str{D_2})}
              {\diamond_{\left[\substack{y_1 \ot \res\ltag \\ y_2 \ot \res\rtag}\right]}\phi'}}
\\[0.5em]
\text{ where }
\begin{cases}
  \psi_3 \equiv \forall x_1 x_2. \phi\defsubst{x_1}{x_2} \Rightarrow
     \diamond_{[ y_1 \ot \res\ltag\ x_1, y_2 \ot \res\rtag\ x_2]}\phi\defsubst{y_1}{y_2} \\
  \psi_4 \equiv \forall x_1\ x_2\ xs_1\ xs_2. \phi\defsubst{x_1}{x_2}
      \Rightarrow \later[y_1 \ot xs_1, y_2 \ot xs_2]{\phi'} \Rightarrow \\
      \quad\quad\quad\phi'\subst{y_1}{\cons{x_1}{xs_1}}\subst{y_2}{\cons{x_2}{xs_2}}
\end{cases}
\end{gather*}
Informally, the rule stipulates the existence of an invariant $\phi$
over states. The first premise insists that the invariant hold on the
initial states, the condition $\psi_3$ states that the transition
functions preserve the invariant, and $\psi_4$ states that the
invariant $\phi$ over pairs of states can be lifted to a stream
property $\phi'$.

Other rules of the logic are given in
Figure~\ref{fig:phol}. 
%
The language construct $\mathsf{munit}$ creates a point distribution
whose entire mass is at its argument. Accordingly, the [\textsf{UNIT}] rule creates a
straightforward coupling.
The [\textsf{MLET}] rule internalizes sequential composition of
couplings (Lemma~\ref{lem:sequential-composition-of-couplings}) into
the proof system. The construct $\mlet{x}{t}{t'}$ composes a
distribution $t$ with a probabilistic computation $t'$ with one free
variable $x$ by sampling $x$ from $t$ and running $t'$.
The [\textsf{MLET-L}] rule supports one-sided reasoning about
$\mlet{x}{t}{t'}$ and relies on the fact that couplings are closed
under convex combinations. Note that one premise of the rule uses a
unary judgement, with a non-relational modality
$\diamond_{[x\leftarrow \res]} \phi$ whose informal meaning is that
$\phi$ holds with probability $1$ in the distribution $\res$.

The following table summarizes the different base logics we consider,the
relational systems we build on top of them, including the ones presented in~\cite{ABGGS17}, and the equivalences between both sides:
\begin{center}
\small
\begin{tabular}{lcl}
Relational logic  & & Base logic \\ \hline
&&\\[-2mm]
  $\begin{array}{l}\text{RHOL~\cite{ABGGS17}} \\
                   \Gamma \mid \Psi \vdash t_1 \sim t_2 \mid \phi \end{array} $
                    & $\stackrel{\text{\cite{ABGGS17}}}{\Longleftrightarrow}$ 	&
  $\begin{array}{l} \text{HOL~\cite{ABGGS17}} \\
  \Gamma \mid \Psi \vdash \phi\defsubst{t_1}{t_2} \end{array} $ \\[1em]

  $\begin{array}{l} \text{Guarded RHOL~\S\ref{sec:grhol}} \\
                   \Delta \mid \Sigma \mid \Gamma \mid \Psi \vdash t_1 \sim t_2 \mid \phi \end{array} $
                    & $\stackrel{\text{Thm~\ref{thm:equiv-rhol-hol}}}{\Longleftrightarrow}$ 	&
  $\begin{array}{l} \text{Guarded HOL~\cite{CBGB16}} \\
    \Delta \mid \Sigma \mid \Gamma \mid \Psi \vdash \phi\defsubst{t_1}{t_2} \end{array} $ \\[1em]

  $\begin{array}{l} \text{Probabilistic Guarded RHOL~\S\ref{sec:grhol}} \\
    \Delta \mid \Sigma \mid \Gamma \mid \Psi \vdash t_1 \sim t_2 \mid
    \diamond_{[y_1 \ot \res\ltag, y_2\ot\res\rtag]}.\phi \end{array} $
                    & \hspace{.3cm}
                      $\stackrel{\text{Thm~\ref{thm:equiv-rhol-hol}}}{\Longleftrightarrow}$
                       \hspace{.3cm}	&
  $\begin{array}{l} \text{Probabilistic Guarded HOL~\S\ref{sec:ghol}} \\
    \Delta \mid \Sigma \mid \Gamma \mid \Psi \vdash
    \diamond_{[y_1 \ot t_1, y_2 \ot t_2]}.\phi \end{array} $
\end{tabular}
\end{center}
%

\begin{figure*}[!tb]
\small
\begin{gather*}
 \infer[\sf UNIT]{\jgrhol{\Delta}{\Sigma}{\Gamma}{\Psi}{\munit{t_1}}{\Distr(C_1)}
          {\munit{t_2}}{\Distr(C_2)}{
            \diamond_{[x_1\leftarrow \res\ltag, x_2\leftarrow \res\rtag]} \phi}} 
      {\jgrhol{\Delta}{\Sigma}{\Gamma}{\Psi}{t_1}{C_1}{t_2}{C_2}{\phi
          [\res\ltag/x_1, \res\rtag/x_2]}}
\\[0.2em]
\infer[\sf MLET]{
      \jgrhol{\Delta}{\Sigma}{\Gamma}{\Psi}{\mlet{x_1}{t_1}{t'_1}}{
      \Distr(D_1)}{\mlet{x_2}{t_2}{t'_2}}{\Distr(D_2)}{\diamond_{\left[\substack{y_1\leftarrow \res_1 \\ y_2\leftarrow \res_2}\right]} \psi}}
    {\begin{array}{c}\jgrhol{\Delta}{\Sigma}{\Gamma}{\Psi}{t_1}{\Distr(C_1)}{t_2}{\Distr(C_2)}{
     \diamond_{[x_1\leftarrow \res_1, x_2\leftarrow \res_2]} \phi} \\
     \jgrhol{\Delta}{\Sigma}{\Gamma,x_1:C_1,x_2:C_2}{\Psi,\phi
     }{t'_1}{\Distr(D_1)} {t'_2}{\Distr(D_2)}{\diamond_{[y_1\leftarrow \res_1, y_2\leftarrow \res_2]} \psi}
    \end{array}}
\\[0.2em]
 \infer[\sf MLET-L]{
   \jgrhol{\Delta}{\Sigma}{\Gamma}{\Psi}{\mlet{x_1}{t_1}{t'_1}}{\Distr(D_1)}{t'_2}{\Distr(D_2)}{
     \diamond_{[y_1\leftarrow \res_1, y_2\leftarrow \res_2]} \psi}}
     {\begin{array}{c}
         \jguhol{\Delta}{\Sigma}{\Gamma}{\Psi}{t_1}{\Distr(C_1)}{\diamond_{[x\leftarrow \res]} \phi}
         \\
     \jgrhol{\Delta}{\Sigma}{\Gamma,x_1:C_1}{\Psi,\phi
     }{t'_1}{\Distr(D_1)}{t'_2}{\Distr(D_2)}{\diamond_{[y_1\leftarrow \res_1, y_2\leftarrow \res_2]} \psi}
     \end{array}}
\end{gather*}     
\shrinkcaption
\caption{Proof rules for probabilistic constructs}\label{fig:phol}
\end{figure*}

\subsection{Examples}\label{sec:proba-ex}
We formalize elementary examples from the literature on security and
Markov chains. None of these examples can be verified in prior
systems. Uniformity of \emph{one-time pad} and lumping of \emph{random
  walks} cannot even be stated in prior systems because the two
related expressions in these examples have different types. The
\emph{random walk vs lazy random walk} (shift coupling) cannot be
proved in prior systems because it requires either asynchronous
reasoning or code rewriting.  Finally, the \emph{biased coin example}
(stochastic dominance) cannot be proved in prior work because it
requires Strassen's formulation of the existence of coupling (rather
than a bijection-based formulation) or code rewriting.  We give
additional details below. 

\subsubsection{One-time pad/probabilistic non-interference}
Non-interference~\cite{GoguenM82} is a baseline information flow
policy that is often used to model confidentiality of computations. In
its simplest form, non-interference distinguishes between public (or
low) and private (or high) variables and expressions, and requires
that the result of a public expression not depend on the value of its
private parameters. This definition naturally extends to probabilistic
expressions, except that in this case the evaluation of an expression
yields a distribution rather than a value. There are deep connections
between probabilistic non-interference and several notions of
(information-theoretic) security from cryptography. In this paragraph,
we illustrate different flavours of security properties for one-time
pad encryption. Similar reasoning can be carried out for proving
(passive) security of secure multiparty computation algorithms in the
3-party or multi-party setting~\cite{BogdanovNTW12,Cramer:2015:SMC}.

One-time pad is a perfectly secure symmetric encryption scheme. Its
space of plaintexts, ciphertexts and keys is the set
$\{0,1\}^\ell$---fixed-length bitstrings of size $\ell$. The
encryption algorithm is parametrized by a key $k$---sampled uniformly
over the set of bitstrings $\{ 0,1 \}^\ell$---and maps every plaintext
$m$ to the ciphertext $c = k \oplus m$, where the operator $\oplus$
denotes bitwise exclusive-or on bitstrings. We let $\mathsf{otp}$
denote the expression $\lambda
m. \mlet{k}{\mathcal{U}_{\{0,1\}^\ell}}{\munit{k\oplus m}}$, where
$\mathcal{U}_{X}$ is the uniform distribution over a finite set $X$.

One-time pad achieves perfect security, i.e.\, the distributions of
ciphertexts is independent of the plaintext. Perfect security can be
captured as a probabilistic non-interference property: 
$$\jgrholnoc{\mathsf{otp}}{\{ 0,1 \}^\ell \rightarrow \Distr(\{ 0, 1
  \}^\ell)}{\mathsf{otp}}{\{ 0,1 \}^\ell \rightarrow \Distr(\{ 0, 1 \}^\ell)}{
  \forall m_1m_2.
  \res\ltag~m_1 \stackrel{\diamond}{=} \res\rtag~m_2
}$$
where $e_1 \stackrel{\diamond}{=} e_2$ is used as a shorthand for
$\diamond_{[y_1\leftarrow e_1, y_2\leftarrow e_2]} y_1 = y_2$. The
crux of the proof is to establish
$$m_1,m_2: \{ 0,1 \}^\ell
\jgrholnoc{\mathcal{U}_{\{0,1\}^\ell}}{\Distr(\{ 0, 1 \}^\ell)}{
  \mathcal{U}_{\{0,1\}^\ell}}{\Distr(\{ 0, 1 \}^\ell)}{
\res\ltag \oplus m_2 \stackrel{\diamond}{=} \res\rtag \oplus m_1
}$$
using the [\textsf{COUPLING}] rule. It suffices to observe that the
assertion induces a bijection, so the image of an arbitrary set $X$
under the relation has the same cardinality as $X$, and hence their
probabilities w.r.t.\, the uniform distributions are equal. One can
then conclude the proof by applying the rules for monadic
sequenciation (\rname{MLET}) and abstraction (rule \rname{ABS} in appendix), using
algebraic properties of $\oplus$.

Interestingly, one can prove a stronger property: rather than proving
that the ciphertext is independent of the plaintext, one can prove
that the distribution of ciphertexts is uniform. This is captured by
the following judgement: 
$$c_1, c_2: \{ 0,1 \}^\ell
\jgrholnoc{\mathsf{otp}}{\{ 0,1 \}^\ell \rightarrow \Distr(\{ 0, 1 \}^\ell)}
{\mathsf{otp}}{\{ 0,1 \}^\ell \rightarrow \Distr(\{ 0, 1 \}^\ell)}{\psi}$$
where
$\psi\defeq \forall m_1\,m_2. m_1=m_2\Rightarrow
  \diamond_{[y_1\leftarrow \res\ltag~m_1, y_2\leftarrow \res\rtag ~m_2]} 
  y_1=c_1  \Leftrightarrow y_2=c_2$.
This style of modelling uniformity as a relational property is
inspired from~\cite{BartheEGHS17}. The proof is similar to the
previous one and omitted. However, it is arguably more natural to
model uniformity of the distribution of ciphertexts by the judgement:
$$\jgrholnoc{\mathsf{otp}}{\{ 0,1 \}^\ell \rightarrow \Distr(\{ 0, 1 \}^\ell)}
{\mathcal{U}_{\{0,1\}^\ell}}{\Distr(\{ 0, 1 \}^\ell)}{
  \forall m.~
\res\ltag~m \stackrel{\diamond}{=} \res\rtag
}$$
This judgement is closer to the simulation-based notion of security
that is used pervasively in cryptography, and notably in Universal
Composability~\cite{canetti2001universally}. Specifically, the
statement captures the fact that the one-time pad algorithm can be
simulated without access to the message.  It is interesting to note
that the judgement above (and more generally simulation-based security)
could not be expressed in prior works, since the two expressions of
the judgement have different types---note that in this specific case,
the right expression is a distribution but in the general case the
right expression will also be a function, and its domain will be a
projection of the domain of the left expression.

The proof proceeds as follows. First, we prove
$$
\jgrholnocnot{\mathcal{U}_{\{0,1\}^\ell}}{
\mathcal{U}_{\{0,1\}^\ell}}{
\forall m.~\diamond_{[y_1\leftarrow \res\ltag, y_2\leftarrow \res\rtag]}
  y_1 \oplus m = y_2}$$
using the [\textsf{COUPLING}] rule. Then, we apply the [\textsf{MLET}]
rule to obtain
$$\jgrholnocnot
  {\begin{array}{l}\mlet{k}{\mathcal{U}_{\{0,1\}^\ell}}{\\ \munit{k\oplus m}} \end{array}}
  {\begin{array}{l}\mlet{k}{\mathcal{U}_{\{0,1\}^\ell}}{\\ \munit{k}}\end{array}}
     {\diamond_{\left[y_1\leftarrow \res\ltag,
      y_2\leftarrow \res\rtag \right]} y_1 = y_2}$$
We have $\mlet{k}{\mathcal{U}_{\{0,1\}^\ell}}{\munit{k}} \equiv
\mathcal{U}_{\{0,1\}^\ell}$; hence by equivalence (rule
\rname{Equiv} in appendix), this entails
$$
\jgrholnocnot{\mlet{k}{\mathcal{U}_{\{0,1\}^\ell}}{\munit{k\oplus m}}}
             {\mathcal{U}_{\{0,1\}^\ell}}{\diamond_{[y_1\leftarrow
                   \res\ltag, y_2\leftarrow \res\rtag]} y_1 = y_2}$$
We conclude by applying the one-sided rule for abstraction.

\subsubsection{Stochastic dominance}
Stochastic dominance defines a partial order between random variables
whose underlying set is itself a partial order; it has many different
applications in statistical biology (e.g.\ in the analysis of the
birth-and-death processes), statistical physics (e.g.\ in percolation
theory), and economics. First-order stochastic dominance, which we
define below, is also an important application of probabilistic
couplings. We demonstrate how to use our proof system for proving
(first-order) stochastic dominance for a simple Markov process which
samples biased coins. While the example is elementary, the proof
method extends to more complex examples of stochastic dominance, and
illustrates the benefits of Strassen's formulation of the coupling
rule over alternative formulations stipulating the existence of
bijections (explained later).

We start by recalling the definition of (first-order) stochastic
dominance for the $\mathbb{N}$-valued case. The definition extends to
arbitrary partial orders.
\begin{definition}[Stochastic dominance]
Let $\mu_1,\mu_2\in \Distr(\mathbb{N})$. We say that $\mu_2$
stochastically dominates $\mu_1$, written $\mu_1\leq_{\mathrm{SD}}
\mu_2$, iff for every $n\in\mathbb{N}$,
$$\Pr_{x\leftarrow \mu_1}[x\geq n] \leq \Pr_{x\leftarrow \mu_2}[x\geq n]$$
\end{definition}
The following result, equivalent to \autoref{cor:fundamental},
characterizes stochastic dominance using probabilistic couplings.
\begin{proposition}
Let $\mu_1,\mu_2\in \Distr(\mathbb{N})$. Then $\mu_1\leq_{\mathrm{SD}}
\mu_2$ iff $\coupl{\mu_1}{\mu_2}{(\leq)}$.
\end{proposition}

We now turn to the definition of the Markov chain. For $p\in [0,1]$,
we consider the parametric $\mathbb{N}$-valued Markov chain
$\mathsf{coins} \defeq \markov(0,h)$, with initial state $0$ and
(parametric) step function:
$$ h \defeq \lambda x. \mlet{b}{\bern{p}}{\munit{x+b}}
$$
where, for $p \in [0,1]$, $\bern{p}$ is the Bernoulli distribution on
$\{0,1\}$ with probability $p$ for $1$ and $1-p$ for $0$. Our goal is
to establish that $\mathsf{coins}$ is monotonic, i.e.\, for every
$p_1,p_2\in [0,1]$, $p_1\leq p_2$ implies $\mathsf{coins}~p_1
\leq_{\mathrm{SD}} \mathsf{coins}~p_2$. We formalize this statement as
$$
\jgrholnoc{\mathsf{coins}}{[0,1] \rightarrow \Distr(\Str{\mathbb{N}})}
          {\mathsf{coins}}{[0,1] \rightarrow \Distr(\Str{\mathbb{N}})}
          {\psi}
          $$
where $\psi\defeq \forall p_1,p_2. p_1\leq p_2 \Rightarrow
\diamond_{[y_1 \ot \res\ltag, y_2 \ot \res\rtag]} \All(y_1, y_2, z_1.z_2.z_1\leq z_2)$.
The crux of the proof is to establish stochastic dominance for the
Bernoulli distribution:
$$
p_1:[0,1],p_2:[0,1]\mid p_1\leq p_2 \jgrholnoc{\bern{p_1}}
          {\Distr(\mathbb{N})}
          {\bern{p_2}}
          {\Distr(\mathbb{N})}
          {\res\ltag\stackrel{\diamond}{\leq} \res\rtag}
$$
%
where we use $e_1 \stackrel{\diamond}{\leq} e_2$ as shorthand for
$\diamond_{[y_1 \ot e_1, y_2 \ot e_2]} y_1\leq y_2$. This is proved
directly by the \rname{COUPLING} rule and checking by simple
calculations that the premise of the rule is valid.

We briefly explain how to conclude the proof. Let $h_1$ and $h_2$ be
the step functions for $p_1$ and $p_2$ respectively. It is clear from
the above that (context omitted):
$$
      x_1\leq x_2 \jgrholnoc{h_1\ x_1}{\Distr(\bool)}
         {h_2\ x_2}{\Distr(\bool)}
         {\diamond_{[y_1 \ot \res\ltag, y_2 \ot \res\rtag]}.
           {y_1\leq y_2}}
$$
and by the definition of $\All$:
$$x_1 \leq x_2 \Rightarrow {\All(xs_1,xs_2,z_1.z_2.z_1\leq z_2)}
\Rightarrow \All(\cons{x_1}{\later xs_1}, \cons{x_2}{\later xs_2}, z_1.z_2.z_1\leq z_2)$$
So, we can conclude by applying the \rname{Markov} rule.

It is instructive to compare our proof with prior formalizations, and
in particular with the proof in \cite{BartheEGHSS15}. Their proof is
carried out in the \textsf{pRHL} logic, whose [\textsf{COUPLING}] rule
is based on the existence of a bijection that satisfies some property,
rather than on our formalization based on Strassen's Theorem. Their
rule is motivated by applications in cryptography, and works well for
many examples, but is inconvenient for our example at hand, which
involves non-uniform probabilities. Indeed, their proof is based on
code rewriting, and is done in two steps. First, they prove equivalence between sampling and
returning $x_1$ from $\bern{p_1}$; and sampling $z_1$ from
$\bern{p_2}$, $z_2$ from $\bern{\rfrac{p_1}{p_2}}$ and returning
$z= z_1 \land z_2$. Then, they find a coupling between $z$ and
$\bern{p_2}$.

\subsubsection{Shift coupling: random walk vs lazy random walk}
The previous example is an instance of a lockstep coupling, in that it
relates the $k$-th element of the first chain with the $k$-th element
of the second chain. Many examples from the literature follow this
lockstep pattern; however, it is not always possible to establish
lockstep couplings. Shift couplings are a relaxation of lockstep
couplings where we relate elements of the first and second chains
without the requirement that their positions coincide.

We consider a simple example that motivates the use of shift
couplings.  Consider the random walk and lazy random walk (which, at
each time step, either chooses to move or stay put), both defined as
Markov chains over $\mathbb{Z}$. For simplicity, assume that both
walks start at position 0. It is not immediate to find a coupling
between the two walks, since the two walks necessarily get
desynchronized whenever the lazy walk stays put. Instead, the trick is
to consider a lazy random walk that moves two steps instead of
one. The random walk and the lazy random walk of step 2 are defined by
the step functions:
$$\begin{array}{rcl}
  \operatorname{step} & \defeq & \lambda x.\mlet{z}{\mathcal{U}_{\{-1,1\}}}{\munit{z+x}} \\
  \operatorname{lstep2} & \defeq & \lambda x.\mlet{z}{\mathcal{U}_{\{-1,1\}}}{\mlet{b}{\mathcal{U}_{\{0,1\}}}{\munit{x+2*z*b}}}
\end{array}$$
After 2 iterations of $\operatorname{step}$, the position has either
changed two steps to the left or to the right, or has returned to the
initial position, which is the same behaviour $\operatorname{lstep2}$
has on every iteration.  Therefore, the coupling we want to find should
equate the elements at position $2i$ in $\operatorname{step}$ with the
elements at position $i$ in $\operatorname{lstep2}$. The details on how
to prove the existence of this coupling are in \autoref{sec:grhol}.

\subsubsection{Lumped coupling: random walks on 3 and 4 dimensions}
A Markov chain is \emph{recurrent} if it has probability 1 of
returning to its initial state, and \emph{transient} otherwise. It is
relatively easy to show that the random walk over $\mathbb{Z}$ is
recurrent. One can also show that the random walk over $\mathbb{Z}^2$
is recurrent.  However, the random walk over $\mathbb{Z}^3$ is
transient.
%
%
%
%

For higher dimensions, we can use a coupling
argument to prove transience. Specifically, we can define a coupling
between a lazy random walk in $n$ dimensions and a random walk in $n
+m$ dimensions, and derive transience of the latter from transience of
the former. We define the (lazy) random walks below, and sketch the
coupling arguments.

Specifically, we show here the particular case of the transience of
the 4-dimensional random walk from the transience of the 3-dimensional
lazy random walk.  We start by defining the stepping functions:
\[\begin{array}{rl}
  \operatorname{step}_4 &: \mathbb{Z}^4 \to \Distr(\mathbb{Z}^4)
  \defeq \lambda z_1. \mlet{x_1}{\unif{U_4}}{\munit{z_1 +_4 x_1}}
  \\
  \operatorname{lstep}_3 &: \mathbb{Z}^3 \to \Distr(\mathbb{Z}^3)
  \defeq \lambda z_2. \mlet{x_2}{\unif{U_3}}{\mlet{b_2}{\bern{\rfrac{3}{4}}}{\munit{z_2 +_3 b_2*x_2 }}}
  \end{array}
\]
where $U_i=\{(\pm 1,0,\dots 0), \dots, (0,\dots,0,\pm 1)\}$ are the
vectors of the basis of $\mathbb{Z}^i$ and their opposites.  Then, the
random walk of dimension 4 is modelled by $\operatorname{rwalk4} \defeq
\markov(0, \operatorname{step_4})$, and the lazy walk of dimension 3 is
modelled by $\operatorname{lwalk3} \defeq
\markov(0, \operatorname{step_3})$.  We want to prove:
$$
\jgrholnoc{\operatorname{rwalk4}}{\Distr(\Str{\mathbb{Z}^{4}})}
       {\operatorname{lwalk3}}{\Distr(\Str{\mathbb{Z}^{3}})}
       {\diamond_{\left[\substack{y_1\leftarrow \res\ltag \\ y_2\leftarrow \res\rtag}\right]} 
  \All(y_1, y_2, z_1.z_2.\operatorname{pr}^{4}_{3}(z_1) = z_2)}
$$
where $\operatorname{pr}^{n_2}_{n_1}$ denotes the standard projection from
$\mathbb{Z}^{n_2}$ to $\mathbb{Z}^{n_1}$. 

We apply the \rname{Markov} rule. The only interesting premise requires proving that the transition function preserves the coupling:
\[p_2=\operatorname{pr}^{4}_{3}(p_1) \vdash \operatorname{step_4} \sim \operatorname{lstep}_3
                    \mid \forall x_1 x_2. x_2=\operatorname{pr}^4_3(x_1) \Rightarrow \diamond_{\left[\substack{y_1\ot \res\ltag\ x_1 \\ y_2\ot \res\rtag\ x_2}\right]} \operatorname{pr}^4_3(y_1)=y_2 \]


To prove this, we need to find the appropriate coupling, i.e., one
that preserves the equality.  The idea is that the step in
$\mathbb{Z}^3$ must be the projection of the step in
$\mathbb{Z}^4$. This corresponds to the following judgement:
\[\left.\begin{array}{rl}
   \lambda z_1. &\mlet{x_1}{\unif{U_4}}
               {\\ &\munit{z_1 +_4 x_1}}
  \end{array}
  \sim
  \begin{array}{rl}
  \lambda z_2. &\mlet{x_2}{\unif{U_3}}
               {\\ &\mlet{b_2}{\bern{\rfrac{3}{4}}}
               {\\ &\munit{z_2 +_3 b_2*x_2}}}
  \end{array}
  \; \right| \;
  \begin{array}{c}
  \forall z_1 z_2. \operatorname{pr}^4_3(z_1) = z_2 \Rightarrow \\ \operatorname{pr}^4_3(\res\ltag\ z_1) \stackrel{\diamond}{=} \res\rtag\ z_2
  \end{array}\]
which by simple equational reasoning is the same as
\[\left.\begin{array}{rl}
   \lambda z_1. &\mlet{x_1}{\unif{U_4}}
               {\\ &\munit{z_1 +_4 x_1}}
  \end{array}
  \sim
  \begin{array}{rl}
  \lambda z_2. &\mlet{p_2}{\unif{U_3} \times \bern{\rfrac{3}{4}}}
               {\\ &\munit{z_2 +_3 \pi_1(p_2)*\pi_2(p_2)}}
  \end{array}
  \; \right| \;
  \begin{array}{c}
  \forall z_1 z_2. \operatorname{pr}^4_3(z_1) = z_2 \Rightarrow \\ \operatorname{pr}^4_3(\res\ltag\ z_1) \stackrel{\diamond}{=} \res\rtag\ z_2
  \end{array}
  \]

We want to build a coupling such that if we sample $(0,0,0,1)$ or
$(0,0,0,-1)$ from $\unif{U_3}$, then we sample $0$ from
$\bern{\rfrac{3}{4}}$, and otherwise if we sample
$(x_1,x_2,x_3,0)$ from $\unif{U_4}$, we sample $(x_1,x_2,x_3)$ from
$U_3$. Formally, we prove this with the \rname{Coupling} rule.
Given $X:U_4 \to \bool$, by simple computation we show that:
\[ \Pr_{z_1 \sim \unif{U_4}}[z_1 \in X] \leq \Pr_{z_2 \sim \unif{U_3} \times \bern{\rfrac{3}{4}}}[z_2 \in \{ y \mid \exists x \in X. {\sf pr}^4_3(x) = \pi_1(y)*\pi_2(y) \}]\]


This concludes the proof. From the previous example, it follows that
the lazy walk in 3 dimensions is transient, since the random walk in
3 dimensions is transient. By simple reasoning, we now conclude that
the random walk in 4 dimensions is also transient.

\section{Probabilistic Guarded Lambda Calculus}
\label{sec:syntax}

To ensure that a function on infinite datatypes is well-defined, one
must check that it is \emph{productive}. This means that any finite
prefix of the output can be computed in finite time. For instance,
consider the following function on streams:
\[
 \mathtt{letrec\ bad\ (x:xs) = x : tail (bad\ xs)}
\]
This function is not productive since only the first element can be computed. We
can argue this as follows:
Suppose that the tail of a stream is available one unit of time after its head,
and that that \texttt{x:xs} is available at time 0. How much time does it take
for \texttt{bad} to start outputting its tail? Assume it takes $k$ units of
time. This means that \texttt{tail(bad\ xs)} will be available at time $k+1$ ,
since \texttt{xs} is only available at time 1. But \texttt{tail(bad\ xs)} is
exactly the tail of \texttt{bad(x:xs)}, and this is a contradiction, since
\texttt{x:xs} is available at time 0 and therefore the tail of
\texttt{bad(x:xs)} should be available at time $k$. Therefore, the tail of
\texttt{bad} will never be available.

The guarded lambda calculus solves the productivity problem by
distinguishing at type level between data that is available now and
data that will be available in the future, and restricting when
fixpoints can be defined.  Specifically, the guarded lambda calculus
extends the usual simply typed lambda calculus with two modalities:
$\later$ (pronounced \textit{later}) and $\square$
(\textit{constant}).  The later modality represents data that will be
available one step in the future, and is introduced and removed by the
term formers $\later$ and $\prev$\! respectively.  This modality is
used to guard recursive occurrences, so for the calculus to remain
productive, we must restrict when it can be eliminated.  This is
achieved via the constant modality, which expresses that all the data
is available at all times.  
In the remainder of this section we present a probabilistic extension
of this calculus.

\paragraph*{Syntax}
Types of the calculus are defined by the grammar
\begin{align*}
  A,B ::= b \mid \nat \mid 
  A \times B \mid A + B  \mid A \to B \mid \Str{A} \mid \square~A \mid \later{A} \mid \Distr(C)
\end{align*}
where $b$ ranges over a collection of base types.
$\Str{A}$ is the type of guarded streams of elements of type
$A$. Formally, the type $\Str{A}$ is isomorphic to $A \times
\later{\Str{A}}$. This isomorphism gives a way to introduce streams
with the function $(\cons{}{}) : A \to \later\Str{A} \to
\Str{A}$ and to eliminate them with the functions $\hd: \Str{A} \to A$
and $\tl: \Str{A} \to \later\Str{A}$.
%
$\Distr (C)$ is the type of distributions over \emph{discrete types}
$C$.  Discrete types are defined by the following grammar, where $b_0$
are discrete base types, e.g., $\zint$.
\begin{align*}
  C,D ::= b_0 \mid \nat \mid 
  C \times D \mid C + D \mid  \Str{C} | \later{C}.
\end{align*}
Note that, in particular, arrow types are not discrete but streams
are. This is due to the semantics of streams as sets of finite
approximations, which we describe in the next subsection. Also note
that $\square\Str{A}$ is not discrete since it makes the full infinite
streams available.

We also need to distinguish between arbitrary types $A, B$ and
constant types $S, T$, which are defined by the following grammar
\begin{align*}
  S, T ::= b_C \mid \nat \mid 
  S \times T \mid S + T  \mid S \to T \mid \square~A
\end{align*}
where $b_C$ is a collection of constant base types.  Note in
particular that for any type $A$ the type $\square~A$ is constant.

The terms of the language $t$ are defined by the following grammar
\begin{align*}
  t &::= ~x
        \mid c
        \mid 0 \mid S t \mid \casenat{t}{t}{t} 
        \mid \mu \mid \munit{t} \mid \mlet{x}{t}{t} \\
        &\mid \langle t, t \rangle \mid \pi_1 t \mid \pi_2 t
        \mid \inj{1} t \mid \inj{2} t \mid \CASE t \OF \inj{1} x . t; \inj{2} y . t
        \mid \lambda x . t \mid t\,t \mid \fix{x}{t}\\
        &\mid \cons{t}{ts} \mid \hd t \mid \tl t
        \mid \boxx{t} \mid \letbox{x}{t}{t} \mid \letconst{x}{t}{t}
        \mid \latern[\xi]t \mid \prev t
\end{align*}
where $\xi$ is a delayed substitution, a sequence of bindings
$\hrt{x_1 \gets t_1, \ldots, x_n \gets t_n}$.
The terms $c$ are constants corresponding to the base types used and
$\munit{t}$ and $\mlet{x}{t}{t}$ are the introduction and sequencing
construct for probability distributions.  The meta-variable $\mu$
stands for base distributions like $\unif{C}$ and $\bern{p}$.

Delayed substitutions were introduced in \cite{BGCMB16} in a dependent type theory to be able to work with types dependent on terms of type $\later{A}$.
In the setting of a simple type theory, such as the one considered in this paper, delayed substitutions are equivalent to having the applicative structure~\cite{McBride:Applicative} $\app$ for the $\later$ modality.
However, delayed substitutions extend uniformly to the level of propositions, and thus we choose to use them in this paper in place of the applicative structure.

\paragraph*{Denotational semantics}

The meaning of terms is given by a denotational model in the category $\trees$ of presheaves over $\omega$, the first infinite ordinal.
This category $\trees$ is also known as the \emph{topos of trees}~\cite{Birkedal-et-al:topos-of-trees}.
In previous work~\cite{CBGB16}, it was shown how to model most of the constructions of the guarded lambda calculus and its internal logic, with the notable exception of the probabilistic features.
Below we give an elementary
presentation of the semantics.

Informally, the idea behind the topos of trees is to represent
(infinite) objects from their finite approximations, which we observe
incrementally as time passes.  Given an object $x$, we can consider a
sequence $\{x_i\}$ of its finite approximations observable at time
$i$. These are trivial for finite objects, such as a natural number,
since for any number $n$, $n_i = n$ at every $i$.  But for infinite
objects such as streams, the $i$th approximation is the prefix of
length $i+1$.

Concretely, the category $\trees$ consists of:
\begin{itemize}
  \item Objects $X$: families of sets $\{X_i\}_{i\in\nat}$ together with \emph{restriction functions} $r_n^X : X_{n+1} \to X_n$.
  We will write simply $r_n$ if $X$ is clear from the context.
  \item Morphisms $X \to Y$ : families of functions $\alpha_n : X_n \to Y_n$ commuting with restriction functions in the sense of $r_n^Y \circ \alpha_{n+1} = \alpha_n \circ r_n^X$.
\end{itemize}

The full interpretation of types of the calculus can be found in
\autoref{fig:sem-types} in the appendix. The main points we want to
highlight are:
\begin{itemize}
  \item Streams over a type $A$ are interpreted as sequences of finite
    prefixes of elements of $A$ with the restriction functions of $A$:
    $$\sem{\Str{A}} \defeq \sem{A}_0 \times \{*\} \xleftarrow{r_0
      \times !} \sem{A}_1 \times \sem{\Str{A}}_0
    \xleftarrow{r_1 \times r_0 \times !}
    \sem{A}_2\times \sem{\Str{A}}_1 \leftarrow \cdots$$

  \item Distributions over a discrete object $C$ are
    defined as a sequence of distributions over each
    $\sem{C}_i$: $$\sem{\Distr(C)} \defeq \Distr(\sem{C}_0)
    \stackrel{\Distr(r_0)}{\longleftarrow} \Distr(\sem{C}_1)
    \stackrel{\Distr(r_1)}{\longleftarrow} \Distr(\sem{C}_2)
    \stackrel{\Distr(r_2)}{\longleftarrow} \ldots,$$ where $\Distr(\sem{C}_i)$ is
    the set of (probability density) functions $\mu : \sem{C}_i \to [0,1]$ such that $\sum_{x_\in X} \mu
    x = 1$, and $\Distr(r_i)$ adds the probability density of all the points
    in $\sem{C}_{i+1}$ that are sent by $r_i$ to the same point in the $\sem{C}_{i}$. In other words,
    $\Distr(r_i)(\mu)(x) = \Pr_{y \ot \mu}[r_i(y) = x]$
\end{itemize}

An important property of the interpretation is that discrete types are interpreted as objects $X$ such that $X_i$ is finite or countably infinite for every $i$.
This allows us to define distributions on these objects without the need for measure theory.
In particular, the type of guarded streams $\Str{A}$ is discrete provided $A$ is, which is clear from the interpretation of the type $\Str{A}$.
Conceptually this holds because $\sem{\Str{A}}_i$ is an approximation of real streams, consisting of only the first $i+1$ elements.

An object $X$ of $\trees$ is \emph{constant} if all its restriction
functions are bijections.  Constant types are interpreted as constant
objects of $\trees$ and for a constant type $A$
the objects $\sem{\square A}$ and $\sem{A}$ are isomorphic in $\trees$.

\paragraph*{Typing rules}

Terms are typed under a dual context $\Delta \mid \Gamma$, where $\Gamma$ is a usual context that binds variables to a type, and $\Delta$ is a constant context containing variables bound to types that are \textit{constant}.
The term $\letconst{x}{u}{t}$ allows us to shift variables between constant and non-constant contexts. The typing rules can be found in \autoref{fig:glambda}.

The semantics of such a dual context $\Delta \mid \Gamma$ is given as the product of types in $\Delta$ and $\Gamma$, except that we implicitly add $\square$ in front of every type in $\Delta$.
In the particular case when both contexts are empty, the semantics of the dual context correspond to the terminal object $1$, which is the singleton set $\{\ast\}$ at each time. 


The interpretation of the well-typed term $\Delta \mid \Gamma \vdash t : A$ is
defined by induction on the typing derivation, and can be found in
\autoref{fig:sem-glc} in the appendix.


\begin{figure*}[!tb]
\small
\begin{gather*}
  \inferrule
  {x : A \in \Gamma}
  {\Delta \mid \Gamma \vdash x : A}
  \qquad
  \inferrule
  {x : A \in \Delta}
  {\Delta \mid \Gamma \vdash x : A}
  \qquad
  \inferrule
  {\Delta \mid \Gamma, x : A \vdash t : B}
  {\Delta \mid \Gamma \vdash \lambda x . t : A \to B}
  \\
  \inferrule
  {\Delta \mid \Gamma \vdash t : A \to B \and \Delta \mid \Gamma \vdash u : A}
  {\Delta \mid \Gamma \vdash t\,u :  B}
  \qquad
  \inferrule
  {\Delta \mid \Gamma, f : \later{A} \vdash t :  A}
  {\Delta \mid \Gamma \vdash \fix{f}{t} : A}
  \qquad
  \inferrule
  {\Delta \mid \cdot \vdash t : \later A}
  {\Delta \mid \Gamma \vdash \prev{t} : A}
  \\
  \inferrule
  {\Delta \mid \cdot \vdash t : A}
  {\Delta \mid \Gamma \vdash \boxx{t} : \square A}
  \qquad
  \inferrule
  {\Delta \mid \Gamma \vdash u : \square B \and
    \Delta, x : B \mid \Gamma \vdash t : A}
  {\Delta \mid \Gamma \vdash \letbox{x}{u}{t} : A}
  \\
  \inferrule
  {\Delta \mid \Gamma \vdash u : B \and 
    \Delta, x : B \mid \Gamma \vdash t : A \and 
    B \text{ constant}}
  {\Delta \mid \Gamma \vdash \letconst{x}{u}{t} : A}
  \\
  \inferrule
  {\Delta \mid \Gamma, x_1 : A_1, \cdots x_n : A_n \vdash t : A
    \and
    \Delta \mid \Gamma \vdash t_i : \later{A_i}}
  {\Delta \mid \Gamma \vdash \later[x_1 \gets t_1, \ldots, x_n \gets t_n]{t} : \later{A}}
  \qquad
  \inferrule
  {\Delta \mid \Gamma \vdash t : A \and A \text{ discrete }}
  {\Delta \mid \Gamma \vdash \munit{t} : \Distr(A)}
  \\
  \inferrule
  {\Delta \mid \Gamma \vdash t : \Distr(A) \and \Delta \mid \Gamma, x : A \vdash u : \Distr(B)}
  {\Delta \mid \Gamma \vdash \mlet{x}{t}{u} : \Distr(B)}
  \qquad
  \inferrule
  {\mu \text{ primitive distribution on type } A}
  {\Delta \mid \Gamma \vdash \mu : \Distr(A)}
\end{gather*}
\shrinkcaption
\caption{A selection of the typing rules of the guarded lambda calculus.
  The rules for products, sums, and natural numbers 
  are standard.}
\label{fig:glambda}
\end{figure*}

\paragraph*{Applicative structure of the later modality}
As in previous work we can define the operator $\app$ satisfying the typing rule
\[
  \inferrule
  {\Delta \mid \Gamma \vdash t : \later{(A \to B)} \and \Delta \mid \Gamma \vdash u : \later{A}}
  {\Delta \mid \Gamma \vdash t \app u : \later{B}}
\]
and the equation
  $(\later{t}) \app (\later{u}) \equiv \later(t\ u)$
as the term
  $t \app u \defeq \latern[\hrt{f \gets t, x \gets u}]{f\,x}$.

%

\paragraph*{Example: Modelling Markov chains}

As an application of $\app$ and an example of how to use guardedness
and probabilities together, we now give the precise
definition of the $\markov$ construct that we used to model Markov chains earlier:
\[
\begin{array}{rl}
  \markov &: C  \to (C \to \Distr(C)) \to \Distr(\Str{C}) \\
  \markov &\defeq \fix{f}{\lambda x. \lambda h. \\
          &\quad \mlet {z}{h\ x}{ \mlet{t}{\operatorname{swap_{\later\Distr}^{\Str{C}}}(f \app  \later z \app \later h)}
            { \munit{\cons{x}{t}}}}}
\end{array}
\]
The guardedness condition gives $f$ the type $\later( C \to (C \to
\Distr(C)) \to \Distr(\Str{C}))$ in the body of the
fixpoint. Therefore, it needs to be applied functorially (via $\app$)
to $\later z$ and $\later h$, which gives us a term of type $\later
\Distr(\Str{C})$. To complete the definition we need to build a term
of type $\Distr(\later\Str{C})$ and then sequence it with $\cons{}{}$
to build a term of type $\Distr(\Str{C})$. To achieve this, we use the
primitive operator $\operatorname{swap_{\later\Distr}^C} :
\later\Distr(C) \to \Distr(\later C)$, which witnesses the isomorphism
between $\later\Distr(C)$ and $\Distr(\later C)$. For this isomorphism
to exist, it is crucial that distributions be total (i.e., we cannot
use subdistributions). Indeed, the denotation for $\later\Distr(C)$ is
the sequence $\{\ast\} \ot \Distr(C_1) \ot \Distr(C_2) \ot \dots$,
while the denotation for $\Distr(\later C)$ is the sequence
$\Distr(\{\ast\}) \ot \Distr(C_1) \ot \Distr(C_2) \ot \dots$, and
$\{\ast\}$ is isomorphic to $\Distr(\{\ast\})$ in ${\sf Set}$ only if
$\Distr$ considers only total distributions.





\section{Guarded higher-order logic}

\label{sec:ghol}
We now introduce Guarded HOL (GHOL), which is a higher-order logic to reason about terms of the guarded lambda calculus.
The 
logic is essentially that of~\cite{CBGB16}, but presented with the dual context formulation analogous to the dual-context typing judgement of the guarded lambda calculus.
Compared to standard intuitionistic higher-order logic, the logic GHOL has two additional constructs, corresponding to additional constructs in the guarded lambda calculus.
These are the later modality ($\later$) \emph{on propositions}, with delayed substitutions,
which expresses that a proposition holds one time unit into the future, and the ``always''
modality $\square$, which expresses that a proposition holds at all times.
Formulas are defined by the grammar:
\[ \phi, \psi ::= \top \mid \phi \wedge \psi \mid \phi \vee \psi \mid \neg \psi
  \mid \forall x. \phi \mid \exists x. \phi \mid \later[x_1 \ot t_1 \dots x_n
  \ot t_n]\phi \mid \square \phi\]
The basic judgement of the logic is $\Delta \mid \Sigma \mid \Gamma \mid \Psi \vdash \phi$ where $\Sigma$ is a logical context for $\Delta$ (that is, a list of formulas well-formed in $\Delta$) and $\Psi$ is another logical context for the dual context $\Delta\mid\Gamma$.
The formulas in context $\Sigma$ must be \emph{constant} propositions.
We say that a proposition $\phi$ is \emph{constant} if it is well-typed in context $\Delta \mid \cdot$ and moreover if every occurrence of the later modality in $\phi$ is under the $\square$ modality.
Selected rules are displayed in Figure~\ref{fig:hol} on page~\pageref{fig:hol}. We highlight \rname{Loeb} induction, which is the key to reasoning about fixpoints: to prove that $\phi$ holds now, one
can assume that it holds in the future. 
The interpretation of the formula $\Delta \mid \Gamma \vdash \phi$ is a subobject of the interpretation $\sem{\Delta \mid \Gamma}$.
Concretely the interpretation $A$ of $\Delta \mid \Gamma \vdash \phi$ is a family $\left\{A_i\right\}_{i=0}^\infty$ of sets such that $A_i \subseteq \sem{\Delta \mid \Gamma}_i$.
This family must satisfy the property that if $x \in A_{i+1}$ then $r_i(x) \in A_i$ where $r_i$ are the restriction functions of $\sem{\Delta \mid \Gamma}$.
The interpretation of formulas is defined by induction on the typing derivation.
In the interpretation of the context $\Delta \mid \Sigma \mid \Gamma \mid \Psi$ the formulas in $\Sigma$ are interpreted with the added $\square$ modality.
Moreover all formulas $\phi$ in $\Sigma$ are typeable in the context $\Delta \mid \cdot \vdash \phi$ and thus their interpretations are subsets of $\sem{\square \Delta}$.
We treat these subsets of $\sem{\Delta \mid \Gamma}$ in the obvious way.

The cases for the semantics of the judgement $\Delta \mid \Gamma \vdash \phi$ can be found in the appendix.
It can be shown that this logic is sound with respect to its model in the topos of trees.

\begin{theorem}[Soundness of the semantics]\label{thm:sound-ghol}
  The semantics of guarded higher-order logic is sound: if $\Delta \mid \Sigma \mid \Gamma \mid \Psi \vdash \phi$ is derivable then for all $n \in \nat$, $\sem{\square\Sigma}_n \cap \sem{\Psi}_n \subseteq \sem{\phi}$.
\end{theorem}

In addition, Guarded HOL is expressive enough to axiomatize standard probabilities over discrete sets. 
This axiomatization can be used to define the $\diamond$ modality directly in Guarded HOL (as opposed to our relational proof system,
were we use it as a primitive).
Furthermore, we can derive from this axiomatization additional rules to reason about couplings, which can be seen
in Figure~\ref{fig:prob-hol}. These rules will be the key to proving the soundness of the probabilistic fragment of the relational proof system, and can
be shown to be sound themselves.

\begin{proposition}[Soundness of derived rules]\label{thm:sound-prob-ghol}
  The additional rules are sound.
\end{proposition}

\begin{figure*}[!tb]
  \small
\begin{gather*}
\infer[\sf AX_U]
       {\jghol{\Delta}{\Sigma}{\Gamma}{\Psi}{\phi}}
       {\phi \in \Psi}
\quad
\infer[\sf AX_G]
       {\jghol{\Delta}{\Sigma}{\Gamma}{\Psi}{\phi}}
       {\phi \in \Sigma}
\quad
\infer[\sf CONV]
      {\jghol{\Delta}{\Sigma}{\Gamma}{\Psi}{t=t'}}
      {\Gamma \vdash t:\tau & \Gamma \vdash t':\tau & t \equiv t'}
\\[0.3em]
\infer[\sf SUBST]
       {\jghol{\Delta}{\Sigma}{\Gamma}{\Psi}{\phi\subst{x}{u}}}
       {\jghol{\Delta}{\Sigma}{\Gamma}{\Psi}{\phi\subst{x}{t}} & \jghol{\Delta}{\Sigma}{\Gamma}{\Psi}{t = u}}
\quad\quad
\infer[\sf Loeb]
        {\jghol{\Delta}{\Sigma}{\Gamma}{\Psi}{\phi}} 
        {\jghol{\Delta}{\Sigma}{\Gamma}{\Psi, \later\phi}{\phi}}
\\[0.3em]
\infer[\sf \later_I]
        {\jghol{\Delta}{\Sigma}{\Gamma}{\Psi}{\later[x_1 \ot t_1,\dots,x_n \ot t_n]\phi}} 
        {\jghol{\Delta}{\Sigma}{\Gamma, x_1:A_1, \dots, x_n:A_n}{\Psi}{\phi} &
          \Delta\mid\Gamma\vdash t_1 : \later A_1 & \dots &
          \Delta\mid\Gamma\vdash t_n : \later A_n}
\\[0.3em]
\infer[\sf \later_E]
        {\jghol{\Delta}{\Sigma}{\Gamma}{\Psi}{\phi\subst{x_1}{\prev t_1}\dots\subst{x_n}{\prev t_n}}}
        {\jghol{\Delta}{\Sigma}{\cdot}{\cdot}{\later[x_1 \ot t_1 \dots x_n \ot t_n]\phi} &
         \Delta \mid \bullet \vdash t_1 : \later A_1 & \dots & \Delta \mid
         \bullet \vdash t_n : \later A_n}
\\[0.3em]
\inferrule
        {\jghol{\Delta}{\Sigma}{\Gamma}{\Psi}{\later[x_1\!\ot\!t_1,\dots,x_n\!\ot\!t_n]\psi} \\
         \Delta \mid \Gamma \vdash t_1 : \later A_1 \; \dots\; \Delta \mid
         \Gamma \vdash t_n : \later A_n \\\\
         \jghol{\Delta}{\Sigma}{\Gamma, x_1:A_1, \dots, x_n:A_n}{\Psi, \psi}{\phi}}
        {\jghol{\Delta}{\Sigma}{\Gamma}{\Psi}{\later[x_1 \ot t_1,\dots,x_n \ot t_n]\phi}}
 {\sf \later_{App}}
\\[0.3em]
\infer[\sf \square_I]
        {\jghol{\Delta}{\Sigma}{\Gamma}{\Psi}{\square\phi}}
        {\jghol{\Delta}{\Sigma}{\cdot}{\cdot}{\phi}}
\quad\quad
\infer[\sf \square_E]
        {\jghol{\Delta}{\Sigma}{\Gamma}{\Psi}{\phi}}  
        {\jghol{\Delta}{\Sigma}{\Gamma}{\Psi}{\square\psi} &
         \jghol{\Delta}{\Sigma,\psi}{\Gamma}{\Psi}{\phi}}
\end{gather*}
\shrinkcaption
\caption{Selected Guarded Higher-Order Logic rules}\label{fig:hol}
\end{figure*}

\begin{figure*}[!tb]
  \small
\begin{gather*}
%
\inferrule*[right = \sf MONO2]
  {\jghol{\Delta}{\Sigma}{\Gamma}{\Psi}{\diamond_{[x_1\leftarrow t_1, x_2 \leftarrow t_2]}\phi} \\
    \jghol{\Delta}{\Sigma}{\Gamma, x_1 : C_1, x_2 : C_2}{\Psi, \phi}{\psi}}
  {\jghol{\Delta}{\Sigma}{\Gamma}{\Psi}{\diamond_{[x_1\leftarrow t_1, x_2\leftarrow t_2]}\psi}}
\\[0.3em]
%
%
\inferrule*[right = \sf UNIT2]
  {\jghol{\Delta}{\Sigma}{\Gamma}{\Psi}{\phi\subst{x_1}{t_1}\subst{x_2}{t_2}}}
  {\jghol{\Delta}{\Sigma}{\Gamma}{\Psi}
  {\diamond_{[x_1\leftarrow \munit{t_1}, x_2\leftarrow \munit{t_2}]}\phi}}
\\[0.3em]
%
%
\inferrule*[right = \sf MLET2]
          {\jghol{\Delta}{\Sigma}{\Gamma}{\Psi}{\diamond_{[x_1\leftarrow t_1, x_2\leftarrow t_2]} \phi}
          \\ \jghol{\Delta}{\Sigma}{\Gamma,x_1:C_1,x_2:C_2}{\Psi,\phi}
            {\diamond_{[y_1\leftarrow t'_1, y_2\leftarrow t'_2]}\psi}}
      {\jghol{\Delta}{\Sigma}{\Gamma}{\Psi}
             {\diamond_{[y_1\leftarrow \mlet{x_1}{t_1}{t'_1}, y_2\leftarrow \mlet{x_2}{t_2}{t'_2}]} \psi}}
\\[0.3em]
\inferrule*[right = \sf MLET{-}L]
          {\jghol{\Delta}{\Sigma}{\Gamma}{\Psi}{\diamond_{[x_1\leftarrow t_1]}\phi}
          \\ \jghol{\Delta}{\Sigma}{\Gamma,x_1:C_1}{\Psi,\phi}
            {\diamond_{[y_1\leftarrow t'_1, y_2\leftarrow t'_2]}\psi}}
      {\jghol{\Delta}{\Sigma}{\Gamma}{\Psi}
          {\diamond_{[y_1\leftarrow \mlet{x_1}{t_1}{t'_1}, y_2\leftarrow t'_2]} \psi}}
\end{gather*}
\shrinkcaption
\caption{Derived rules for probabilistic constructs}\label{fig:prob-hol}
\end{figure*}


\section{Relational proof system}
\label{sec:grhol}

We complete the formal description of the system by describing the
proof rules for the non-probabilistic fragment of the relational proof
system (the rules of the probabilistic fragment were described in
Section~\ref{sec:overview:grhol}).

\subsection{Proof rules}

The rules for core $\lambda$-calculus constructs are
identical to those of~\cite{ABGGS17}; for convenience, we present a selection
of the main rules in Figure \ref{fig:sel-rhol} in the appendix.


We briefly comment on the two-sided rules for the new constructs
(Figure~\ref{fig:s-rhol}).
The notation $\Omega$ abbreviates a context $\Delta\mid\Sigma\mid\Gamma\mid\Psi$.
The rule \rname{Next} relates two terms that have a $\later$ term
constructor at the top level. We require that both have one term in
the delayed substitutions and that they are related pairwise.  Then
this relation is used to prove another relation between the main
terms. This rule can be generalized to terms with more than one term
in the delayed substitution.
The rule \rname{Prev} proves a relation between terms from the same delayed relation by
applying $\mathrm{prev}$ to both terms.
The rule \rname{Box} proves a relation between two boxed terms if the
same relation can be proven in a constant context.
Dually, \rname{LetBox} uses a relation between two boxed terms to prove
a relation between their unboxings.
\rname{LetConst} is similar to \rname{LetBox}, but it requires instead a relation
between two constant terms, rather than explicitly $\square$-ed terms.
The rule \rname{Fix} relates two fixpoints following the \rname{Loeb} rule from Guarded HOL.
Notice that in the premise, the fixpoints need to appear in the delayed substitution so that
the inductive hypothesis is well-formed.
The rule \rname{Cons} 
proves relations on streams
from relations between their heads and tails, while
\rname{Head} and \rname{Tail} behave as converses of \rname{Cons}.
%

Figure~\ref{fig:a-rhol} contains the one-sided versions of the rules.
We only present the left-sided versions as the right-sided versions
are completely symmetric. The rule \rname{Next-L} relates at $\phi$ a
term that has a $\later$ with a term that does not have a
$\later$. First, a unary property $\phi'$ is proven on the term $u$ in
the delayed substitution, and it is then used as a premise to prove
$\phi$ on the terms with delays removed. Rules for proving unary
judgements can be found in the appendix.
Similarly, \rname{LetBox-L} proves a unary property on the term that gets unboxed
and then uses it as a precondition.
The rule \rname{Fix-L} builds a fixpoint just on the left, and relates it with
an arbitrary term $t_2$ at a property $\phi$. Since $\phi$ may contain the
variable $\res\rtag$ which is not in the context, it has to be replaced when
adding $\later\phi$ to the logical context in the premise of the rule.
The remaining rules are similar to their two-sided counterparts.

\begin{figure*}[!htb]
\small
\begin{gather*}
\mbox{}\hspace{-5mm}\inferrule*[right=\sf Next]
      {\jgrhol{\Delta}{\Sigma}{\Gamma, x_1:A_1, x_2:A_2}{\Psi, \phi'\defsubst{x_1}{x_2}}{t_1}{A_1}{t_2}{A_2}{\phi} \\
        \jgrholsc{\Delta}{\Sigma}{\Gamma}{\Psi}{u_1}{\later{A_1}}{u_2}{\later{A_2}}
        {\dsubst{\res\ltag, \res\rtag}{\res\ltag, \res\rtag}{\phi'}}}
      {\jgrholsc{\Delta}{\Sigma}{\Gamma}{\Psi}
              {\nextt{x_1\!\leftarrow\!u_1}{t_1}}{\later{A_1}} 
              {\nextt{x_2\!\leftarrow\!u_2}{t_2}}{\later{A_2}}
              {\triangleright[x_1\!\leftarrow\!u_1,x_2\!\leftarrow\!u_2,
                \res\ltag\!\leftarrow\!\res\ltag,\res\rtag\!\leftarrow\!\res\rtag].\phi}} 
\\[0.3em]
 \inferrule*[right=\sf Prev]
      {\jgrhol{\Delta}{\Sigma}{\cdot}{\cdot}{t_1}{\later A_1}{t_2}{\later A_2}{\dsubst{\res\ltag,\res\rtag}{\res\ltag,\res\rtag}{\phi}}}
      {\jgrholsc{\Delta}{\Sigma}{\Gamma}{\Psi}{\prev{t_1}}{A_1}{\prev{t_2}}{A_2}{\phi}} 
\\[0.3em]
 \mbox{}\hspace{-2mm}\inferrule*[Right=\sf Box]
      {\jgrhol{\Delta}{\Sigma}{\cdot}{\cdot}{t_1}{A_1}{t_2}{A_2}{\phi}}
      {\jgrholsc{\Delta}{\Sigma}{\Gamma}{\Psi}{\boxx{t_1}}{\square A_1}{\boxx{t_2}}{\square A_2}
              {\square \phi\defsubst{\letbox{x_1}{\res\ltag}{x_1}}{\letbox{x_2}{\res\rtag}{x_2}}}} 
\\[0.3em]
 \inferrule*[right=\sf LetBox]
      {\jgrholsc{\Delta}{\Sigma}{\Gamma}{\Psi}{u_1}{\square B_1}{u_2}{\square B_2}{\square \phi\defsubst{\letbox{x_1}{\res\ltag}{x_1}}{\letbox{x_2}{\res\rtag}{x_2}}} \\
       \jgrhol{\Delta, x_1 : B_1, x_2 : B_2}{\Sigma, \phi\defsubst{x_1}{x_2}}{\Gamma}{\Psi}{t_1}{A_1}{t_2}{A_2}{\phi'}}
      {\jgrholsc{\Delta}{\Sigma}{\Gamma}{\Psi}{\letbox{x_1}{u_1}{t_1}}{A_1}{\letbox{x_2}{u_2}{t_2}}{A_2}{\phi'}} 
\\[0.3em]
  \inferrule*[right=\sf LetConst]
      {B_1,B_2,\phi\ \text{constant} \\ FV(\phi)\cap FV(\Gamma) = \emptyset \\
        \jgrholsc{\Delta}{\Sigma}{\Gamma}{\Psi}{u_1}{B_1}{u_2}{B_2}{\phi} \\
       \jgrhol{\Delta, x_1 : B_1, x_2 : B_2}{\Sigma, \phi\defsubst{x_1}{x_2}}{\Gamma}{\Psi}{t_1}{A_1}{t_2}{A_2}{\phi'}}
      {\jgrholsc{\Delta}{\Sigma}{\Gamma}{\Psi}{\letconst{x_1}{u_1}{t_1}}{A_1}{\letconst{x_2}{u_2}{t_2}}{A_2}{\phi'}} 
\\[0.3em]
  \inferrule*[right=\sf Fix]
      {\jgrhol{\Delta}{\Sigma}{\Gamma, f_1:\later{A_1}, f_2:\later{A_2}}{\Psi, \dsubst{\res\ltag,\res\rtag}{f_1,f_2}{\phi}}
              {t_1}{A_1}{t_2}{A_2}{\phi}}
      {\jgrholsc{\Delta}{\Sigma}{\Gamma}{\Psi}{\fix{f_1}{t_1}}{A_1}{\fix{f_2}{t_2}}{A_2}{\phi}} 
\\[0.3em]
  \mbox{}\hspace{-5.2mm}\inferrule*[Right=\sf Cons]
      {\jgrholsc{\Delta}{\Sigma}{\Gamma}{\Psi}{x_1}{A_1}{x_2}{A_2}{\phi_h} \\
       \jgrholsc{\Delta}{\Sigma}{\Gamma}{\Psi}{xs_1}{\later{\Str{A_1}}}{xs_2}{\later{\Str{A_2}}}{\phi_t} \\
       \jgholsc{\Delta}{\Sigma}{\Gamma}{\Psi}{\forall x_1,x_2, s_1, s_2. \phi_h\defsubst{x_1}{x_2} \Rightarrow
          \phi_t\defsubst{s_1}{s_2} \Rightarrow \phi\defsubst{\cons{x_1}{s_1}}{\cons{x_2}{s_2}}}}
      {\jgrholsc{\Delta}{\Sigma}{\Gamma}{\Psi}{\cons{x_1}{s_1}}{\Str{A_1}}{\cons{x_2}{s_2}}{\Str{A_2}}{\phi}}
\\[0.3em]
  \inferrule*[right=\sf Head]
      {\jgrholsc{\Delta}{\Sigma}{\Gamma}{\Psi}{t_1}{\Str{A_1}}{t_1}{\Str{A_1}}{\phi\defsubst{hd\ \res\ltag}{hd\ \res\rtag}}}
      {\jgrholsc{\Delta}{\Sigma}{\Gamma}{\Psi}{hd\ t_1}{A_1}{hd\ t_2}{A_2}{\phi}} 
\\[0.3em]
   \inferrule*[right=\sf Tail]
      {\jgrholsc{\Delta}{\Sigma}{\Gamma}{\Psi}{t_1}{\Str{A_1}}{t_2}{\Str{A_2}}{\phi\defsubst{tl\ \res\ltag}{tl\ \res\rtag}}}
      {\jgrholsc{\Delta}{\Sigma}{\Gamma}{\Psi}{tl\ t_1}{\later{\Str{A_1}}}{tl\ t_2}{\later{\Str{A_2}}}{\phi}} 
\end{gather*}
\shrinkcaption
\caption{Two-sided rules for Guarded RHOL}\label{fig:s-rhol}
\end{figure*}

\begin{figure*}[!htb]
\small
\begin{gather*}
\inferrule*[right=\sf Next{-}L]
      {\jgrhol{\Delta}{\Sigma}{\Gamma, x_1:B_1}{\Psi, \phi'\subst{\res}{x_1}}{t_1}{A_1}{t_2}{A_2}{\phi} \\
       \jguholsc{\Delta}{\Sigma}{\Gamma}{\Psi}{u_1}{\later{B_1}}{\dsubst{\res}{\res}{\phi'}}}
      {\jgrholsc{\Delta}{\Sigma}{\Gamma}{\Psi}
              {\later[x_1 \ot u_1]{t_1}}{\later{A_1}} 
              {t_2}{A_2}
              {\later[x_1 \ot u_1,\res\ltag\leftarrow\res\ltag]{\phi}}}
\\[0.3em]
  \inferrule*[right=\sf Prev{-}L ]
      {\jgrhol{\Delta}{\Sigma}{\cdot}{\cdot}{t_1}{\later A_1}{t_2}{A_2}{\dsubst{\res\ltag}{\res\ltag}{\phi}}}
      {\jgrhol{\Delta}{\Sigma}{\Gamma_1; \Gamma_2}{\Psi_1 ; \Psi_2}{\prev{t_1}}{A_1}{t_2}{A_2}{\phi}}
\\[0.3em]
 \inferrule*[right=\sf Box{-}L]
      {\jgrhol{\Delta}{\Sigma}{\Gamma_2}{\Psi_2}{t_1}{A_1}{t_2}{A_2}{\phi} \\ FV(t_1)\not\subseteq FV(\Gamma_2)\\
      FV(\Psi_2)\subseteq FV(\Gamma_2)}
      {\jgrhol{\Delta}{\Sigma}{\Gamma_1; \Gamma_2}{\Psi_1; \Psi_2}{\boxx{t_1}}{\square A_1}{t_2}{A_2}
              {\square \phi\subst{\res\ltag}{\letbox{x_1}{\res\ltag}{x_1}}}} 
\\[0.3em]
 \inferrule*[right=\sf LetBox{-}L]
      {\jguholsc{\Delta}{\Sigma}{\Gamma}{\Psi}{u_1}{\square B_1}{\square \phi\subst{\res}{\letbox{x_1}{\res\ltag}{x_1}}} \\
       \jgrhol{\Delta, x_1 : B_1}{\Sigma, \phi\subst{\res}{x_1}}{\Gamma}{\Psi}{t_1}{A_1}{t_2}{A_2}{\phi'}}
      {\jgrholsc{\Delta}{\Sigma}{\Gamma}{\Psi}{\letbox{x_1}{u_1}{t_1}}{A_1}{t_2}{A_2}{\phi'}} 
\\[0.3em]
  \inferrule*[right=\sf LetConst{-}L]
      {B_1,\phi\ \text{constant} \\ FV(\phi)\cap FV(\Gamma) = \emptyset \\
        \jguholsc{\Delta}{\Sigma}{\Gamma}{\Psi}{u_1}{B_1}{\phi} \\\\
       \jgrhol{\Delta, x_1 : B_1}{\Sigma, \phi\subst{\res}{x_1}}{\Gamma}{\Psi}{t_1}{A_1}{t_2}{A_2}{\phi'}}
      {\jgrholsc{\Delta}{\Sigma}{\Gamma}{\Psi}{\letconst{x_1}{u_1}{t_1}}{A_1}{t_2}{A_2}{\phi'}}
\\[0.3em]
  \inferrule*[right=\sf Fix{-}L]
      {\jgrhol{\Delta}{\Sigma}{\Gamma, f_1:\later{A_1}}{\Psi, \dsubst{\res\ltag}{f_1}{(\phi\subst{\res\rtag}{t_2})}}
              {t_1}{A_1}{t_2}{A_2}{\phi}}
      {\jgrhol{\Delta}{\Sigma}{\Gamma}{\Psi}{\fix{f_1}{t_1}}{A_1}{t_2}{A_2}{\phi}}
\\[0.3em]
  \inferrule*[right=\sf Cons{-}L]
      {\jgrholsc{\Delta}{\Sigma}{\Gamma}{\Psi}{x_1}{A_1}{t_2}{A_2}{\phi_h} \\
       \jgrholsc{\Delta}{\Sigma}{\Gamma}{\Psi}{xs_1}{\later{\Str{A_1}}}{t_2}{A_2}{\phi_t} \\
       \jgholsc{\Delta}{\Sigma}{\Gamma}{\Psi}{\forall x_1,x_2, xs_1. \phi_h\defsubst{x_1}{x_2} \Rightarrow
          \phi_t\defsubst{xs_1}{x_2} \Rightarrow \phi\defsubst{\cons{x_1}{xs_1}}{x_2}}}
      {\jgrholsc{\Delta}{\Sigma}{\Gamma}{\Psi}{\cons{x_1}{xs_1}}{\Str{A_1}}{t_2}{A_2}{\phi}} 
%
%
\\[0.3em]
  \inferrule*[right=\sf Head{-}L]
      {\jgrholsc{\Delta}{\Sigma}{\Gamma}{\Psi}{t_1}{\Str{A_1}}{t_1}{A_2}{\phi\subst{\res\ltag}{hd\ \res\ltag}}}
      {\jgrholsc{\Delta}{\Sigma}{\Gamma}{\Psi}{hd\ t_1}{A_1}{t_2}{A_2}{\phi}}
\\[0.3em]
   \inferrule*[right=\sf Tail{-}L]
      {\jgrholsc{\Delta}{\Sigma}{\Gamma}{\Psi}{t_1}{\Str{A_1}}{t_2}{A_2}{\phi\subst{\res\ltag}{tl\ \res\ltag}}}
      {\jgrholsc{\Delta}{\Sigma}{\Gamma}{\Psi}{tl\ t_1}{\later{\Str{A_1}}}{t_2}{A_2}{\phi}} 
\end{gather*}
\shrinkcaption
\caption{One-sided rules for Guarded RHOL}\label{fig:a-rhol}
\end{figure*}

\subsection{Metatheory}
We review some of the most interesting metatheoretical properties of
our relational proof system, highlighting the equivalence with Guarded HOL.
\begin{theorem}[Equivalence with Guarded HOL] \label{thm:equiv-rhol-hol}
  For all contexts $\Delta,\Gamma$; types $\sigma_1,\sigma_2$; terms $t_1,t_2$; sets of assertions $\Sigma,\Psi,$;
  and assertions $\phi$:
    \[\jgrhol{\Delta}{\Sigma}{\Gamma}{\Psi}{t_1}{\sigma_1}{t_2}{\sigma_2}{\phi}
    \quad\Longleftrightarrow\quad
    \jghol{\Delta}{\Sigma}{\Gamma}{\Psi}{\phi\subst{\res\ltag}{t_1}\subst{\res\rtag}{t_2}}\]
\end{theorem}
The forward implication follows by induction on the given derivation.
The reverse implication is immediate from the rule which allows to
fall back on Guarded HOL in relational proofs. (Rule \rname{SUB} in the appendix).
The full proof is in the appendix. The consequence of this theorem is
that the syntax-directed, relational proof system we have built 
on top of Guarded HOL does not lose expressiveness.

The intended semantics of a judgement
$\jgrhol{\Delta}{\Sigma}{\Gamma}{\Psi}{t_1}{A_1}{t_2}{A_2}{\phi}$ is
that, for every valuation $\delta \models \Delta$, $\gamma\models \Gamma$,
if $\sem{\Sigma}(\delta)$ and $\sem{\Psi}(\delta,\gamma)$, then
$$\sem{\phi}(\delta,\gamma[\res\ltag \ot \sem{t_1}(\delta,\gamma),
  \res\rtag\ot\sem{t_2}(\delta,\gamma)])$$
Since Guarded HOL is sound with respect to its semantics in
the topos of trees, and our relational proof system is equivalent to
Guarded HOL, we obtain that our relational proof system is also sound in
the topos of trees.

\begin{corollary}[Soundness and consistency]\label{cor:rhol:sound}
If $\jgrhol{\Delta}{\Sigma}{\Gamma}{\Psi}{t_1}{\sigma_2}{t_2}{\sigma_2}{\phi}$, then for every valuation
$\delta \models \Delta$, $\gamma\models\Gamma$:
\[\begin{array}{c}
  \sem{\Delta \vdash \square \Sigma}(\delta) \wedge \sem{\Delta \mid \Gamma \vdash \Psi}(\delta,\gamma) \Rightarrow
    \\ \sem{\Delta \mid \Gamma,\res\ltag:\sigma_1, \res\ltag:\sigma_2 \vdash \phi}
         (\delta, \gamma[\res\ltag \ot \sem{\Delta \mid \Gamma \vdash t_1}(\delta,\gamma)][\res\rtag \ot \sem{\Delta \mid \Gamma \vdash t_2}(\delta,\gamma)])
\end{array}\]
In particular, there is no proof of
$\jgrhol{\Delta}{\emptyset}{\Gamma}{\emptyset}{t_1}{\sigma_1}{t_2}{\sigma_2}{\bot}$.
\end{corollary}




\subsection{Shift couplings revisited}
\label{sec:examples}
We give further details on how to prove the example with shift couplings from Section~\ref{sec:proba-ex}.
(Additional examples of relational reasoning on non-probabilistic streams can be
found in the appendix.)
Recall the step functions:
$$\begin{array}{rcl}
  \operatorname{step} & \defeq & \lambda x.\mlet{z}{\mathcal{U}_{\{-1,1\}}}{\munit{z+x}} \\
  \operatorname{lstep2} & \defeq & \lambda x.\mlet{z}{\mathcal{U}_{\{-1,1\}}}{\mlet{b}{\mathcal{U}_{\{0,1\}}}{\munit{x+2*z*b}}}
\end{array}$$
We axiomatize the predicate $\All_{2,1}$, which relates the element at position $2i$ in one stream to the element at position $i$ in another stream, as follows.
\[\begin{array}{l}
    \forall x_1 x_2 xs_1 xs_2 y_1. \phi\subst{x_1}{z_1}\subst{x_2}{z_2} \Rightarrow \\
    \quad  \later[ys_1\ot xs_1]{\later[zs_1 \ot ys_1, ys_2 \ot xs_2]{\All_{2,1}(zs_1, ys_2, z_1.z_2.\phi)}} \Rightarrow \\
    \quad\quad   \All_{2,1}(\cons{x_1}{\cons{y_1}{xs_1}},\cons{x_2}{xs_2}, z_1.z_2. \phi)
\end{array}
\]
In fact, we can assume that, in general, we have a family of
$\All_{m_1, m_2}$ predicates relating two streams at positions
$m_1\cdot i$ and $m_2\cdot i$ for every $i$.

We can now express the existence of a shift coupling by the statement:
{\small
\[p_1 = p_2 \vdash \markov(p_1, \operatorname{step}) \sim \markov(p_2,
  \operatorname{lstep2}) \mid
  \diamond_{\left[\substack{ y_1 \ot \res\ltag\\ y_2 \ot \res\rtag}\right]}\All_{2,1}(y_1,y_2,z_1.z_2.z_1=z_2) \]}
%
%
For the proof, we need to introduce an asynchronous rule for Markov chains:
{\small\[
\inferrule*[right=\sf Markov-2-1]
      {\jgrholsc{\Delta}{\Sigma}{\Gamma}{\Psi}{t_1}{C_1}{t_2}{C_2}{\phi} \\\\
       \begin{array}{c}
         \jgrholsc{\Delta}{\Sigma}{\Gamma}{\Psi}{(\lambda x_1. \mlet{x'_1}{h_1\ x_1}{h_1\ x'_1})}{C_1\to\Distr(C_1)}{h_2}{C_2\to \Distr(C_2)}
              {\\ \forall x_1 x_2. \phi\subst{z_1}{x_1}\subst{z_2}{x_2} \Rightarrow \diamond_{[ z_1 \ot \res\ltag\ x_1, z_2 \ot \res\rtag\ x_2]}\phi}
       \end{array}}
      {\begin{array}{c}\jgrholsc{\Delta}{\Sigma}{\Gamma}{\Psi}{\markov(t_1,h_1)}{\Distr(\Str{C_1})}{\markov(t_2,h_2)}{\Distr(\Str{C_2})}
              {\\ \diamond_{[y_1 \ot \res\ltag, y_2 \ot \res\rtag]}\All_{2,1}(y_1,y_2,z_1.z_2.\phi)}
        \end{array}}
\]}
%
%
This asynchronous rule for Markov chains shares the motivations of the
rule for loops proposed in~\cite{BartheGHS17}. Note that one can define
a rule \rname{Markov-m-n} for arbitrary $m$ and $n$ to prove a judgement
of the form $\All_{m,n}$ on two Markov chains.


We show the proof of the shift coupling. By equational reasoning, we get:
\[
\begin{array}{rl}
  & \lambda x_1. \mlet{x'_1}{h_1\ x_1}{h_1\ x'_1}\;
  \equiv \;
 \lambda x_1. \mlet{z_1}{\unif{\{-1,1\}}}
              {h_1\ (z_1 + x_1)} \; \equiv \\
  &\equiv \; \lambda x_1. \mlet{z_1}{\unif{\{-1,1\}}}
                {\mlet{z'_1}{\unif{\{-1,1\}}}
                 {\munit{z'_1 + z_1 + x'_1}}}
\end{array}
\]
and the only interesting premise of \rname{Markov-2-1} is:
\[\left.\begin{array}{rl}
   \lambda x_1. &\mlet{z_1}{\unif{\{-1,1\}}}
               {\\ &\mlet{z'_1}{\unif{\{-1,1\}}}
               {\\ &\munit{z'_1 + z_1 + x'_1}}}
  \end{array}
  \sim
  \begin{array}{rl}
  \lambda x_2. &\mlet{z_2}{\unif{\{-1,1\}}}
               {\\ &\mlet{b_2}{\unif{\{1,0\}}}
               {\\ &\munit{x_2+2*b_2*z_2}}}
  \end{array}
  \;\right|\;
  \begin{array}{c}
    \forall x_1 x_2. x_1 = x_2 \Rightarrow \\
    \res\ltag\ x_1 \stackrel{\diamond}{=} \res\rtag\ x_2
  \end{array}\]
Couplings between $z_1$ and $z_2$ and between $z_1'$ and $b_2$ can be found by simple computations.
This completes the proof.
  




\section{Related work}
\label{sec:rw}
Our probabilistic guarded $\lambda$-calculus and the associated logic
Guarded HOL build on top of the guarded $\lambda$-calculus and its internal 
logic~\cite{CBGB16}.  The guarded $\lambda$-calculus has been extended
to guarded dependent type theory~\cite{BGCMB16}, which can be
understood as a theory of guarded refinement types and as a foundation
for proof assistants based on guarded type theory. These systems do
not reason about probabilities, and do not support syntax-directed
(relational) reasoning, both of which we support.

Relational models for higher-order programming languages are often
defined using logical relations. \cite{PlotkinA93} showed how to use
second-order logic to define and reason about logical relations for
the second-order lambda calculus.  Recent work has extended this approach
to logical relations for higher-order programming languages with
computational effects such as nontermination, general references, and
concurrency~\cite{DreyerAB11,caresl,ipm,Krogh-Jespersen17}.  The
logics used in \emph{loc. cit.} are related to our work in two ways:
(1) the logics in \emph{loc. cit.} make use of the later modality for
reasoning about recursion, and (2) the models of the logics in
\emph{loc. cit.} can in fact be defined using guarded type theory.
Our work is more closely related to Relational Higher Order
Logic~\cite{ABGGS17}, which applies the idea of logic-enriched type
theories~\cite{AczelG00,AczelG06} to a relational setting. There
exist alternative approaches for reasoning about relational properties
of higher-order programs; for instance,~\cite{GrimmMFHMPRSB17} have
recently proposed to use monadic reification for reducing relational
verification of $F^*$ to proof obligations in higher-order logic.


A series of work develops reasoning methods for probabilistic
higher-order programs for different variations of the lambda calculus.
One line of work has focused on operationally-based techniques for
reasoning about contextual equivalence of programs.  The methods are
based on probabilistic
bisimulations~\cite{DBLP:conf/esop/CrubilleL14,DBLP:conf/popl/SangiorgiV16}
or on logical relations~\cite{DBLP:conf/fossacs/BizjakB15}.  Most of
these approaches have been developed for languages with discrete
distributions, but recently there has also been work on languages with
continuous
distributions~\cite{DBLP:conf/icfp/BorgstromLGS16,DBLP:conf/esop/CulpepperC17}.
Another line of work has focused on denotational models, starting with
the seminal work in~\cite{Jones:powerdomain-evaluations}.  Recent work
includes support for relational reasoning about equivalence of
programs with continuous distributions for a total programming
language~\cite{DBLP:conf/lics/StatonYWHK16}.  Our approach is most
closely related to prior work based on relational refinement types for
higher-order probabilistic programs. These were initially considered
by~\cite{BFGSSZ14} for a stateful fragment of $F^*$, and later
by~\cite{BGGHRS15,BFGGGHS16} for a pure language.  Both systems are
specialized to building probabilistic couplings; however, the latter
support approximate probabilistic couplings, which yield a natural
interpretation of differential privacy~\cite{DR14}, both in its
vanilla and approximate forms (i.e.\, $\epsilon$- and
$(\epsilon,\delta)$-privacy). Technically, approximate couplings are
modelled as a graded monad, where the index of the monad tracks the privacy budget ($\epsilon$ or $(\epsilon,\delta)$).
Both systems are strictly syntax-directed, and cannot reason about
computations that have different types or syntactic structures, while
our system can.



\section{Conclusion}
\label{sec:conclusion}

We have developed a probabilistic extension of the (simply typed)
guarded $\lambda$-calculus, and proposed a syntax-directed proof
system for relational verification. Moreover, we have verified a
series of examples that are beyond the reach of prior work. Finally,
we have proved the soundness of the proof system with respect to the
topos of trees.

There are several natural directions for future work. One first
direction is to enhance the expressiveness of the underlying simply
typed language. For instance, it would be interesting to introduce
clock variables and some type dependency as in~\cite{BGCMB16}, and extend the proof system accordingly.
This would allow us, for example, to type the function taking the $n$-th element of a \emph{guarded} stream,
which cannot be done in the current system.
Another exciting direction is to
consider approximate couplings, as in~\cite{BGGHRS15,BFGGGHS16},
and to develop differential privacy for infinite streams---preliminary
work in this direction, such as~\cite{KellarisPXP14}, considers very
large lists, but not arbitrary streams. A final direction would be to
extend our approach to continuous distributions to support
other application domains.


\subsubsection*{Acknowledgments.}
We would like to thank the anonymous reviewers for their time and their helpful input.
This research was supported in part by the ModuRes Sapere Aude Advanced Grant from The Danish Council for Independent Research for the Natural Sciences (FNU),
 by a research grant (12386, Guarded Homotopy Type Theory) from
the VILLUM foundation, and by NSF under grant 1718220.

\bibliography{refs}
\appendix
\newpage

\section{Additional proof rules}

\begin{figure*}[!htb]
\small
\begin{mathpar}
\infer[\sf ABS]
      {\jgrhol{\Delta}{\Sigma}{\Gamma}{\Psi}{\lambda x_1:\tau_1. t_1}{\tau_1 \to \sigma_1}{\lambda x_2:\tau_2. t_2}{\tau_2\to \sigma_2}{\forall x_1,x_2. \phi' \Rightarrow \phi\subst{\res\ltag}{\res\ltag\ x_1}\subst{\res\rtag}{\res\rtag\ x_2}}}
      {\jgrhol{\Delta}{\Sigma}{\Gamma,x_1:\tau_1,x_2:\tau_2}{\Psi,\phi'}{t_1}{\sigma_1}{t_2}{\sigma_2}{\phi}}
\and
\infer[\sf APP]{\jgrhol{\Delta}{\Sigma}{\Gamma}{\Psi}{t_1 u_1}{\sigma_1}{t_2 u_2}{\sigma_2}{\phi\subst{x_1}{u_1}\subst{x_2}{u_2}}}
      {\begin{array}{c}
\jgrhol{\Delta}{\Sigma}{\Gamma}{\Psi}{t_1}{\tau_1\to \sigma_1}{t_2}{\tau_2\to \sigma_2}{
          \forall x_1,x_2. \phi'\subst{\res\ltag}{x_1}\subst{\res\rtag}{x_2}\Rightarrow \phi\subst{\res\ltag}{\res\ltag\ x_1}\subst{\res\rtag}{\res\rtag\ x_2}}\\
\jgrhol{\Delta}{\Sigma}{\Gamma}{\Psi}{u_1}{\tau_1}{u_2}{\tau_2}{
          \phi'}
\end{array}}
\and
\infer[\sf VAR]{\jgrhol{\Delta}{\Sigma}{\Gamma}{\Psi}{x_1}{\sigma_1}{x_2}{\sigma_2}{\phi}}
      {\jlc{\Delta\mid\Gamma}{x_1}{\sigma_1} & \jlc{\Delta\mid\Gamma}{x_2}{\sigma_2} & 
       \jghol{\Delta}{\Sigma}{\Gamma}{\Psi}{\phi\subst{\res\ltag}{x_1}\subst{\res\rtag}{x_2}}}
\and
\infer[\sf SUB]{\jgrhol{\Delta}{\Sigma}{\Gamma}{\Psi}{t_1}{\sigma_1}{t_2}{\sigma_2}{\phi}}
      {\jgrhol{\Delta}{\Sigma}{\Gamma}{\Psi}{t_1}{\sigma_1}{t_2}{\sigma_2}{\phi'} & 
       \Delta \mid \Sigma \mid \Gamma \mid \Psi \vdash_{\sf GHOL} \phi'\defsubst{t_1}{t_2} \Rightarrow \phi\defsubst{t_1}{t_2}}
\and
\infer[\sf UHOL-L]
      {\jgrhol{\Delta}{\Sigma}{\Gamma}{\Psi}{t_1}{\sigma_1}{t_2}{\sigma_2}{\phi}}
      {\jguhol{\Delta}{\Sigma}{\Gamma}{\Psi}{t_1}{\sigma_1}{\phi\defsubst{\res}{t_2}}
      \\  \jlc{\Delta\mid\Gamma}{t_2}{\sigma_2}}
\and
\infer[\sf ABS{-}L]
      {\jgrhol{\Delta}{\Sigma}{\Gamma}{\Psi}{\lambda x_1:\tau_1. t_1}{\tau_1 \to \sigma_1}{t_2}{\sigma_2}{\forall x_1. \phi' \Rightarrow \phi \subst{\res\ltag}{\res\ltag\ x_1}}}
      {\jgrhol{\Delta}{\Sigma}{\Gamma,x_1:\tau_1}{\Psi, \phi'}{t_1}{\sigma_1}{t_2}{\sigma_2}{\phi}}
\and
\infer[\sf APP{-}L]
      {\jgrhol{\Delta}{\Sigma}{\Gamma}{\Psi}{t_1 u_1}{\sigma_1}{u_2}{\sigma_2}{\phi\subst{x_1}{u_1}}}
      {\begin{array}{c}
          \jgrhol{\Delta}{\Sigma}{\Gamma}{\Psi}{t_1}{\tau_1\to \sigma_1}{u_2}{\sigma_2}{\forall x_1. \phi'\subst{\res\ltag}{x_1} \Rightarrow \phi\subst{\res\ltag}{\res\ltag\ x_1}}\\
          \jguhol{\Delta}{\Sigma}{\Gamma}{\Psi}{u_1}{\sigma_1}{\phi'}
\end{array}}
\and
\infer[\sf VAR{-}L]
      {\jgrhol{\Delta}{\Sigma}{\Gamma}{\Psi}{x_1}{\sigma_1}{t_2}{\sigma_2}{\phi}}
      {\phi\subst{\res\ltag}{x_1} \in \Psi &  \res\rtag\not\in\ FV(\phi) &
       \jlc{\Delta\mid\Gamma}{t_2}{\sigma_2}}
\and      
  \inferrule*[Right=\sf Equiv]
      {\jgrholsc{\Delta}{\Sigma}{\Gamma}{\Psi}{t'_1}{A_1}{t'_2}{A_2}{\Phi} \\ t_1 \equiv t_1' \\ t_2 \equiv t_2' \\
      \Delta \mid \Gamma \vdash t_1 : A_1 \\ \Delta \mid \Gamma \vdash t_2 : A_2}
      {\jgrholsc{\Delta}{\Sigma}{\Gamma}{\Psi}{t_1}{A_1}{t_2}{A_2}{\Phi}}
\end{mathpar}
\caption{Selected RHOL rules}\label{fig:sel-rhol}
\end{figure*}

\section{Denotational semantics}

\subsection{Types and terms in context}

The meaning of terms is given by the denotational model in the category $\trees$ of presheaves over $\omega$, the first infinite ordinal.
This category $\trees$ is also known as the \emph{topos of trees}~\cite{Birkedal-et-al:topos-of-trees}.
In previous work~\cite{CBGB16} it was shown how to model most of the constructions of the guarded lambda calculus and the associated logic, with the notable exception of the probabilistic features.
Below we give an elementary and self-contained presentation of the semantics.

Concretely, objects $X$ of $\trees$ are families of sets $X_i$ indexed over $\nat$ together with functions $r_n^X : X_{n+1} \to X_n$.
These are called \emph{restriction functions}.
We will write simply $r_n$ if $X$ is clear from the context.
Moreover if $x \in X_i$ and $j \leq i$ we will write $x\restriction_j$ for the element $r_j(\cdots (r_{i-1}(x))\cdots) \in X_j$.
Morphisms $X \to Y$ are families of functions $\alpha_n : X_n \to Y_n$ commuting with restriction functions in the sense of $r_n^Y \circ \alpha_{n+1} = \alpha_n \circ r_n^X$.
One can see the restriction function $r_n : X_{n+1} \to X_n$ as mapping elements of $X_{n+1}$ to their approximations at time $n$.

Semantics of types can be found on \autoref{fig:sem-types}, where
$G\left(\sem{A}\right)$ consists of sequences $\{x_n\}_{n \in \nat}$
such that $x_i \in \sem{A}_i$ and $r_i(x_{i+1}) = x_i$ for all $i$,
i.e., $\square\sem{A}$ is the set of so-called global sections of
$\sem{A}$.
 
The semantics of a dual context $\Delta \mid \Gamma$ is given as the product of types in $\Delta$ and $\Gamma$, except that we implicitly add $\square$ in front of every type in $\Delta$.
In the particular case when both contexts are empty, the semantics of the dual context correspond to the terminal object $1$, which is the singleton set $\{\ast\}$ at each stage. 
A term in context $\Delta \mid \Gamma \vdash t : \tau$ is interpreted as a family of functions $\sem{t}_n : \sem{\Delta \mid \Gamma}_n \to \sem{\tau}_n$ commuting with restriction functions of $\sem{\Delta \mid \Gamma}$ and $\sem{\tau}$.
Semantics of products,
coproducts, 
and natural numbers is pointwise as in sets, so we
omit writing it.  The cases for the other constructs are in
\autoref{fig:sem-glc} where $\mathsf{munit}$ and $\mathsf{mlet}$
are the standard unit and bind operations on discrete probabilities,
i.e.\,
\begin{align*}
  \begin{array}{rcl}
    \munit{c} & = & \lambda y.~\carac{c=y} \\
    \mathsf{mlet}~x = \mu~\mathsf{in}~M & = & \lambda y.~\sum_{c\in C} \mu(c) \cdot M(c)(y)
  \end{array}
\end{align*}
The functions $\pi_0$ and $\pi_1$ are the first and second projections, respectively.

\begin{figure*}[!htb]
  
\begin{align*}
  \sem{b} &\defeq \text{ chosen object of } \trees\\
  \sem{\nat} &\defeq \nat \xleftarrow{id} \nat \xleftarrow{id} \nat \xleftarrow{id} \cdots\\
  \sem{A\times B} &\defeq \sem{A}_0\times\sem{B}_0 \xleftarrow{r_0 \times r_0} \sem{A}_1\times\sem{B}_1 \xleftarrow{r_1 \times r_1} \sem{A}_2\times\sem{B}_2 \xleftarrow{r_2 \times r_2} \cdots \\
  \sem{A\to B} &\defeq \left(\sem{B}^{\sem{A}}\right)_0 \xleftarrow{\pi} \left(\sem{B}^{\sem{A}}\right)_1 \xleftarrow{\pi} \left(\sem{B}^{\sem{A}}\right)_2 \xleftarrow{\pi} \cdots \\
  \sem{\Str{A}} &\defeq \sem{A}_0 \times \{*\} \xleftarrow{r_0 \times !} \sem{A}_1 \times \left(\sem{A}_0 \times \{*\}\right)
                 \xleftarrow{r_1 \times r_0 \times !} \sem{A}_2\times\left(\sem{A}_1\times\left(\sem{A}_0 \times \{*\}\right)\right) \leftarrow \cdots \\
  \sem{\later A} &\defeq \{*\} \xleftarrow{!} \sem{A}_0 \xleftarrow{r_0} \sem{A}_1 \xleftarrow{r_1} \cdots \\
  \sem{\square A} &\defeq G(\sem{A}) \xleftarrow{id} G(\sem{A}) \xleftarrow{id} \cdots\\
  \sem{\Distr(C)} &\defeq \Distr(\sem{C}_0) \stackrel{\Distr(r_0)}{\longleftarrow} \Distr(\sem{C}_1) \stackrel{\Distr(r_1)}{\longleftarrow} \Distr(\sem{C}_2) \stackrel{\Distr(r_2)}{\longleftarrow} \ldots
\end{align*}
\caption{Semantics of types in the topos of trees}
\label{fig:sem-types}
\end{figure*}

\begin{figure*}[!htb]
\begin{align*}
  &\sem{\Delta \mid \Gamma \vdash \lambda x:A. t : A \to B}_i(\delta,\gamma) 
      \defeq (f_0,\ldots,f_i)\\
      &\quad\quad\text{ where } f_i(x) = \sem{\Delta\mid \Gamma, x : A \vdash t : B}\left(\delta,(\gamma\restriction_i, x)\right)\\
  &\sem{\Delta \mid \Gamma \vdash t_1\ t_2 : B}_i(\delta,\gamma)
     \defeq f_i\left(\sem{\Delta \mid\Gamma \vdash t_2 : A}(\delta,\gamma)\right)\\
   &\quad\quad\text{ where } \sem{\Delta \mid \Gamma \vdash t_1 : A \to B}_i(\delta,\gamma) = (f_0,\ldots,f_i)\\
  &\sem{\Delta \mid \Gamma \vdash \later[x_1\ot t_1, \dots, x_n\ot t_n]t : \later A}_0(\delta,\gamma) \defeq *\\ 
  &\sem{\Delta \mid \Gamma \vdash \later[x_1\ot t_1, \dots, x_n\ot t_n]t : \later A}_{i+1}(\delta,\gamma) \defeq \\
     &\quad\quad\sem{\Delta \mid \Gamma, \{x_k : A_k\}_{k=1}^n \vdash t : A}_i\left(\delta, (\gamma\restriction_i, \left\{\sem{\Delta \mid \Gamma \vdash t_k : \later A_k}_{i+1}\right\}_{k=1}^n(\delta,\gamma))\right)\\
  &\sem{\Delta \mid \Gamma \vdash \prev t : A}_i(\delta,\gamma) \defeq \sem{\Delta \mid \cdot \vdash t : \later A}_{i+1}(\delta)\\
  &\sem{\Delta \mid \Gamma \vdash \boxx t : \square A}_i(\delta,\gamma) \defeq \{\sem{\Delta \mid \cdot \vdash t: A}_j(\delta)\}_{j=0}^{\infty}\\
  &\sem{\Delta \mid \Gamma \vdash \letbox{x}{u}{t} : A}_i(\delta,\gamma) \defeq \\
    &\quad\quad\sem{\Delta, x : B \mid \Gamma \vdash t : A}_i\left(\left(\delta,\sem{\Delta \mid \Gamma \vdash u : \square B}_i(\delta,\gamma)\right),\gamma\right)\\
  &\sem{\Delta \mid \Gamma \vdash \letconst{x}{u}{t} : A}_i(\delta,\gamma)\defeq\\
    &\quad\quad\sem{\Delta, x : B \mid \Gamma \vdash t : A}_i\left(\left(\delta,\varepsilon^{-1}_i\left(\sem{\Delta \mid \Gamma \vdash u : \square B}_i(\delta,\gamma)\right)\right),\gamma\right)\\
  &\sem{\Delta \mid \Gamma \vdash \hd t : A}_i(\delta,\gamma) \defeq \pi_0\left(\sem{\Delta \mid \Gamma \vdash t : \Str{A}}_i(\delta,\gamma)\right)\\
  &\sem{\Delta \mid \Gamma \vdash \tl t : \later{\Str{A}}}_i(\delta,\gamma) \defeq \pi_1\left(\sem{\Delta \mid \Gamma \vdash t : \Str{A}}_i(\delta,\gamma)\right)\\
  &\sem{\Delta \mid \Gamma \vdash \cons{t}{u} : \Str{A}}_i(\delta,\gamma) \defeq
   \left(\sem{\Delta \mid \Gamma \vdash t : A}_i(\delta,\gamma), \sem{\Delta \mid \Gamma \vdash u : \Str{A}}_i(\delta,\gamma)\right)\\
  &\sem{\Delta \mid \Gamma \vdash \munit{t} : \Distr(C)}_i (\delta,\gamma)
   \defeq \munit{\sem{\Delta \mid \Gamma \vdash t:C}_i (\delta,\gamma)}  \\
  &\sem{\Delta \mid \Gamma \vdash \mlet{x}{t}{u} : \Distr(C)}_i (\delta,\gamma)
   \defeq
  \mathsf{mlet}~v= \sem{\Delta \mid \Gamma \vdash t : \Distr(D)}_i
  (\delta,\gamma)~\mathsf{in} \\
   \mbox{}\hspace{0.5cm}
    &\quad\quad\sem{\Delta \mid \Gamma,x:D \vdash u : \Distr(C)}_i (\delta,\gamma[x:=v])
\end{align*}
\caption{Semantics for the Guarded $\lambda$-calculus}
\label{fig:sem-glc}
\end{figure*}



\subsection{Equational theory of the calculus}


The denotational semantics validates the following equational theory in addition to the standard equational theory of the simply typed lambda calculus with sums and natural numbers.\\
\textbf{Rules for fixed points, always modality and streams}
\begin{align*}
  \begin{split}
    \fix{f}{t} \quad &\equiv \quad t\subst{f}{\later(\fix{f}{t})} \\
    \prev{(\later t)} \quad &\equiv \quad t \\
    \letbox{x}{(\boxx{u})}{t} \quad &\equiv \quad t\subst{x}{u} \\
    \letconst{x}{u}{t} \quad&\equiv\quad t\subst{x}{u} \\
  \end{split}
  \begin{split}
    \hd\ (\cons{x}{xs}) \quad&\equiv\quad x \\
    \tl\ (\cons{x}{xs}) \quad&\equiv\quad xs \\
    \cons{\hd t}{\tl t} \quad&\equiv\quad t
  \end{split}
\end{align*}
\textbf{Rules for delayed substitutions}
\begin{align*}
  \latern[\xi\hrt{x\gets t}]{u} &\equiv \latern[\xi]{u} & & \text{ if } x \text{ not in } u\\
  \latern[\xi\hrt{x\gets t, y \gets s}\xi']{u} &\equiv\latern[\xi\hrt{y\gets s, x \gets t}\xi']{u}\\
  \latern[\xi\hrt{x\gets \latern[\xi]{t}}]{u} &\equiv \latern[\xi]{\left(u\subst{x}{t}\right)}\\
  \latern[\hrt{x\gets t}]x &\equiv t
\end{align*}
\textbf{Monad laws for distributions}
\begin{align*}
  \mlet{x}{\munit{t}}{u} &\equiv u[t/x]\\
  \mlet{x}{t}{\munit{x}} &\equiv t\\
  \mlet{x_2}{(\mlet{x_1}{t_1}{t_2})}{u} &\equiv \mlet{x_1}{t_1}{(\mlet{x_2}{t_2}{u})}
\end{align*}
In particular, notice that fix does not reduce as usual, but instead the whole term is delayed before the substitution is performed.


\subsection{Logical judgements}

The cases for the semantics of the judgement $\Delta \mid \Gamma \vdash \phi$ of the non-probabilistic fragment are as follows (we omit writing the contexts if they are clear):
\begin{align*}
  \sem{\top}_i &\defeq \sem{\Delta \mid \Gamma}_i\\
  \sem{\phi \wedge \psi}_i &\defeq \sem{\phi}_i \cap \sem{\psi}_i\\
  \sem{\phi \vee \psi}_i &\defeq \sem{\phi}_i \cup \sem{\psi}_i\\
  \sem{\phi \To \psi}_i &\defeq
  \left\{ x \isetsep \forall j \leq i, x\restriction_j \in \sem{\phi}_j \To x\restriction_j \in \sem{\psi}_j\right\}\\
  \sem{\forall x:A. \phi}_i &\defeq \left\{(\delta,\gamma) \isetsep \forall j \leq i, \forall x \in \sem{A}_j, \left(\delta, \left(\gamma\restriction_j\right), x\right) \in \sem{\phi}\right\}\\
  \sem{\later[x_1 \ot t_1, \dots, x_n \ot t_n]\phi}_i &\defeq 
    \left\{ (\delta, \gamma) \isetsep i > 0 \To \left(\delta,\gamma\restriction_{i-1}, \left\{\sem{t_k}_{i}(\delta,\gamma)\right\}_{k=1}^n\right) \in \sem{\phi}_{i-1}\right\}\\
  \sem{\square \phi}_i &\defeq \left\{ x \isetsep \forall j, x \in \sem{\phi}_j\right\}
\end{align*}

\section{Additional background}

One consequence of Strassen's theorem is
that couplings are closed under convex combinations.
\begin{lemma}[Convex combinations of couplings]\label{lem:convex-combination-of-couplings}
Let $(\mu_i)_{i\in I}$ and $(\nu_i)_{i\in I}$ bet two families of
distributions on $C_1$ and $C_2$ respectively, and let $(p_i)_{i\in I}
\in [0,1]$ such that $\sum_{i\in I} p_i=1$. If
$\coupl{\mu_i}{\nu_i}{R}$ for all $i\in I$ then $\coupl{(\sum_{i\in I} p_i
\mu_i)}{(\sum_{i\in I} p_i \nu_i)}{R}$, where the convex
combination $\sum_{i\in I} p_i \mu_i$ is defined by the clause
$(\sum_{i\in I} p_i \mu_i)(x)=\sum_{i\in I} p_i \mu_i(x)$.
\end{lemma}
One obtains an asymmetric version of the lemma by observing that if
$\mu_i=\mu$ for every $i\in I$, then $(\sum_{i\in I} p_i \mu_i)=\mu$.

One can also show that couplings are closed under relation
composition.
\begin{lemma}[Couplings for relation composition]
Let $\mu_1\in\Distr(C_1)$, $\mu_2\in\Distr(C_2)$, $\mu_3
\in\Distr(C_3)$. Moreover, let $R\subseteq C_1\times C_2$ and
$S\subseteq C_2\times C_3$. If $\coupl{\mu_1}{\mu_2}{R}$ and
$\coupl{\mu_2}{\mu_3}{R}$ then $\coupl{\mu_1}{\mu_3}{R}$.
\end{lemma}

\section{Proofs of the theorems}

\subsection{Proof of Theorem~\ref{thm:sound-ghol}}
\label{sec:soundness-of-ghol}

The semantics of the guarded higher-order logic without the probabilistic fragment has been explained in previous work~\cite{Birkedal-et-al:topos-of-trees,CBGB16}.
Thus we focus on showing soundness of the additional rules for the diamond modality, which will be useful for proving soundness of the relational proof system.
Moreover we only describe soundness for the binary diamond modality, the soundness of the rules for the unary modality being entirely analogous.

\noindent\textbf{Soundness of the rule \textsf{MONO2}}
\begin{mathpar}
  \infer[\sf MONO2]
  {\jghol{\Delta}{\Sigma}{\Gamma}{\Psi}{\diamond_{[x_1\gets t_1, x_2\gets t_2]}\psi}}
  {\jghol{\Delta}{\Sigma}{\Gamma}{\Psi}{\diamond_{[x_1\gets t_1, x_2 \gets t_2]}\phi} \and
    \jghol{\Delta}{\Sigma}{\Gamma, x_1 : C_1, x_2 : C_2}{\Psi, \phi}{\psi}
  }
\end{mathpar}

Let $n \in \nat$ and $(\delta, \gamma) \in \sem{\Delta \mid \Sigma \mid \Gamma \mid \Psi}_n$.
Then from the first premise we have $(\delta, \gamma) \in \sem{\diamond_{[x_1\gets t_1, x_2 \gets t_2]}\phi}_n$ and thus there exists an $\left\{(v, u) \isetsep (\delta, \gamma, v, u) \in \sem{\phi}_n\right\}$ coupling for the distributions $\sem{t_1}_n(\delta,\gamma)$ and $\sem{t_2}_n(\delta,\gamma)$.
But since $(\delta, \gamma) \in \sem{\Delta \mid \Sigma \mid \Gamma \mid \Psi}_n$ we have from the second premise of the rule that
\begin{align*}
  \left\{(v, u) \isetsep (\delta, \gamma, v, u) \in \sem{\phi}_n\right\} \subseteq
  \left\{(v, u) \isetsep (\delta, \gamma, v, u) \in \sem{\psi}_n\right\}
\end{align*}
and thus any $\left\{(v, u) \isetsep (\delta, \gamma, v, u) \in \sem{\phi}_n\right\}$ coupling is also an $\left\{(v, u) \isetsep (\delta, \gamma, v, u) \in \sem{\psi}_n\right\}$ coupling, which means there exists an $\left\{(v, u) \isetsep (\delta, \gamma, v, u) \in \sem{\psi}_n\right\}$ coupling for $\sem{t_1}_n(\delta,\gamma)$ and $\sem{t_2}_n(\delta,\gamma)$ as required.

\noindent\textbf{Soundness of the rule \textsf{UNIT2}}
\begin{mathpar}
  \infer[\sf UNIT2]
  {\jghol{\Delta}{\Sigma}{\Gamma}{\Psi}{ \diamond_{[x_1\leftarrow \munit{t_1}, x_2\leftarrow \munit{t_2}]} \phi}}
  {\jghol{\Delta}{\Sigma}{\Gamma}{\Psi}{ \phi[t_1/x_1][t_2/x_2]}}
\end{mathpar}
Let $n \in \nat$ and $(\delta, \gamma) \in \sem{\Delta \mid \Sigma \mid \Gamma \mid \Psi}_n$.
We need to show the existence of a
\begin{align*}
  \left\{(v, u) \isetsep (\delta, \gamma, v, u) \in \sem{\phi}_n\right\}
\end{align*}
coupling for the point-mass distributions concentrated at $\sem{t_1}_n(\delta,\gamma)$ and $\sem{t_2}_n(\delta,\gamma)$.
The premise of the rule establishes the membership
\begin{align*}
  \left(\sem{t_1}_n(\delta,\gamma), \sem{t_2}_n(\delta,\gamma)\right)
  \in
  \left\{(v, u) \isetsep (\delta, \gamma, v, u) \in \sem{\phi}_n\right\}
\end{align*}
and thus the point-mass distribution concentrated at $\left(\sem{t_1}_n(\delta,\gamma), \sem{t_2}_n(\delta,\gamma)\right)$ is a required coupling.

\noindent\textbf{Soundess of the rule \textsf{MLET2}}
\begin{mathpar}
\infer[\sf MLET2]
      {\jghol{\Delta}{\Sigma}{\Gamma}{\Psi}{
          \diamond_{[y_1\leftarrow \mlet{x_1}{t_1}{t'_1}, y_2\leftarrow
              \mlet{x_2}{t_2}{t'_2}]} \psi}}{
          \jghol{\Delta}{\Sigma}{\Gamma}{\Psi}{\diamond_{[x_1\leftarrow t_1, x_2\leftarrow t_2]} \phi}
          \and
          \jghol{\Delta}{\Sigma}{\Gamma,x_1:C_1,x_2:C_2}{\Psi,\phi}{
            \diamond_{[y_1\leftarrow t'_1, y_2\leftarrow t'_2]}\psi}}
\end{mathpar}
Let $n \in \nat$ and $(\delta, \gamma) \in \sem{\Delta \mid \Sigma \mid \Gamma \mid \Psi}_n$.
Then from the first premise we have that there exists an
\begin{align*}
  \left\{(v, u) \isetsep (\delta, \gamma, v, u) \in \sem{\phi}_n\right\}
\end{align*}
coupling for the distributions $\sem{t_1}_n(\delta,\gamma)$ and $\sem{t_2}_n(\delta,\gamma)$.

From the second premise we get that for every $v, u$ such that $(\delta, \gamma, v, u) \in \sem{\phi}_n$
there exists an 
\begin{align*}
  \left\{(v', u') \isetsep (\delta, \gamma, v, u, v', u') \in \sem{\psi}_n\right\}
\end{align*}
coupling for $\sem{t_1'}_n(\delta,\gamma, v)$ and $\sem{t_2'}_n(\delta,\gamma, u)$.
Since $x_1$ and $x_2$ are fresh for $\psi$ the relation
\begin{align*}
  \left\{(v', u') \isetsep (\delta, \gamma, v, u, v', u') \in \sem{\psi}_n\right\}
\end{align*}
is independent of $v, u$.

Thus Lemma~\ref{lem:sequential-composition-of-couplings} instantiated with
\begin{align*}
  \mu_1 &= \sem{t_1}_n(\delta,\gamma)\\
  \mu_2 &= \sem{t_2}_n(\delta,\gamma)\\
  M_1 &= \sem{t_1'}_n(\delta,\gamma, -)\\
  M_2 &= \sem{t_2'}_n(\delta,\gamma, -)\\
  R &= \left\{(v, u) \isetsep (\delta, \gamma, v, u) \in \sem{\phi}_n\right\}\\
  S &= \left\{(v', u') \isetsep (\delta, \gamma, v, u, v', u') \in \sem{\psi}_n\right\}
\end{align*}
concludes the proof.

\noindent\textbf{Soundness of the rule \textsf{MLET-L}}
\begin{mathpar}
\infer[\sf MLET-L]
      {\jghol{\Delta}{\Sigma}{\Gamma}{\Psi}{
          \diamond_{[y_1\leftarrow \mlet{x_1}{t_1}{t'_1}, y_2\leftarrow t'_2]} \psi}}{
          \jghol{\Delta}{\Sigma}{\Gamma}{\Psi}{\diamond_{[x_1\leftarrow t_1]}\phi}
          \and
          \jghol{\Delta}{\Sigma}{\Gamma,x_1:C_1}{\Psi,\phi}{
            \diamond_{[y_1\leftarrow t'_1, y_2\leftarrow t'_2]}\psi}}
\end{mathpar}
Let $n \in \nat$ and $(\delta, \gamma) \in \sem{\Delta \mid \Sigma \mid \Gamma \mid \Psi}_n$.
Then from the first premise we have that the support of the distribution $\sem{t_1}_n(\delta,\gamma)$ is included in
\begin{align*}
  \left\{v \isetsep (\delta, \gamma, v) \in \sem{\phi}_n\right\}.
\end{align*}

From the second premise we get that for every $v$ such that $(\delta, \gamma, v) \in \sem{\phi}_n$ there exists an
\begin{align*}
  \left\{(v', u') \isetsep (\delta, \gamma, v, v', u') \in \sem{\psi}_n\right\}
\end{align*}
coupling for $\sem{t_1'}_n(\delta,\gamma, v)$ and $\sem{t_2'}_n(\delta,\gamma)$.
Since $x_1$ is fresh for $\psi$ the relation
\begin{align*}
  R \defeq \left\{(v', u') \isetsep (\delta, \gamma, v, v', u') \in \sem{\psi}_n\right\}
\end{align*}
is independent of $v$.

Let $\mathcal{I} = \left\{v \isetsep (\delta, \gamma, v) \in \sem{\phi}_n\right\}$ and for any $v \in \mathcal{I}$ let $p_v = \sem{t_1}_n(\delta,\gamma)(v)$, $\mu_v = \sem{t_1'}_n(\delta,\gamma,v)$, and $\nu_v = \sem{t_2'}_n(\delta,\gamma)$.
Then we have $\sum_{v \in \mathcal{I}} p_v = 1$ from the first premise of the rule and $\coupling{R}{\mu_v}{\nu_v}$ for all $v \in \mathcal{I}$ from the second premise.
Lemma~\ref{lem:convex-combination-of-couplings} concludes the proof.


\subsection{Proof of Theorem \ref{thm:equiv-rhol-hol}}

The inverse implication follows immediately from the \rname{SUB} rule and
the fact that we can always prove a judgement of the shape
\[ \jgrhol{\Gamma}{\Sigma}{\Gamma}{\Psi}{t_1}{A_1}{t_2}{A_2}{\top} \]
for well-typed $t_1$ and $t_2$.

We will prove the direct implication by induction on the
derivation. We will just prove the two-sided rules.  The proofs for the one
sided rule are similar.
\\ \\
\noindent {\bf Case.} $\small\inferrule*[Right=\sf Next]
      {\jgrhol{\Delta}{\Sigma}{\Gamma, x_1:A_1, x_2:A_2}{\Psi, \Phi'\defsubst{x_1}{x_2}}{t_1}{A_1}{t_2}{A_2}{\Phi} \\
       \jgrhol{\Delta}{\Sigma}{\Gamma}{\Psi}{u_1}{\later{A_1}}{u_2}{\later{A_2}}{\dsubst{\res\ltag, \res\rtag}{\res\ltag, \res\rtag}{\Phi'}}}
      {\jgrhol{\Delta}{\Sigma}{\Gamma}{\Psi}
              {\nextt{x_1 \leftarrow u_1}{t_1}}{\later{A_1}} 
              {\nextt{x_2 \leftarrow u_2}{t_2}}{\later{A_2}}
              {\triangleright[x_1\leftarrow u_1,x_2\leftarrow x_2,\res\ltag\leftarrow\res\ltag,\res\rtag\leftarrow\res\rtag]\Phi}}$
\\ \\
\noindent
By I.H. $\jghol{\Delta}{\Sigma}{\Gamma, x_1:A_1, x_2:A_n}{\Psi, \Phi'\defsubst{x_1}{x_2}}{\Phi\defsubst{t_1}{t_2}}$, \hfill{}(H1)
\\
and $\jghol{\Delta}{\Sigma}{\Gamma}{\Psi}{\later[\res\ltag \ot u_1, \res\rtag \ot u_2]{\Phi'}}$ \hfill{}(H2)
\\
To show: $\jghol{\Delta}{\Sigma}{\Gamma}{\Psi}{\later[x_1 \ot u_1, x_2 \ot u_2, \res\ltag \ot \nextt{x_1 \leftarrow u_1}{t_1} , \res\rtag\ot \nextt{x_2 \leftarrow u_2}{t_2}]{\Phi}}$. \hfill{}(G)
\\
By \rname{CONV} we can change the goal (G) into
\\
$\jghol{\Delta}{\Sigma}{\Gamma}{\Psi}{\later[x_1 \ot u_1, x_2 \ot u_2]{\Phi\defsubst{t_1}{t_2}}}$ \hfill{}(G')
\\
and (H2) into:
\\
$\jghol{\Delta}{\Sigma}{\Gamma}{\Psi}{\later[x_1 \ot u_1, x_2 \ot u_2]{\Phi'\defsubst{x_1}{x_2}}}$ \hfill{}(H2')
\\
Finally, by applying \rname{$\later_{App}$} to (H1) and (H2) we get (G')
\\


\noindent {\bf Case.} $\small \inferrule*[Right=\sf Prev]
      {\jgrhol{\Delta}{\Sigma}{\cdot}{\cdot}{t_1}{\later A_1}{t_2}{\later A_2}{\dsubst{\res\ltag,\res\rtag}{\res\ltag,\res\rtag}{\Phi}}}
      {\jgrhol{\Delta}{\Sigma}{\Gamma}{\Psi}{\prev{t_1}}{A_1}{\prev{t_2}}{A_2}{\Phi}}$
\\ \\
\noindent
We just apply \rname{$\later_E$}.
\\

\noindent {\bf Case.} $\small\inferrule*[Right=\sf Box]
      {\jgrhol{\Delta}{\Sigma}{\cdot}{\cdot}{t_1}{A_1}{t_2}{A_2}{\Phi}}
      {\jgrhol{\Delta}{\Sigma}{\Gamma}{\Psi}{\boxx{t_1}}{\square A_1}{\boxx{t_2}}{\square A_2}
              {\square \Phi\defsubst{\letbox{x_1}{\res\ltag}{x_1}}{\letbox{x_2}{\res\rtag}{x_2}}}}$
\\ \\
\noindent
By I.H. $\jghol{\Delta}{\Sigma}{\cdot}{\cdot}{\Phi\defsubst{t_1}{t_2}}$
\\
To show: $\jghol{\Delta}{\Sigma}{\Gamma}{\Psi}{\square \Phi\defsubst{\letbox{x_1}{\boxx{t_1}}{x_1}}{\letbox{x_2}{\boxx{t_2}}{x_2}}}$
\\
By \rname{CONV} we can change the goal into:
\\
$\jghol{\Delta}{\Sigma}{\Gamma}{\Psi}{\square \Phi\defsubst{t_1}{t_2}}$
\\
And then we can prove it by \rname{$\square_I$}.
\\

\noindent {\bf Case.}$\small\inferrule*[Right=\sf LetBox]
      {\jgrhol{\Delta}{\Sigma}{\Gamma}{\Psi}{u_1}{\square B_1}{u_2}{\square B_2}{\square \Phi\defsubst{\letbox{x_1}{\res\ltag}{x_1}}{\letbox{x_2}{\res\rtag}{x_2}}} \\
       \jgrhol{\Delta, x_1 : B_1, x_2 : B_2}{\Sigma, \Phi\defsubst{x_1}{x_2}}{\Gamma}{\Psi}{t_1}{A_1}{t_2}{A_2}{\Phi'}}
      {\jgrhol{\Delta}{\Sigma}{\Gamma}{\Psi}{\letbox{x_1}{u_1}{t_1}}{A_1}{\letbox{x_2}{u_2}{t_2}}{A_2}{\Phi'}}$
\\ \\
\noindent
By I.H. $\jghol{\Delta}{\Sigma}{\Gamma}{\Psi}{\square \Phi\defsubst{\letbox{x_1}{u_1}{x_1}}{\letbox{x_2}{u_2}{x_2}}}$ \hfill{}(H1)
\\
and $\jghol{\Delta, x_1 : B_1, x_2 : B_2}{\Sigma, \Phi\defsubst{x_1}{x_2}}{\Gamma}{\Psi}{\Phi'\defsubst{t_1}{t_2}}$. \hfill{}(H2)
\\
To show: $\jghol{\Delta}{\Sigma}{\Gamma}{\Psi}{\Phi'\defsubst{\letbox{x_1}{u_1}{t_1}}{\letbox{x_2}{u_2}{t_2}}}$ \hfill{}(G)
\\
We instantiate (H2) into:
\[\begin{array}{c}
  \jghol{\Delta}{\Sigma, \Phi\defsubst{\letbox{x_1}{u_1}{x_1}}{\letbox{x_2}{u_2}{x_2}}}{\Gamma}{\Psi}
        {\\ \Phi'\defsubst{t_1\subst{x_1}{\letbox{x_1}{u_1}{x_1}}}{t_2\subst{x_2}{\letbox{x_2}{u_2}{x_2}}}}
        \end{array}\]
And by the equality $t\subst{x}{\letbox{x}{u}{x}} \equiv \letbox{x}{u}{t}$, we get:
\[\begin{array}{c}
   \jghol{\Delta}{\Sigma, \Phi\defsubst{\letbox{x_1}{u_1}{x_1}}{\letbox{x_2}{u_2}{x_2}}}{\Gamma}{\Psi}
         {\\ \Phi'\defsubst{\letbox{x_1}{u_1}{t_1}}{\letbox{x_2}{u_2}{t_2}}} \end{array} \]
and then, by applying \rname{$\square_E$} to (H1) and the previous judgement we get (G).
\\

\noindent {\bf Case.} $\small \inferrule*[Right=\sf LetConst]
      {B_1,B_2,\Phi\ \text{constant} \\ FV(\Phi)\cap FV(\Gamma) = \emptyset \\
        \jgrhol{\Delta}{\Sigma}{\Gamma}{\Psi}{u_1}{B_1}{u_2}{B_2}{\Phi} \\
       \jgrhol{\Delta, x_1 : B_1, x_2 : B_2}{\Sigma, \Phi\defsubst{x_1}{x_2}}{\Gamma}{\Psi}{t_1}{A_1}{t_2}{A_2}{\Phi'}}
      {\jgrhol{\Delta}{\Sigma}{\Gamma}{\Psi}{\letconst{x_1}{u_1}{t_1}}{A_1}{\letconst{x_2}{u_2}{t_2}}{A_2}{\Phi'}}$
\\ \\
\noindent
By I.H. $\jghol{\Delta}{\Sigma}{\Gamma}{\Psi}{\Phi\subst{u_1}{u_2}}$, \hfill{}(H1)
\\
and $\jghol{\Delta, x_1 : B_1, x_2 : B_2}{\Sigma, \Phi\defsubst{x_1}{x_2}}{\Gamma}{\Psi}{\Phi'\defsubst{t_1}{t_2}}$. \hfill{}(H2)
\\
To show: $\jghol{\Delta}{\Sigma}{\Gamma}{\Psi}{\Phi'\defsubst{\letconst{x_1}{u_1}{t_1}}{\letconst{x_2}{u_2}{t_2}}}$ \hfill{}(G)
\\
From (H1) and the fact that $\Phi, B_1$ and $B_2$ are constant, we get:
\\
$\jghol{\Delta}{\Sigma}{\Gamma}{\Psi}{\square\Phi\defsubst{u_1}{u_2}}$
\\
The rest of the prove is analogous to the previous case.
\\
     
\noindent {\bf Case.} $\small \inferrule*[Right=\sf Fix]
      {\jgrhol{\Delta}{\Sigma}{\Gamma, f_1:\later{A_1}, f_2:\later{A_2}}{\Psi, \dsubst{\res\ltag,\res\rtag}{f_1,f_2}{\Phi}}
              {t_1}{A_1}{t_2}{A_2}{\Phi}}
      {\jgrhol{\Delta}{\Sigma}{\Gamma}{\Psi}{fix f_1. t_1}{A_1}{fix f_2. t_2}{A_2}{\Phi}}$
\\ \\
\noindent
By I.H. $\jghol{\Delta}{\Sigma}{\Gamma, f_1:\later{A_1}, f_2:\later{A_2}}{\Psi,\later[\res\ltag\ot f_1, \res\rtag\ot f_2]{\Phi}}{\Phi\defsubst{t_1}{t_2}}$.
\\
To show: $\jghol{\Delta}{\Sigma}{\Gamma}{\Psi}{\Phi\subst{\fix{f_1}{t_1}}{fix{f_2}{t_2}}}$ \\
Instantiating the I.H. with $f_1 = \later \fix{f_1}{t_1}$ and $f_2 = \later \fix{f_2}{t_2}$ we get:
\\
$\jghol{\Delta}{\Sigma}{\Gamma}{\Psi,\later[\res\ltag \ot \later \fix{f_1}{t_1}, \res\rtag \ot \later \fix{f_2}{t_2}]{\Phi}}
       {\Phi\defsubst{t_1\subst{f_1}{\later \fix{f_1}{t_1}}}{t_2\subst{f_2}{\later \fix{f_2}{t_2}}}}$.
\\
Since $t\subst{f}{\later \fix{f}{t}} \equiv \fix{f}{t}$, by \rname{CONV}:
\\
$\jghol{\Delta}{\Sigma}{\Gamma}{\Psi,\later[\res\ltag \ot \later \fix{f_1}{t_1}, \res\rtag \ot \later \fix{f_2}{t_2}]{\Phi}}{\Phi\defsubst{\fix{f_1}{t_1}}{\fix{f_2}{t_2}}}$.
\\
and since $\later[\res\ltag \ot \later \fix{f_1}{t_1}, \res\rtag \ot \later \fix{f_2}{t_2}]{\Phi} \Leftrightarrow \later \Phi\defsubst{\fix{f_2}{t_2}}{\fix{f_2}{t_2}}$,
\\
$\jghol{\Delta}{\Sigma}{\Gamma}{\Psi,\later\Phi\defsubst{\fix{f_1}{t_1}}{fix{f_2}{t_2}}}{\Phi\defsubst{fix{f_1}{t_1}}{fix{f_2}{t_2}}}$,
\\
and finally, by \rname{L\"ob} we get our goal.
\\

\noindent {\bf Case.} $\small\inferrule*[Right=\sf Cons]
      {\jgrhol{\Delta}{\Sigma}{\Gamma}{\Psi}{x_1}{A_1}{x_2}{A_2}{\Phi_h} \\
       \jgrhol{\Delta}{\Sigma}{\Gamma}{\Psi}{xs_1}{\later{\Str{A_1}}}{xs_2}{\later{\Str{A_2}}}{\Phi_t} \\
       \jhol{\Gamma}{\Psi}{\forall x_1,x_2, xs_1, xs_2. \Phi_h\defsubst{x_1}{x_2} \Rightarrow
          \Phi_t\defsubst{xs_1}{xs_2} \Rightarrow \Phi\defsubst{\cons{x_1}{xs_1}}{\cons{x_2}{xs_2}}}}
      {\jgrhol{\Delta}{\Sigma}{\Gamma}{\Psi}{\cons{x_1}{xs_1}}{\Str{A_1}}{\cons{x_2}{xs_2}}{\Str{A_2}}{\Phi}}$
\\ \\
Apply the I.H., \rname{$\forall_E$} and \rname{$\To_E$}.
\\

\noindent {\bf Case.}$\small\inferrule*[Right=\sf Head]
      {\jgrhol{\Delta}{\Sigma}{\Gamma}{\Psi}{t_1}{\Str{A_1}}{t_1}{\Str{A_1}}{\Phi\defsubst{hd\ \res\ltag}{hd\ \res\rtag}}}
      {\jgrhol{\Delta}{\Sigma}{\Gamma}{\Psi}{hd\ t_1}{A_1}{hd\ t_2}{A_2}{\Phi}}$
\\ \\
Trivial by I.H.
\\

\noindent {\bf Case.} $\small\inferrule*[Right=\sf Tail]
      {\jgrhol{\Delta}{\Sigma}{\Gamma}{\Psi}{t_1}{\Str{A_1}}{t_2}{\Str{A_2}}{\Phi\defsubst{tl\ \res\ltag}{tl\ \res\rtag}}}
      {\jgrhol{\Delta}{\Sigma}{\Gamma}{\Psi}{tl\ t_1}{\later{Str_{A_1}}}{tl\ t_2}{\later{Str_{A_2}}}{\Phi}}$
\\ \\
Trivial by I.H.
\\

\noindent {\bf Case.} $\small\inferrule*[Right=\sf Equiv]
      {\jgrhol{\Delta}{\Sigma}{\Gamma}{\Psi}{t'_1}{A_1}{t'_2}{A_2}{\Phi} \\ t_1 \equiv t_1' \\ t_2 \equiv t_2' \\
      \Delta \mid \Gamma \vdash t_1 : A_1 \\ \Delta \mid \Gamma \vdash t_2 : A_2}
      {\jgrhol{\Delta}{\Sigma}{\Gamma}{\Psi}{t_1}{A_1}{t_2}{A_2}{\Phi}}$
\\ \\
Trivial by I.H. and \rname{Conv}.
\\


Most of the proofs for the probabilistic fragment are a consequence of the proof of \autoref{thm:sound-ghol}.
The only interesting case is \rname{Markov}. We do the proof directly in RHOL by showing we can derive it from \rname{Fix}.
We have the premises:
\begin{enumerate}

\item $\jgrhol{\Delta}{\Sigma}{\Gamma}{\Psi}{t_1}{C_1}{t_2}{C_2}{\phi}$

\item $\jgrhol{\Delta}{\Sigma}{\Gamma}{\Psi}{h_1}{C_1 \to \Distr(C_1)}{h_2}{C_2 \to \Distr(C_2)}{\psi_3}$

\item $\jghol{\Delta}{\Sigma}{\Gamma}{\Psi}{\psi_4}$
\end{enumerate}

where:
\begin{align*}
\psi_3 &\equiv \forall x_1 x_2. \phi\defsubst{x_1}{x_2} \Rightarrow \diamond_{[ y_1 \ot \res\ltag\ x_1, y_2 \ot \res\rtag\ x_2]}\phi\defsubst{y_1}{y_2} \\
\psi_4 &\equiv \forall x_1\ x_2\ xs_1\ xs_2. \phi\defsubst{x_1}{x_2} \Rightarrow \later[y_1 \ot xs_1, y_2 \ot xs_2]{\Phi}
       \Rightarrow \Phi\subst{y_1}{\cons{x_1}{xs_1}}\subst{y_2}{\cons{x_2}{xs_2}}
\end{align*}

If we inline the definition of unfold, we have to prove:

\[\begin{aligned}[t]
        \fix{f}{&\lambda x_1. \lambda h_1. \mlet{z_1}{h_1\ x_1}
                                                   { \mlet{t_1}{\operatorname{swap_{\later\Distr}^C}(f_1\app \later z_1 \app \later h_1)}
                                                   { \munit{\cons{x_1}{t_1}}}}
                                      }\sim 
       \\ \fix{f}{&\lambda x_2. \lambda h_2. \mlet{z_2}{h_2\ x_2}
                                                   { \mlet{t_2}{\operatorname{swap_{\later\Distr}^C}(f_2\app  \later z_2 \app \later h_2)}
                                                   { \munit{\cons{x_2}{t_2}}}}
                                      }
        \\\mid\; &\forall x_1 x_2 h_1 h_2. \phi\defsubst{x_1}{x_2} \Rightarrow \psi_3\defsubst{h_1}{h_2} \Rightarrow \diamond_{[y_2 \ot\res\ltag, y_2 \ot\res\rtag]}\Phi
\end{aligned}
\]

We apply \rname{FIX}, \rname{MLET} twice, and then \rname{MUNIT}. The main judgements we have to prove are: 

\begin{enumerate}[label*=(\alph*)]

\item $\jgrhol{\Delta}{\Sigma}{\Gamma}{\Psi}{h_1\ x_1}{\Distr(C_1)}{h_2\ x_2}{\Distr(C_2)}
              {\diamond_{[\res\ltag\ot\res\ltag, \res\rtag\ot\res\rtag]}\phi}$

\item $\jgrhol{\Delta}{\Sigma}{\Gamma}{\Psi}{\operatorname{swap_{\later\Distr}^C}(f_1\app  \later z_1 \app \later h_1)}{\Distr(\later C_1)}
              {\operatorname{swap_{\later\Distr}^C}(f_2\app \later z_2 \app \later h_2)}{\Distr(\later C_2)}
              {\diamond_{[z_1 \ot \res\ltag, z_2 \ot \res\rtag]} \later[y_1 \ot z_1, y_2 \ot z_2]{\Phi}}$

\item $\jgrhol{\Delta}{\Sigma}{\Gamma, x_1,x_2,t_1,t_2}{\Psi, \phi\defsubst{x_1}{x_2}, \later[y_1 \ot t_1, y_2 \ot t_2]{\Phi}}{\cons{y_1}{t_1}}{\Distr(\Str{C_1})}{\cons{y_2}{t_2}}{\Distr(\Str{C_2})}
              {\Phi\subst{y_1}{\res\ltag}\subst{y_2}{\res\rtag}}$
\end{enumerate}

The judgement (a) is a direct consequence of premises (1) and (2), (b) is proven from the inductive hypothesis,
and (d) is a direct consequence of (3). This completes the proof.

\section{Examples}

\subsection{Proof of ZipWith}

This example, taken from \cite{BGCMB16}, proves a
property about the $\operatorname{ZipWith}$ function, which takes two
streams of type A, a function on pairs of elements, and ``zips'' the
two streams by applying that function to the elements that are at the
same position on the two streams.  We want to show that if the
function on the elements is commutative, zipping two streams with that
function is commutative as well.

We can define the zipWith function as:
\begin{align*}
  \operatorname{zipWith} &: (\nat \to \nat \to \nat) \to \Str{\nat} \to \Str{\nat} \to \Str{\nat} \\
  \operatorname{zipWith} &\defeq \fix{\operatorname{zipWith}}{\lambda f. \lambda xs. \lambda ys. \cons{(f\ (\hd\ xs)\ (\hd\ ys))}{(\operatorname{zipWith} \app (\tl\ xs) \app (\tl\ ys))}}
\end{align*}
We prove (omitting types of expressions):
$$\jgrholnocnot{\operatorname{zipWith}}{\operatorname{zipWith}}{\Phi}$$
where
\begin{align*}
 \Phi &\defeq \forall f_1 f_2. (f_1 = f_2 \wedge \forall x y. f_1 x y = f_1 y x) \Rightarrow
 \forall xs_1 xs_2. \forall ys_1 ys_2. (xs_1 = ys_2 \wedge xs_2 = ys_1)\\
 & \Rightarrow
                  \res\ltag\ f_1\ xs_1\ ys_1 = \res\rtag\ f_2\ xs_2\ ys_2 \\
 A    &\defeq  (\nat \to \nat \to \nat) \to \Str{\nat} \to \Str{\nat} \to \Str{\nat}
\end{align*}
The proof proceeds by applying two-sided rules all the way. We invite
interested readers to compare this proof with the one given
in~\cite{BGCMB16} to see how the two approaches differ.

We show how to derive the statement backwards. The derivation begins
with the \rname{Fix} rule. Its premise is (omitting constant contexts):
\[
\jrhol{\operatorname{zipWith}_1,\operatorname{zipWith}_2:\later A}{\later[\res\ltag \leftarrow \operatorname{zipWith}_1,\res\rtag \leftarrow \operatorname{zipWith}_2] \Phi}{\lambda f_1.(\cdots)}{A}
      {\lambda f_2.(\cdots)}{A}{\Phi}
      \]
Then we apply the \rname{ABS} rule three times to introduce into the
context the logical relations on $f_1,f_2,xs_1,xs_2,ys_1$, and $ys_2$. The premise
we need to prove is then:
\[
\begin{array}{c}
\jrhol{\Gamma}{\Psi}{\cons{(f_1\ (hd\ xs_1)\ (hd\ ys_1))}{(\operatorname{zipWith}_1 \app (tl\ xs_1) \app (tl\ ys_1))}}{\Str{\nat}}
      {\\ \cons{(f_2\ (hd\ xs_2)\ (hd\ ys_2))}{(\operatorname{zipWith}_2 \app (tl\ xs_2) \app (tl\ ys_2))}}{\Str{\nat}}{\res\ltag = \res\rtag}
\end{array}
\]
where
\begin{align*}
  \Gamma &\defeq \operatorname{zipWith}_1,\operatorname{zipWith}_2 :\later A; f_1,f_2:(\nat\to\nat\to\nat); xs_1,xs_2,ys_1,ys_2: \Str{\nat} \\
  \Psi   &\defeq \later[\res\ltag \leftarrow \operatorname{zipWith}_1,\res\rtag \leftarrow \operatorname{zipWith}_2] \Phi, (f_1 = f_2 \wedge \forall x y. f x y = f y x), xs_1 = ys_2, xs_2 = ys_1
\end{align*}
Now we can apply the \rname{Cons} rule, which has three premises:
\begin{enumerate}
 \item $\jrhol{\Gamma}{\Psi}{f_1\ (hd\ xs_1)\ (hd\ ys_1)}{\nat}
       {f_2\ (hd\ xs_2)\ (hd\ ys_2)}{\nat}{\res\ltag = \res\rtag}$

 \item $\jrhol{\Gamma}{\Psi}{\operatorname{zipWith}_1 \app (tl\ xs_1) \app (tl\ ys_1)}{\later \Str{\nat}}
       {\operatorname{zipWith}_2 \app (tl\ xs_2) \app (tl\ ys_2)}{\later \Str{\nat}}{\res\ltag = \res\rtag}$

 \item $\jhol{\Gamma}{\Psi}{\forall x y xs ys. x=y \Rightarrow xs = ys \Rightarrow \cons{x}{xs} = \cons{y}{ys}}$
\end{enumerate}
Premise (3) is easily provable in HOL. To prove premise (1) we first apply the \rname{App} rule twice, and we have to prove the judgments:
\begin{itemize}
  \item $\jrhol{\Gamma}{\Psi}{f_1}{\nat\to\nat\to\nat}
       {f_2}{\nat\to\nat\to\nat}{ \forall vs_1 vs_2 ws_1 ws_2.
                                   \\ vs_1 = hd\ ys_2 \wedge vs_2 = hd\ ys_1 \Rightarrow
                                     ws_1 = hd\ xs_2 \wedge ws_2 = hd\ xs_1 \Rightarrow \res\ltag\ vs_1\ ws_1 = \res\rtag\ vs_2\ ws_2}$

  \item $\jrhol{\Gamma}{\Psi}{hd\ xs_1}{\nat}
       {hd\ xs_2}{\nat}{\res\ltag = hd\ ys_2 \wedge \res\rtag = hd\ ys_1}$

  \item $\jrhol{\Gamma}{\Psi}{hd\ ys_1}{\nat}
       {hd\ ys_2}{\nat}{\res\ltag = hd\ xs_2 \wedge \res\rtag = hd\ xs_1}$
\end{itemize}
The three can be proven in HOL from the conditions imposed on $f_1,f_2$ and the equalities $xs_1 = ys_2$, $xs_2 = ys_1$.

All that remains to prove is premise (2) of the \rname{Cons} application, which, by expanding the definition of $\app$ and
using the equational theory of delayed substitutions, can be desugared to:

\[\jrhol{\Gamma}{\Psi}{\latern[\xi_1]{(g_1\ t_1\ u_1)}}{\later \Str{\nat}}
          {\latern[\xi_2]{(g_2\ t_2\ u_2)}}{\later \Str{\nat}}
          {\latern[\xi_1, \xi_2, [\res\ltag\leftarrow\res\ltag, \res\rtag \leftarrow \res\rtag]]{(\res\ltag = \res\rtag)}}\]
where, for $i=1,2$:
\[ \xi_i = [g_i \leftarrow \operatorname{zipWith}_i, t_i\leftarrow(tl\ xs_i), u_i \leftarrow (tl\ ys_i)]\]

We apply the \rname{Next} rule, and we have the four following premises:
\begin{itemize}
  \item $\jrhol{\Gamma}{\Psi}{\operatorname{zipWith}_1}{\later A}
          {\operatorname{zipWith}_2}{\later A}
          {\later[\res\ltag\leftarrow\res\ltag, \res\rtag\leftarrow\res\rtag] (\res\ltag = \res\rtag \wedge \forall x y. \res\ltag x y = \res\ltag y x)}$

  \item $\jrhol{\Gamma}{\Psi}{tl\ xs_1}{\later \Str{\nat}}
          {tl\ xs_2}{\later \Str{\nat}}
          {\later[\res\ltag\leftarrow\res\ltag, \res\rtag\leftarrow\res\rtag] (\res\ltag = tl\ ys_2 \wedge \res\rtag = tl\ ys_1)}$

  \item $\jrhol{\Gamma}{\Psi}{tl\ ys_1}{\later \Str{\nat}}
          {tl\ ys_2}{\later \Str{\nat}}
          {\later[\res\ltag\leftarrow\res\ltag, \res\rtag\leftarrow\res\rtag] (\res\ltag = tl\ xs_2 \wedge \res\rtag = tl\ xs_1)}$

  \item $\jrhol{\Gamma; g_1,g_2: A; t_1,t_2,u_1,u_2: \Str{\nat}}
               {\Psi, g_1 = g_2 \wedge \forall x y. g_1 x y = g_1 y x, t_1 = tl\ ys_2 \wedge t_2 = tl\ ys_1,
                     \\ u_1 = tl\ xs_2 \wedge u_2 = tl\ xs_1}
               {g_1\ t_1\ u_1}{\Str{\nat}}
               {g_2\ t_2\ u_2}{\Str{\nat}}
               {\res\ltag = \res\rtag}$

\end{itemize}

To prove the first premise we instantiate the inductive hypothesis we got
from \rname{Fix}. To prove the second and the third premises we use the
equalities $xs_1 = ys_2$, $xs_2 = xs_1$. Finally, the fourth premise is a
simple derivation in HOL that follows from the same equalities plus
the refinements of $g_1,g_2,t_1,t_2,u_1,u_2$. This concludes the
proof.

\subsection{Proof of approximation series}

We now continue with another example that, while still being fully synchronous
(i.e., uses only two-sided rules), goes beyond reasoning about equality of
streams, and showcases the flexibility of streams to represent different kinds
of information and structures.

For instance, streams can be used to represent series of numbers. In
this example, we illustrate an instance of a property about series
that can be proven in our system.  Consider the series $x_0, x_1,
\ldots$ for any $p \geq \frac{1}{2}$ and any $a \geq 0$, where $x_0$
is given and:
\[ x_{i+1} = px_i + (1-p) \frac{a}{x_i} \]
It can be easily shown that if $x_0 \geq \sqrt{a}$, then this series
converges \emph{monotonically} from the top to $\sqrt{a}$. In
particular, $\displaystyle\lim_{i \to \infty} x_i = \sqrt{a}$. (For $p
= \frac{1}{2}$, this is the standard Newton-Raphson series for
square-root computation \cite{Scott11})

The interesting relational property is that for smaller $p$, this
series converges faster. Concretely, define $f(p, a, x_0, i)$ as the
$i$th element of the above series (for the given $p$, $a$ and
$x_0$). Then, the relational property to prove is that:
\[ \forall p_1\, p_2\, a\, x_0\, i.\, (\dfrac{1}{2}\leq p_1 \leq p_2 \conj x_0 \geq \sqrt{a}) \Rightarrow |f(p_1,a,x_0,i) - \sqrt{a}| \leq |f(p_2,a,x_0,i)- \sqrt{a}|\]

We outline the proof of this property. First, note that because
convergence is from the top, $|f(p_2,a,x_0,i) - \sqrt{a}| =
f(p_2,a,x_0,i) - \sqrt{a}$. Therefore, the property above is the same
as:
\[ \forall p_1\, p_2\, a\, x_0\, i.\, (\dfrac{1}{2}\leq p_1 \leq p_2\conj x_0 \geq \sqrt{a}) \Rightarrow f(p_2,a,x_0,i) - f(p_1,a,x_0,i) \geq 0\]
This is easy to establish by induction on $i$.

(Note the importance of the assumption $p \geq \frac{1}{2}$: Without
this assumption, convergence is not monotonic, and this relational
property may not hold. If we start with $x_0 \leq \sqrt{a}$ instead of
$x_0 \geq \sqrt{a}$, we need $p \leq \frac{1}{2}$ for convergence to
be monotonic, this time from below.)

Now we see how we can encode and prove this as a relational property
of a pair of streams.  We can define a stream whose elements are the
elements of one of this series:
\begin{align*}
  \operatorname{approx\_sqrt} &: \real \to \real \to \real \to \Str{\real} \\
  \operatorname{approx\_sqrt} &\defeq \fix{f}{\lambda p. \lambda a. \lambda x.
   \cons{x}{(f \app \later p \app \later a \app \later (p*x + (1-p)*a/x))}}
\end{align*}
We prove:
\[\jgrholnoc{\operatorname{approx\_sqrt_1}}{\real\to\real\to\real\to \Str{\real}}{\operatorname{approx\_sqrt_2}}{\real\to\real\to\real\to \Str{\real}}
         {\Phi}
\]
where
\begin{align*}
\Phi &\defeq \forall p_1 p_2. (\dfrac{1}{2}\leq p_1 \leq p_2) \Rightarrow \forall a_1 a_2. 0 \leq a_1 = a_2 \Rightarrow \forall x_1 x_2. (0 \leq x_1 \leq x_2 \wedge a_1\leq x_1*x_1)
          \\ &\Rightarrow \All(\res\ltag\ p_1\ a_1\ x_1, \res\rtag\ p_2\ a_2\ x_2, \lambda n_1 n_2. 0 \leq n_1 \leq n_2 \wedge a_1 \leq n_1 * n_1) 
\end{align*}
and $\All$ is defined axiomatically as follows:
\[\forall s_1,s_2,n_1,n_2. \phi n_1 n_2 \Rightarrow \later[s'_1\ot s_1, s'_2 \ot s_2]{\All(s'_1,s'_2,\lambda x_1 x_2. \phi)} \Rightarrow \All(\cons{n_1}{s_1}, \cons{n_2}{s_2}, \lambda x_1 x_2. \phi)\]

The meaning of the judgement is that, if we have two approximation series for
the square root of $a$ (formally, we write $a = a_1 = a_2$), with
initial guesses $x_1 \leq x_2$, and parameters $1/2 \leq p_1 \leq
p_2$, then, at every position, the first series is going to be closer
to the root than the second one. Note that we have removed the square
roots in the specification by squaring.

Let $A =\defeq \real\to\real\to\real\to \Str{\real}$.  We will show
how to derive the judgment backwards. The proof starts by applying
\rname{Fix} which has the premise (omitting constant contexts):
\[\jrhol{f_1,f_2 : \later A}{\later[\res\ltag, \res\rtag \leftarrow f_1, f_2]\Phi}{\lambda p_1. \lambda a_1. \lambda x_1. \dots}{A}{\lambda p_2. \lambda a_2. \lambda x_2. \dots}{A}
         {\Phi}
\]
and after applying \rname{Abs} three times:
\[\begin{array}{c}
    \jrhol{f_1,f_2 : \later A; p_1,p_2,a_1,a_2,x_1,x_2 : \real}
        {\\ \later[\res\ltag, \res\rtag \leftarrow f_1, f_2]\Phi, (\dfrac{1}{2}\leq p_1 \leq p_2) , 0 \leq a_1 = a_2 , 0 \leq x_1 \leq x_2, a_1 \leq x_1*x_1}
        {\\ (\lambda y_1.\cons{x_1}{(f_1\app \later p_1\app \later a_1\app \later y_1)})(p_1*x_1 + (1-p_1)*a_1/x_1)}{\Str{\real}}
        {\\ (\lambda y_2.\cons{x_2}{(f_2\app \later p_2\app \later a_2\app \later y_2)})(p_2*x_2 + (1-p_2)*a_2/x_2)}{\Str{\real}}
        {\\ All(\res\ltag, \res\rtag, \lambda n_1 n_2. n_1 \leq n_2)}
\end{array}
\]

Let $\Gamma$ and $\Psi$ denote the typing and logical contexts in the previous judgement. Now we apply \rname{App}, which has two premises:
\begin{itemize}
  \item $\jrhol{\Gamma}{\Psi}
        {\lambda y_1.\cons{x_1}{(f_1\app \later p_1\app \later a_1\app \later y_1)}}{\real\to \Str{\real}}
        {\lambda y_2.\cons{x_2}{(f_2\app \later p_2\app\later a_2\app\later y_2)}}{\real\to \Str{\real}}
        {\\ \forall y_1, y_2. (0 \leq y_1 \leq y_2 \wedge a_1 \leq y_1 * y_1) \Rightarrow All(\res\ltag\ y_1, \res\rtag\ y_2, \lambda n_1 n_2. 0\leq n_1\leq n_2 \wedge a_1 \leq n_1 * n_1)}$
        
  \item $\jrhol{\Gamma}{\Psi}
        {p_1*x_1 + (1-p_1)*a_1/x_1}{\Str{\real}}
        {p_2*x_2 + (1-p_2)*a_2/x_2}{\Str{\real}}
        {\\ 0 \leq \res\ltag \leq \res\rtag \wedge a_1 \leq \res\ltag * \res\ltag}$
\end{itemize}

The second premise can be established in Guarded HOL as an arithmetic
property in our theory of reals. To prove the first one, we start by applying the \rname{Abs}
rule, followed by the \rname{Cons} rule, which has three premises:
\begin{enumerate}
  \item $\jrhol{\Gamma, y_1,y_2 : \real}{\Psi, (y_1 \leq y_2 \wedge y_1*y_1 \geq a_1)}
        {x_1}{\real}
        {x_2}{\real}
        {0 \leq \res\ltag \leq \res\rtag \wedge a \leq \res\ltag * \res\ltag}$
  \item $\jrhol{\Gamma, y_1,y_2 : \real}{\Psi, (y_1 \leq y_2 \wedge y_1*y_1 \geq a_1)}
        {(f_1\app\later p_1\app\later a_1\app\later y_1)}{\later \Str{\real}}
        {\\ (f_2\app\later p_2\app\later a_2\app\later y_2)}{\later \Str{\real}}
        {\later[\res\ltag \ot \res\ltag, \res\rtag\ot\res\rtag]All(\res\ltag, \res\rtag, \lambda n_1 n_2. 0 \leq n_1 \leq n_2 \wedge a \leq n_1 * n_1)}$
  \item $\jhol{\Gamma, y_1,y_2 : \real}{\Psi, (y_1 \leq y_2 \wedge y_1*y_1 \geq a_1)}
        {\forall h_1 h_2 t_1 t_2. 0 \leq h_1 \leq h_2 \Rightarrow a_1 \leq h_1*h_1 \Rightarrow\\
          \later[\res\ltag\ot t_1, \res\rtag \ot t_2] All(\res\ltag, \res\rtag, \lambda n_1 n_2. 0\leq n_1 \leq n_2 \wedge a \leq n_1*n_1) \Rightarrow \\
          All(\cons{h_1}{t_1}, \cons{h_2}{t_2}, \lambda n_1 n_2. 0\leq n_1 \leq n_2 \wedge a \leq n_1*n_1))}$

\end{enumerate}

Premise (1) is just the refinement on $x_1,x_2$, while premise (3) is
the axiomatization of $All$. To prove
premise (2) one instantiates the induction hypothesis given by the
\rname{Fix} rule. In order to do so, we first rewrite the two terms we are comparing to their desugared form:
\[\later[f_1' \ot f_1, p_1' \ot \later p_1, a_1' \ot \later a_1, y_1' \ot \later y_1] f_1'\ p_1'\ a_1'\ y_1'\]
and
\[\later[f_2' \ot f_2, p_2' \ot \later p_2, a_2' \ot\later a_2, y_2' \ot\later y_2] f_2'\ p_2'\ a_2'\ y_2'\]
We can also add by \rname{SUB} the same substitutions to the $\later$
in the conclusion, since the substituted variables do not appear in
the formula. Then we can apply the \rname{Next} rule, which has the premises:
\begin{itemize}
  \item $\jrhol{\Gamma, y_1,y_2 : \real}{\Psi, (y_1 \leq y_2 \wedge y_1*y_1 \geq a_1)}
        {f_1}{\later A}
        {f_2}{\later A}
        {\later[\res\ltag\ot\res\ltag, \res\rtag\ot\res\ltag]\Phi}$
  \item $\jrhol{\Gamma, y_1,y_2 : \real}{\Psi, (y_1 \leq y_2 \wedge y_1*y_1 \geq a_1)}
        {\later p_1}{\real}
        {\later p_2}{\real}
        {\\ \later[\res\ltag\ot\res\ltag, \res\rtag\ot\res\ltag] \dfrac{1}{2}\leq \res\ltag \leq \res\rtag}$
  \item $\jrhol{\Gamma, y_1,y_2 : \real}{\Psi, (y_1 \leq y_2 \wedge y_1*y_1 \geq a_1)}
        {\later a_1}{\real}
        {\later a_2}{\real}
        {\later[\res\ltag\ot\res\ltag, \res\rtag\ot\res\ltag] 0 \leq \res\ltag = \res\rtag}$
  \item $\jrhol{\Gamma, y_1,y_2 : \real}{\Psi, (y_1 \leq y_2 \wedge y_1*y_1 \geq a_1)}
        {\later y_1}{\real}
        {\later y_2}{\real}
        {\\ \later[\res\ltag\ot\res\ltag, \res\rtag\ot\res\ltag] 0 \leq \res\ltag \leq \res\rtag \wedge a_1' \leq \res\ltag * \res\ltag}$
  \item $\jrhol{\Gamma, y_1,y_2,p_1',p_2',a_1',a_2',y_1',y_2' : \real; f_1', f_2' : A}
        {\Psi, (y_1 \leq y_2 \wedge y_1*y_1 \geq a_1), \Phi\defsubst{f_1'}{f_2'},\\  \dfrac{1}{2}\leq p_1' \leq p_2', 0 \leq a_1' = a_2', 0 \leq y_1' \leq y_2' \wedge a_1' y_1' * y_1'}
        {\\ f_1'\ p_1'\ a_1'\ y_1'}{\Str{\real}}
        {f_2'\ p_2'\ a_2'\ y_2'}{\Str{\real}}{All(\res\ltag, \res\rtag, \lambda n_1 n_2. 0 \leq n_1 \leq n_2 \wedge a \leq n_1 * n_1)}$
 
\end{itemize}

The first four can be proven simply by instantiating and then delaying
one of the axioms. The last one is proven by applying \rname{App} three
times. This concludes the proof.

\subsection{Proof of Cassini's identity}

We continue building on the idea from the previous example of using
streams to represent series of numbers. This time, we prove a
classical identity of the Fibonacci sequence. Since the example
requires to observe the stream at different times, we will also have
to deal with some asynchronicity on the delayed substitutions.

Let $F_n$ be the $n$th Fibonacci number. Cassini's identity states
that $F_{n-1} \cdot F_{n+1} - F_{n}^2 = (-1)^{n}$. Cassini's identity
can be stated as a stream problem as follows. First, let $F$ be the
Fibonnaci stream ($1,1,2,3,5,\ldots$) and $A$ be the stream
$1,-1,1,-1,\ldots$ Let $\oplus$ and $\otimes$ be infix functions that
add and multiply two streams pointwise. Cassini's identity can then be
informally written as:
\[ F \otimes  \tl(\tl\; F) = \tl(F \otimes F) \oplus A \]

In order to formalize Cassini's identity in our system, we first define:
\begin{align*}
  \oplus &: \Str{\nat} \to \Str{\nat} \to \Str{\nat}   &\otimes &: \Str{\nat} \to \Str{\nat} \to \Str{\nat} \\                            
  \oplus &\defeq \begin{array}{l}\fix{f}{\lambda s.\lambda t\\ \cons{(\hd\ x + \hd\ y)}{(f \app (\tl\ x) \app (\tl\ y))}}\end{array} \quad \quad
  &\otimes &\defeq \begin{array}{l}\fix{f}{\lambda s.\lambda t\\ \cons{(\hd\ x * \hd\ y)}{(f \app (\tl\ x) \app (\tl\ y))}}\end{array}
\end{align*}
Then we define $F$ and $A$ as the fixpoints of the equations:
\begin{align*}
  F &\defeq \fix{F}{\cons{1}{\later[ F' \ot F] (\cons{1}{\later [ T \ot \tl\ F' ] ( F' \oplus T)})}} \\
  A &\defeq \fix{A}{\cons{1}{\later(\cons{-1}{A})}}
\end{align*}

We prove (using prefix notation for $\oplus$ and $\otimes$):
\[\jgrholnoc
           {\later [ T_1 \ot \tl\ F] \otimes \app (\later F) \app \tl\ T_1 }{\later\later \Str{\nat}}
           {\oplus \app \tl(F \otimes F) \app (\later A)}{\later \Str{\nat}}
           { \res\ltag = \later \res\rtag}
\]

The proof combines applications of two-sided rules and one-sided
rules; in particular, we use the rule \rname{NEXT-L} to proceed with
the proof for a judgement where the left expression is delayed twice
and the right expression is delayed once.

By conversion, in the logic we can prove the following equalities:
\[
  \Psi \defeq \left\{\begin{aligned}
  F &= \cons{1}{\later(\cons{1}{\later [ T \ot tl\ F ] ( F \oplus T)})}, \\
  A &= \cons{1}{\later(\cons{-1}{\later A})}
  \end{aligned}\right\}
\]

Using these equalities, and desugaring the applications, the judgment we want to prove is (omitting constant contexts):

\[\begin{array}{c}\jrhol{F,A: \Str{\nat}}{\Psi}
           {\later [ T_1 \ot tl\ F] \later [T_1' \ot tl\ T_1] (F \otimes T_1') }{\later\later \Str{\nat}}
           {\\ \later [T_2 \ot tl(F \otimes F)] (T_2 \oplus A)}{\later \Str{\nat}}
           {\\ \later[\res\ltag' \ot \res\ltag, T_1 \ot tl\ F] \later [\res\ltag''\ot \res\ltag, \res\rtag'\ot\res\rtag, T_1' \ot tl\ T_1, T_2' \ot tl(F \otimes F)] \res\ltag'' = \res\rtag'}
\end{array}
\]

Notice that on the left, since we want to apply tail twice to $F$, we
need to delay the term twice so that $F$ and $tl\ tl\ F$ have the same
type. On the right, we just need to delay the term once. As for the
logical conclusion, $\res\ltag$ needs to be delayed twice, while
$\res\rtag$ only once. The way to do this is by having $\res\ltag$
appear on the two substitutions but $\res\rtag$ only on the inner one.

We start by applying \rname{NEXT-L}, which has the two following premises:
\begin{itemize}
  \item $\juhol{F,A: \Str{\nat}}{\Psi}{tl\ F }{\later \Str{\nat}}{\later[\res' \ot \res] tl\ F = \later \res'}$

  \item $\jrhol{F,A, T_1: \Str{\nat}}{\Psi, tl\ F = \later T_1}{\later [T_1' \ot tl\ T_1] (F \otimes T_1') }{\later \Str{\nat}}
              {\\ \later [T_2 \ot tl(F \otimes F)] (T_2 \oplus A)}{\later \Str{\nat}}
              {\later [\res\ltag''\ot \res\ltag, \res\rtag'\ot\res\rtag, T_1' \ot tl\ T_1, T_2' \ot tl(F \otimes F)] \res\ltag'' = \res\rtag'}$

\end{itemize}

The first premise is trivial. We continue by applying \rname{NEXT} to the second, which has the following premises:

\begin{itemize}
  \item $\jrhol{F,A, T_1: \Str{\nat}}{\Psi, tl\ F = \later T_1}{tl\ T_1}{\later \Str{\nat}}{tl(F \otimes F)}{\later \Str{\nat}}
        {\\ \later[\res\ltag' \ot \res\ltag, \res\rtag' \ot \res\rtag] T_1 = \later \res\ltag' \wedge \res\ltag'\otimes\res\ltag' = \res\rtag'}$

  \item $\jrhol{F,A, T_1', T_2: \Str{\nat}}{\Psi,tl\ F = \later T_1 , tl\ T_1 = \later T_1', T_1'\otimes T_1' = T_2}{F \otimes T_1'}{\Str{\nat}}{T_2 \oplus A}{\Str{\nat}}{\res\ltag = \res\rtag}$

\end{itemize}

Again, the first premise is trivial. We apply \rname{APP} twice to the second, and we have to prove:

\begin{itemize}
  \item $\jrhol{F,A, T_1', T_2: \Str{\nat}}{\Psi,tl\ F = \later T_1 , tl\ T_1 = \later T_1', T_1'\otimes T_1' = T_2}{F}{\Str{\nat}}{A}{\Str{\nat}}{\\ \res\ltag = F \wedge \res\rtag = A}$

  \item $\jrhol{F,A, T_1', T_2: \Str{\nat}}{\Psi,tl\ F = \later T_1 , tl\ T_1 = \later T_1', T_1'\otimes T_1' = T_2}{T_1'}{\Str{\nat}}{T_2}{\Str{\nat}}
        {\\ F = \cons{1}{\later(\cons{1}{\later T_1})} \wedge \res\ltag \otimes \res\rtag = \res\rtag}$

  \item $\jrhol{F,A, T_1', T_2: \Str{\nat}}{\Psi,tl\ F = \later T_1 , tl\ T_1 = \later T_1', T_1'\otimes T_1' = T_2}
         {\\ \otimes}{\Str{\nat} \to \Str{\nat} \to \Str{\nat}}{\oplus}{\Str{\nat} \to \Str{\nat} \to \Str{\nat}}
         {\forall X_1 X_2 Y_1 Y_2. X_1 = F \wedge X_2 = A \Rightarrow F = \cons{1}{\later(\cons{1}{\later Y_1})} \wedge Y_1 \otimes Y_1 = Y_2 \Rightarrow \res\ltag\ X_1\ Y_1 = \res\rtag\ X_2\ Y_2}$
\end{itemize}

The two first premises are easy to prove. We will show how to prove the last one. 
For this, we need a stronger induction hypothesis for $\hat{\oplus}$ and
$\hat{\otimes}$. We propose the following:
\[\begin{array}{c}
  \forall g_1,g_2,b_1,G,B. G = \conshat{g_1}{\conshat{g_2}{(G\hat{\oplus} (\hat{tl} G))}} \wedge b_1 = g_1^2 + g_1 g_2 - g_2^2 \wedge B = \conshat{b_1}{\conshat{-b_1}{B}}
   \\  \Rightarrow G \hat{\otimes} \hat{tl}(\hat{tl} G) = \hat{tl} (G \hat{\otimes} G) \hat{\oplus} B 
\end{array}
\]

We then use the \rname{SUB} rule to strengthen the inductive hypothesis, and now the new judgement to prove is:

\[\begin{array}{c}
  \jrhol{F,A, T_1', T_2: \Str{\nat}}{\Psi,tl\ F = \later T_1 , tl\ T_1 = \later T_1', T_1'\otimes T_1' = T_2}
         {\\ \otimes}{\Str{\nat} \to \Str{\nat} \to \Str{\nat}}{\oplus}{\Str{\nat} \to \Str{\nat} \to \Str{\nat}}
         {\\ \forall X_1 X_2 Y_1 Y_2. (\exists g_1, g_2, b_1, G, B. G = \cons{g_1}{\later(\cons{g_2}{\later [ G' \ot tl\ G ] ( G \oplus G')})}
           \wedge b_1 = g_1^2 + g_1 g_2 - g_2^2 \wedge \\ B = \cons{b_1}{\later(\cons{-b_1}{\later B})} \wedge
          X_1 = G \wedge X_2 = B \wedge X_1 = \cons{1}{\later(\cons{1}{\later Y_1})} \wedge Y_1 \otimes Y_1 = Y_2) \Rightarrow \res\ltag\ X_1\ Y_1 = \res\rtag\ X_2\ Y_2}
\end{array}
\]

Let $\Gamma'$, $\Psi'$ and $\Phi_{IH}$ denote respectively the typing context, logical context and logical conclusion of the previous judgement. The premise
of the FIX rule is:

\[\begin{array}{c}
  \jrhol{\Gamma; f_1,f_2 : \later(\Str{\nat} \to \Str{\nat} \to \Str{\nat})}{\Psi',\later[\res\ltag \ot f_1, \res\rtag \ot f_2] \Phi_{IH}}
         {\\ \fix{f_1}{\lambda X_1. \lambda Y_1. \dots}}{\Str{\nat} \to \Str{\nat} \to \Str{\nat}}{\fix{f_2}{\lambda X_2. \lambda Y_2. \dots}}{\Str{\nat} \to \Str{\nat} \to \Str{\nat}}{\Phi_{IH}}
\end{array}
\]

Let $\Phi_E$ denote the existential clause in $\Phi_{IH}$. After applying \rname{ABS} twice, we have:

\[\begin{array}{c}
   \jrhol{\Gamma; f_1,f_2 : \later(\Str{\nat} \to \Str{\nat} \to \Str{\nat}); X_1,X_2,Y_1,Y_2 : \Str{\nat}}{\Psi',\later[\res\ltag \ot f_1, \res\rtag \ot f_2] \Phi_{IH}, \Phi_E}
         {\\ \cons{(hd\ X_1)*(hd\ Y_1)}{f_1 \app (tl\ X_1) \app (tl\ Y_1)}}{\Str{\nat}}{\cons{(hd\ X_2)+(hd\ Y_2)}{f_2 \app (tl\ X_2) \app (tl\ Y_2)}}{\Str{\nat}}{\res\ltag = \res\rtag}
\end{array}\]

And then we apply \rname{Cons} to prove equality on the heads and the tails:

\begin{itemize}
\item  $\jrhol{\Gamma; f_1,f_2 : \later(\Str{\nat} \to \Str{\nat} \to \Str{\nat}); X_1,X_2,Y_1,Y_2 : \Str{\nat}}{\Psi',\later[\res\ltag \ot f_1, \res\rtag \ot f_2] \Phi_{IH}, \Phi_E}
         {\\ (hd\ X_1)*(hd\ Y_1)}{\nat}{(hd\ X_2)+(hd\ Y_2)}{\nat}{\res\ltag = \res\rtag}$
\item  $\jrhol{\Gamma; f_1,f_2 : \later(\Str{\nat} \to \Str{\nat} \to \Str{\nat}); X_1,X_2,Y_1,Y_2 : \Str{\nat}}{\Psi',\later[\res\ltag \ot f_1, \res\rtag \ot f_2] \Phi_{IH}, \Phi_E}
         {\\ f_1 \app (tl\ X_1) \app (tl\ Y_1)}{\later \Str{\nat}}{f_2 \app (tl\ X_2) \app (tl\ Y_2)}{\later \Str{\nat}}{\res\ltag = \res\rtag}$
\end{itemize}

To prove the first one we notice that $hd X_1 * hd Y_1 = g_1 * (g_1 + g_2) = g_1^2 + g_2*g_1 =  g_2^2 + g_1^2 + g_1 * g_2 - g_2^2 = hd X_2 * hd Y_2$. To prove the second one
we need to check that $tl X_1,tl Y_1, tl X_2,tl Y_2$ satisfy the precondition of the inductive hypothesis. In particular, we need to check that
\[-b_1 = -g_1^2 - g_1 g_2 + g_2^2 = g_2^2 + g_2 (g_1 + g_2) - (g_1 + g_2)^2\]
which is can be proven by arithmetic computation.

\section{Unary fragment}

In this section we introduce a unary system to prove properties
about a single term of the guarded lambda calculus. We will start
by adding some definitions Guarded HOL for the unary diamond monad,
following by the derivation rules for both the non-probabilistic
and the probabilistic system, plus the metatheory and an example.

\subsection{Unary fragment of GHOL}

The unary semantics of the diamond monad are:

\[\sem{\diamond_{[x\leftarrow t]}\phi}_i\defeq \left\{(\delta,\gamma) \isetsep
    \operatorname{Pr}_{v\ot \left(\sem{t}_i(\delta,\gamma)\right)} [(\delta, (\gamma, v))\in \sem{\phi}_i] = 1 \right\} \]

The rules are on Figure~\ref{fig:u-prob-ghol}
    
\begin{figure*}[!htb]
  \small
\begin{mathpar}
  \infer[\sf MONO1]
  {\jghol{\Delta}{\Sigma}{\Gamma}{\Psi}{\diamond_{[x\gets t]}\psi}}
  {\jghol{\Delta}{\Sigma}{\Gamma}{\Psi}{\diamond_{[x\gets t]}\phi} \and
    \jghol{\Delta}{\Sigma}{\Gamma, x : C}{\Psi, \phi}{\psi}
  }

\and

  \infer[\sf UNIT1]{\jghol{\Delta}{\Sigma}{\Gamma}{\Psi}{
    \diamond_{[x\leftarrow \munit{t}]} \phi}}
       {\jghol{\Delta}{\Sigma}{\Gamma}{\Psi}{
           \phi[t/x]}}
\and
 
\infer[\sf MLET1]
      {\jghol{\Delta}{\Sigma}{\Gamma}{\Psi}{
          \diamond_{[y\leftarrow \mlet{x}{t}{t'}]} \psi}}{
          \jghol{\Delta}{\Sigma}{\Gamma}{\Psi}{\diamond_{[x\leftarrow t]} \phi}
          \and
          \jghol{\Delta}{\Sigma}{\Gamma,x:C}{\Psi,\phi}{
            \diamond_{[y\leftarrow t']}\psi}}

\end{mathpar}
\caption{Rules for the unary diamond modality}\label{fig:u-prob-ghol}
\end{figure*}

\subsection{Guarded UHOL}

We start by defining the Guarded UHOL system, which allows us to prove
logical properties of a term of the Guarded Lambda Calculus.  More
concretely, judgements have the form:
\[ \jguhol{\Delta}{\Sigma}{\Gamma}{\Psi}{t}{\sigma}{\phi} \]
where $t$ is a term well-typed in the dual context $\Delta \mid \Gamma$ and $\phi$ is a logical formula well-typed in the context
$\Delta \mid \Gamma, \res : \sigma$ and that can refer to $t$ via the special variable $\res$. The logical contexts $\Sigma$ and
$\Psi$ consist respectively of refinements over the contexts $\Delta$ and $\Gamma$.

\subsection{Derivation rules}

The rule \rname{Next} corresponds to the introduction of the later
modality. A refinement $\Phi_i$ is proven on every term in the
substitution, and using those as a premise, a refinement $\Phi$ is
proven on $t$. In the notation $\later[\res \ot \res]\Phi$ the first
$\res$ is the variable bound by the delayed substitution inside $\Phi$
while the second $\res$ is the distinguished variable in the refinement
that refers to the term that is being typed. In other words, $t$ satisfies
$\later[\res \ot \res]\Phi$ if $\later[\res \ot t]\Phi$.
The rule \rname{Prev} corresponds to the elimination of the later modality.
If we can prove $\later \phi$ in a constant context, then we can also prove
$\phi$.
The rule \rname{Box} applies the constant modality on a formula that can
be proven on a constant context.
The rule \rname{LetBox} removes the constant modality from a formula $\Phi$
by using it as a constant premise to prove another formula $\Phi'$.
The rule \rname{LetConst} shifts constant terms between contexts.
The rule \rname{Fix} introduces a fixpoint and proves a refinement on it by
Loeb induction.
The rule \rname{Cons} proves a property on a stream from a refinement on its
head and its tail.
The rule \rname{ConsHat} is the analogue of \rname{Cons} to build constant streams.
In particular, the $\hat{::}$ operator can be defined as $\lambda x. \lambda s. \letbox{(y,t)}{(x,s)}{\boxx(\cons{y}{\later{t}})}$.
Conversely the rules \rname{Head} and \rname{Tail} respectively prove a property
on the head and the tail of a stream from a property on the full stream.

\begin{figure*}[!htb]
\small
\begin{mathpar}
\inferrule*[Right=\sf Next]
      {\jguhol{\Delta}{\Sigma}{\Gamma, x_1:A_1, \dots, x_n:A_n}{\Psi, \Phi_1\subst{\res}{x_1}, \dots, \Phi_n\subst{\res}{x_n}}{t}{A}{\Phi} \\
       \jguhol{\Delta}{\Sigma}{\Gamma}{\Psi}{t_1}{\later{A_1}}{\dsubst{\res}{\res}{\Phi_1}}
       \\ \dots \\
       \jguhol{\Delta}{\Sigma}{\Gamma}{\Psi}{t_n}{\later{A_n}}{\dsubst{\res}{\res}{\Phi_n}}}
      {\jguhol{\Delta}{\Sigma}{\Gamma}{\Psi}{\nextt{x_1 \leftarrow t_1, \dots, x_n \leftarrow t_n}{t}}{\later{A}}{\dsubst{x_1,\dots,x_n,\res}{t_1,\dots,t_n,\res}{\Phi}}}

 \inferrule*[Right=\sf Prev]
      {\jguhol{\Delta}{\Sigma}{\cdot}{\cdot}{t}{\later A}{\dsubst{\res}{\res}{\Phi}}}
      {\jguhol{\Delta}{\Sigma}{\Gamma}{\Psi}{\prev{t}}{A}{\Phi}} 

 \inferrule*[Right=\sf Box]
      {\jguhol{\Delta}{\Sigma}{\cdot}{\cdot}{t}{A}{\Phi}}
      {\jguhol{\Delta}{\Sigma}{\Gamma}{\Psi}{\boxx{t}}{\square A}{\square \Phi\subst{\res}{\letbox{x}{\res}{x}}}} 

 \inferrule*[Right=\sf LetBox]
      {\jguhol{\Delta}{\Sigma}{\Gamma}{\Psi}{u}{\square B}{\square \Phi\subst{\res}{\letbox{x}{\res}{x}}} \\
       \jguhol{\Delta, x : B}{\Sigma, \Phi\subst{\res}{x}}{\Gamma}{\Psi}{t}{A}{\Phi'}}
      {\jguhol{\Delta}{\Sigma}{\Gamma}{\Psi}{\letbox{x}{u}{t}}{A}{\Phi'}}

  \inferrule*[Right=\sf LetConst]
      {\jguhol{\Delta}{\Sigma}{\Gamma}{\Psi}{u}{B}{\Phi} \\
       \jguhol{\Delta, x : B}{\Sigma, \Phi\subst{\res}{x}}{\Gamma}{\Psi}{t}{A}{\Phi'} \\
       B,\Phi\ \text{constant} \\ FV(\Phi)\cap FV(\Gamma) = \emptyset}
      {\jguhol{\Delta}{\Sigma}{\Gamma}{\Psi}{\letconst{x}{u}{t}}{A}{\Phi'}}

  \inferrule*[Right=\sf Fix]
      {\jguhol{\Delta}{\Sigma}{\Gamma, f:\later{A}}{\dsubst{\res}{f}{\Phi}}{t}{A}{\Phi}}
      {\jguhol{\Delta}{\Sigma}{\Gamma}{\Psi}{fix f. t}{A}{\Phi}} 

  \inferrule*[Right=\sf Cons]
      {\jguhol{\Delta}{\Sigma}{\Gamma}{\Psi}{x}{A}{\Phi_h} \\
       \jguhol{\Delta}{\Sigma}{\Gamma}{\Psi}{xs}{\later{\Str{A}}}{\Phi_t} \\
       \jhol{\Gamma}{\Psi}{\forall x,xs. \Phi_h\subst{\res}{x} \Rightarrow \Phi_t\subst{\res}{xs} \Rightarrow \Phi\subst{\res}{\cons{x}{xs}}}}
      {\jguhol{\Delta}{\Sigma}{\Gamma}{\Psi}{\cons{x}{xs}}{\Str{A}}{\Phi}}

  \inferrule*[Right=\sf ConsHat]
      {\jguhol{\Delta}{\Sigma}{\Gamma}{\Psi}{x}{A}{\Phi_h} \\
       \jguhol{\Delta}{\Sigma}{\Gamma}{\Psi}{xs}{\square \Str{A}}{\square \Phi_t} \\
       \jhol{\Gamma}{\Psi}{\forall x,xs. \Phi_h\subst{\res}{x} \Rightarrow \Phi_t\subst{\res}{xs} \Rightarrow \Phi\subst{\res}{\conshat{x}{xs}}} \\
      A,\Phi_h \text{ constant} }
      {\jguhol{\Delta}{\Sigma}{\Gamma}{\Psi}{\conshat{x}{xs}}{\square \Str{A}}{\square \Phi}}

  \inferrule*[Right=\sf Head]
      {\jguhol{\Delta}{\Sigma}{\Gamma}{\Psi}{t}{\Str{A}}{\Phi\subst{\res}{hd\ \res}}}
      {\jguhol{\Delta}{\Sigma}{\Gamma}{\Psi}{hd\ t}{A}{\Phi}} 
      
   \inferrule*[Right=\sf Tail ]
      {\jguhol{\Delta}{\Sigma}{\Gamma}{\Psi}{t}{\Str{A}}{\Phi\subst{\res}{tl\ \res}}}
      {\jguhol{\Delta}{\Sigma}{\Gamma}{\Psi}{tl\ t}{\later{Str_A}}{\Phi}}

\end{mathpar}
\caption{Guarded Unary Higher-Order Logic rules}\label{fig:uhol}
\end{figure*}

The intended meaning for a judgment $\jguhol{\Delta}{\Sigma}{\Gamma}{\Psi}{t}{\tau}{\phi}$ is:
``For every valuations $\delta$, $\gamma$ of $\Delta$ and $\Gamma$,
\[\sem{\Delta\mid\Gamma\vdash \square\Sigma}(\delta,\gamma) \wedge \sem{\Delta\mid\Gamma\vdash \Psi}(\delta,\gamma) \To
 \sem{\Delta\mid\Gamma, \res:\tau\vdash \Sigma}(\delta,\langle \gamma, \sem{\Delta\mid\Gamma\vdash t}(\delta,\gamma)\rangle) \text{''} \]

\subsection{Metatheory}

We now the most interesting metatheoretical properties of Guarded UHOL. In particular, Guarded UHOL
is equivalent to Guarded HOL:

\begin{theorem}[Equivalence with Guarded HOL] \label{thm:equiv-uhol-hol}
  For every contexts $\Delta,\Gamma$, type $\sigma$, term $t$, sets of assertions $\Sigma,\Psi$
  and assertion $\phi$, the following are equivalent:
  \begin{itemize}
  \item
    $\jguhol{\Delta}{\Sigma}{\Gamma}{\Psi}{t}{\sigma}{\phi}$
  \item
    $\jghol{\Delta}{\Sigma}{\Gamma}{\Psi}{\phi\subst{\res}{t}}$
  \end{itemize}
\end{theorem}
The proof is analogous to the relational case

The previous result allows us to lift the soundness result from Guarded HOL to
Guarded UHOL.

\begin{corollary}[Soundness and consistency]\label{cor:uhol:sound}
If $\jguhol{\Delta}{\Sigma}{\Gamma}{\Psi}{t}{\sigma}{\phi}$, then for every valuations
$\delta \models \Delta$, $\gamma\models\Gamma$:
\[\sem{\Delta \vdash \Sigma}(\delta) \wedge \sem{\Delta \mid \Gamma \vdash \Psi}(\delta,\gamma) \Rightarrow
    \sem{\Delta \mid \Gamma,\res:\sigma \vdash \phi}(\delta, \gamma[\res \ot \sem{\Delta \mid \Gamma \vdash t}(\delta,\gamma)])\]
In particular, there is no proof of
$\jguhol{\Delta}{\emptyset}{\Gamma}{\emptyset}{t}{\sigma}{\bot}$ in Guarded UHOL.
\end{corollary}




\subsection{Probabilistic extension}

We comment on the rules, starting from the rules of the unary
logic. There are three new rules for the probabilistic case, and they
all establish that an expression $u$ of type $\Distr(D)$ satisfies the
assertion $\diamond_{[y\leftarrow \res]} \phi$, i.e.\, for every
element $v$ in the support of (the interpretation of) $u$, the
interpretation of $\phi$ with the valuation $[y \mapsto v]$ is
true. This intuition is captured by the rule $[\textsf{SUPP}]$, which
can be used in particular in case $u$ is a primitive distribution. The
rule [\textsf{UNIT}] considers the case where $u$ is of the form
$\munit{t}$; in this case, it is clearly sufficient to know that
$\phi[t/y]$ is valid. The rule [\textsf{MLET}] simply captures the
fact that the support of $\mlet{x}{t}{t'}$ is the disjoint union of
the support of $t'$ under all the assignments of $x$ to values in the
support of $t$.

\begin{figure*}[!htb]
\small

\textbf{Guarded UHOL}
\begin{mathpar}
\infer[\sf UNIT]
      {\jguhol{\Delta}{\Sigma}{\Gamma}{\Psi}{\munit{t}}{\Distr(C)}{
          \diamond_{[y\leftarrow \res]} \Phi}}
      {\jguhol{\Delta}{\Sigma}{\Gamma}{\Psi}{t}{C}{\Phi
          [\res/y]}}
    \and    
 \infer[\sf MLET]
    {\jguhol{\Delta}{\Sigma}{\Gamma}{\Psi}{\mlet{x}{t}{t'}}{
        \Distr(D)}{\diamond_{[y\leftarrow \res]} \psi}}
    {\jguhol{\Delta}{\Sigma}{\Gamma}{\Psi}{t}{\Distr(C)}{\diamond_{[x\leftarrow r]} \Phi}
     \and
     \jguhol{\Delta}{\Sigma}{\Gamma,x:C}{\Psi,\phi
     }{t'}{\Distr(D)}{\diamond_{[y\leftarrow \res]} \psi}}

 \infer[\sf SUPP]{\jguhol{\Delta}{\Sigma}{\Gamma}{\Psi}{u}{\Distr(D)}{
     \diamond_{[y\leftarrow \res]}\phi}}{\jghol{\Delta}{\Sigma}{\Gamma}{\Psi}{
     \Pr_{z\sim u}[\phi[z/y]]=1}}
 
\end{mathpar}

\caption{Proof rules for probabilistic constructs -- unary case}\label{fig:puhol}
\end{figure*}

Finally, we prove an embedding lemma for Guarded UHOL. The proof can be
carried by induction on the structure of derivations, or using the
equivalence between Guarded UHOL and Guarded HOL (Theorem~\ref{thm:equiv-uhol-hol}).
\begin{lem}[Embedding lemma]\label{lem:emb-uhol-rhol} Assume that:
\begin{itemize}
\item $\jguhol{\Delta}{\Sigma}{\Gamma}{\Psi}{t_1}{\sigma_1}{\phi}$
\item $\jguhol{\Delta}{\Sigma}{\Gamma}{\Psi}{t_2}{\sigma_2}{\phi'}$
\end{itemize}
Then
$\jgrhol{\Delta}{\Sigma}{\Gamma}{\Psi}{t_1}{\sigma_1}{t_2}{\sigma_2}{
  \phi\subst{\res}{\res\ltag}\land \phi'\subst{\res}{\res\rtag}}$.
\end{lem}

\subsection{Unary example: Every two}

We define the $every2$ function, which receives a stream and returns another stream consisting
of the elements at even positions in the input stream. Note that this function, while productive,
cannot be built with the type $Str \to Str$, since we need to take twice the tail of the argument,
which would have type $\later\later Str$, and then a $Str$ cannot be built. Instead, we need to use the constant modality
as follows:

\begin{align*}
every2 &: \square Str \to Str \\
every2 &\defeq \fix{every2}{\lambda s. \cons{\hat{hd}(\hat{tl}\ s)}{(every2 \app {\rm next} (\hat{tl}(\hat{tl}\ s)))}}
\end{align*}
Where the $\hat{hd}$ and $\hat{tl}$ functions are not the native ones, but rather they are defined as:
\begin{align*}
  \hat{hd} &: \square Str \to \nat         &\hat{tl} &: \square Str \to \square Str \\              
  \hat{hd} &\defeq \lambda s. \letbox{x}{s}{hd\ x} \quad\quad     &\hat{tl} &\defeq \lambda s. \letbox{x}{s}{\boxx{(\prev{(tl\ x)})}}
\end{align*}
The property we want to prove is:
\[
\jguhol{\cdot}{\cdot}{ones : \square Str}{\Psi}{every2}{\square Str \to Str}{\forall s. s = ones \Rightarrow \res\ s = (\letbox{x}{s}{x})}
\]
where $ones$ is the constant stream containing only the number 1 defined as:
\[
ones \defeq \boxx{(\fix{f}{\cons{1}{f}})}
\]
For which we can prove the following properties:
\[
\Psi \defeq \hat{hd}\ ones = 1, \hat{tl}\ ones = ones
\]
In the rest of the proof we omit the empty contexts $\Delta$ and $\Sigma$.
We start by applying the \rname{Fix} rule, which has the premise:
\[
\begin{array}{c}
\juhol{ones : \square Str, every2:\later{(\square Str \to Str)}}{\Psi, \later[\res \ot every2]{\forall s. s = ones \Rightarrow \res\ s = \letbox{x}{s}{x}}}
                     {\\ \lambda s. (\cdots)}{\square Str \to Str}{\forall s. s = ones \Rightarrow (\res\ s) = \letbox{x}{s}{x}}
\end{array}
\]
We apply the \rname{Abs} rule inmediately after:
\[
\begin{array}{c}
\juhol{ones : \square Str, every2:\later{(\square Str \to Str)}, s: \square Str}
       {\\ \Psi, \later[\res \ot every2]{\forall s. s = ones \Rightarrow \res\ s = \letbox{x}{s}{x}}, s = ones}
                     {\\ \cons{\hat{hd}(\hat{tl}\ s)}{(every2 \app \later (\hat{tl}(\hat{tl}\ s)))}}{Str}{\res = \letbox{x}{s}{x}}
\end{array}
\]

By \rname{SUB} and the equivalence $\letbox{x}{s}{x} \equiv \letbox{x}{ones}{x}$, we can change the
conclusion of the judgement.
Now we use the \rname{Cons} rule, which has three premises:
\begin{enumerate}
  \item $\juhol{ones : \square Str, every2:\later{(\square Str \to Str)}, s: \square Str}
               {\\ \Psi, \later[\res \ot every2]{\forall s. s = ones \Rightarrow \res\ s = \letbox{x}{s}{x}}, s = ones}
                             { \hat{hd}(\hat{tl}\ s)}{\nat}{\res = 1}$
\\
  \item $\juhol{ones : \square Str, every2:\later{(\square Str \to Str)}, s: \square Str}
               {\\ \Psi, \later[\res \ot every2]{\forall s. s = ones \Rightarrow \res\ s = \letbox{x}{s}{x}}, s = ones}
                             {\\(every2 \app \later(\hat{tl}(\hat{tl}\ s)))}{Str}{\dsubst{\res}{\res}{\res = \letbox{x}{ones}{x}}}$
\\
  \item $\jhol{ones : \square Str, every2:\later{(\square Str \to Str)}, s: \square Str}
             {\\ \Psi, \dsubst{\res}{every2}{\forall s. s = ones \Rightarrow \res\ s = \letbox{x}{s}{x}}, s = ones}
                             {\\ \forall y, ys. y = 1 \Rightarrow \later[zs \ot ys]{(zs = \letbox{x}{ones}{x})} \Rightarrow \cons{y}{ys} = (\letbox{x}{ones}{x})}$
    
\end{enumerate}

Premises (1) is a consequence of the properties of $ones$. To prove premise (3)
we reduce the letbox with the box inside $ones$, and do some reasoning using the definition of the fixpoint. To prove the
premise (2) we first desugar the term we are typing:
\[every2 \app \later(\hat{tl}(\hat{tl}\ s))) \defeq \later[g \leftarrow every2, t \leftarrow \later(\hat{tl}(\hat{tl}\ s))]{g t} \]
and then we apply \rname{Next} which has the following premises:
\begin{itemize}
  \item $\juhol{ones : \square Str, every2:\later{(\square Str \to Str)}, s: \square Str}
               {\\ \Psi, \dsubst{\res}{every2}{\forall s. s = ones \Rightarrow \res\ s = \letbox{x}{s}{x}}, s = ones}
                             {\\ every2}{\later{(\square Str \to Str)}}{\dsubst{\res}{\res}{(\forall s = ones \Rightarrow \res\ s = \letbox{x}{s}{x})}}$
\\
  \item $\juhol{ones : \square Str, every2:\later{(\square Str \to Str)}, s: \square Str}
               {\\ \Psi, \dsubst{\res}{every2}{(\forall s. s = ones \Rightarrow \res\ s = \letbox{x}{s}{x})}, s = ones}
                             {\\ \later(\hat{tl}(\hat{tl}\ s))}{\later\square Str}{\dsubst{\res}{\res}{(\res = ones)}}$
\\
  \item $\juhol{ones : \square Str, every2:\later{(\square Str \to Str)}, s: \square Str, g: \square Str \to Str, t: \square Str}
              {\\ \Psi, \dsubst{\res}{every2}{(\forall s. s = ones \Rightarrow \res\ s = \letbox{x}{s}{x})}, s = ones,
                \\ \forall s. s = ones \Rightarrow g\ s = \letbox{x}{s}{x}, t = ones}
                             {g\ t}{Str}{\res = (\letbox{x}{ones}{x})}$
    
\end{itemize}
The first premise is just an application of the \rname{Var} rule.  The
second premise can be proven as a consequence of the properties of
$ones$.  Finally, the third premise can be proven with some simple
logical reasoning in HOL.  This concludes the proof.



\end{document}